\documentclass[12pt,a4paper]{article}
\usepackage[T1]{fontenc}
\usepackage[utf8]{inputenc}
\usepackage{authblk}
\usepackage{bbding} 

\usepackage{amsmath}
\usepackage{graphicx,float,color}
\usepackage{hyperref}
\hypersetup{colorlinks = true, linkcolor = blue, urlcolor  = blue, citecolor = blue, anchorcolor = blue}
\usepackage{multirow}
\usepackage[numbers,sort&compress]{natbib}
\usepackage{braket}
\usepackage{amsfonts}
\usepackage{amssymb} 
\usepackage{epic}
\usepackage{eepic}
\usepackage{epsfig}
\usepackage{latexsym}
\usepackage{color}
\usepackage{url}  
\usepackage{caption}
\usepackage[caption=false]{subfig}
\usepackage{float}
\usepackage{bm}
\usepackage{geometry}
\usepackage{slashed}
\usepackage[detect-all]{siunitx}

\usepackage{authblk}
\usepackage{physics}
\usepackage{amsmath} 
\usepackage{amsfonts}
\usepackage{color}
\usepackage{url}
\usepackage{graphicx,float,color}
\usepackage{verbatim}

\def\lp{\ell_{\mathrm{Pl}}}

\def\snr{\mathrm{SNR}}
\def\uew{\mathrm{UniEw}}
\def\rshift{q_0}

\newcommand*{\ee}{\mathrm{e}}
\newcommand*{\ii}{\mathrm{i}}

\usepackage{lipsum}
\makeatletter
\def\ps@pprintTitle{%
 \let\@oddhead\@empty
 \let\@evenhead\@empty
 \def\@oddfoot{}%
 \let\@evenfoot\@oddfoot}
\makeatother
\usepackage{etoolbox}

\begin{document}
\title{\textbf{Model-independent search for the quasinormal modes of gravitational wave echoes}}    

\author[1,2]{Di Wu\thanks{wudi@ihep.ac.cn}}
\author[1]{Pengyuan Gao\thanks{pygao@ihep.ac.cn}}
\author[1]{Jing Ren\thanks{Corresponding author: renjing@ihep.ac.cn}}
\author[3,4,5]{Niayesh Afshordi\thanks{nafshordi@pitp.ca}}

\affil[1]{\normalsize Institute of High Energy Physics, Chinese Academy of Sciences, Beijing 100049, China}
\affil[2]{\normalsize School of Physics Sciences, University of Chinese Academy of Sciences, Beijing 100039, China}
\affil[3]{\normalsize Waterloo Centre for Astrophysics, University of Waterloo, Waterloo, ON, N2L 3G1, Canada}
\affil[4]{\normalsize Department of Physics and Astronomy, University of Waterloo, Waterloo, ON, N2L 3G1, Canada}
\affil[5]{\normalsize Perimeter Institute of Theoretical Physics, 31 Caroline St. N., Waterloo, ON, N2L 2Y5, Canada}
\maketitle
\begin{abstract}
Postmerger gravitational wave echoes provide a unique opportunity to probe the near-horizon structure of astrophysical black holes, which may be modified due to nonperturbative quantum gravity phenomena. However, since the waveform is subject to large theoretical uncertainties, it is necessary to develop search methods that are less reliant on specific models for detecting echoes from observational data.  
A promising strategy is to identify the characteristic quasinormal modes (QNMs) associated with echoes, {\it in frequency space}, which complements existing searches of quasiperiodic pulses in time. 
In this study, we build upon our previous work targeting these modes by incorporating relative phase information to optimize the Bayesian search algorithm. Using a new phase-marginalized likelihood, the performance can be significantly improved for well-resolved QNMs. 
This enables an efficient search for QNMs of various shapes, utilizing a simple search template that is independent of specific models.
To demonstrate the robustness of the search algorithm, we construct four complementary benchmarks for the echo waveform that span a diverse range of different theoretical possibilities for the near-horizon structure. We then validate our Bayesian search algorithms by injecting the benchmark models into different realizations of Gaussian noise. Using two types of phase-marginalized likelihoods, we find that the search algorithm can efficiently detect the corresponding QNMs. Therefore, our search strategy provides a concrete Bayesian and model-independent approach to ``quantum black hole seismology.'' 
\\

\end{abstract}

  \hypersetup{linkcolor=black}
  \tableofcontents
\section{Introduction}

Observations have yet to map the immediate vicinity of astrophysical black holes, and near-horizon corrections could be crucial in resolving associated theoretical problems through potential links to quantum gravity effects. From an observational standpoint, an additional structure outside the gravitational radius is likely to modify the ingoing boundary condition of the horizon, allowing for substantial reflection. This is expected to generate gravitational wave echoes in the postmerger stage of compact binary coalescence, providing a unique opportunity to probe the near-horizon structure around astrophysical black holes~\cite{Cardoso:2016rao,Cardoso:2016oxy}. For later discussion, we use the term ``ultracompact objects'' (UCOs) to refer to black hole mimickers that can produce echoes (for a review on UCOs, see, e.g. \cite{Cardoso:2019rvt, Abedi:2020ujo}).
Current gravitational wave observations have the potential to probe quantum gravity corrections around macroscopic UCOs through detection of echoes, thanks to the logarithmic dependence of its time delay on the position of the effective interior surface. 
While model-dependent search methods have been extensively explored~\cite{Nakano:2017fvh, Mark:2017dnq, Lo:2018sep, Wang:2019szm, Wang:2019rcf, Maggio:2019zyv,  Xin:2021zir, Ma:2022xmp, Vellucci:2022hpl} and applied to real data~\cite{Abedi:2016hgu, Nielsen:2018lkf, Uchikata:2019frs, Wang:2020ayy, LIGOScientific:2020tif, Westerweck:2021nue, Abedi:2021tti}, the challenge lies in the presence of waveform uncertainties  arising from the unknown details associated with UCOs. Therefore, it is necessary to develop model-independent search methods that target the characteristic features of echoes without relying on {\it ad hoc}  or model-specific details~\cite{LIGOWhite}. 

Echoes result from an effective reflection of gravitational waves inside the photon-sphere barrier but outside the gravitational radius. In the case of a weak reflection, echoes damp quickly with time, resulting in a quasiperiodic signal with only a few pulses. Such signals can be efficiently detected by morphology-independent search methods developed for bursts, such as the BayesWave algorithm~\cite{Tsang:2018uie,Tsang:2019zra} and the coherent WaveBurst pipeline~\cite{Abedi:2021tti,Miani:2023mgl}. In the case of a strong reflection, echoes are characterized by a large number of small pulses, rendering burst-based searches inefficient. However, in this limit, echoes in the frequency space exhibit well-separated narrow resonances that correspond to the quasiperiodic and long-lived quasinormal modes (QNMs) of UCOs, reminiscent of similar long-lived modes in the asteroseismology program (e.g., \cite{brown1994asteroseismology}) that probe the inner structure of stars \cite{Oshita:2020dox,Oshita:2020abc}. These characteristic QNMs then serve as robust search targets for echoes in frequency space, complementing the existing searches of quasiperiodic pulses in time. 

A uniform comb model with triangular teeth has been adopted to search for long-lived and quasiperiodic QNMs associated with echoes~\cite{Conklin:2017lwb,Holdom:2019bdv}. Recently, a Bayesian search algorithm was developed along the same line, employing a phase-marginalized likelihood~\cite{Ren:2021xbe}. In these studies, the focus is on the amplitude in the frequency space, with the phase information being discarded. Although this approach largely mitigates the theoretical uncertainties associated with the phase, it comes at the cost of reduced search sensitivity. Specifically, the absence of phase information makes it impossible to detect a signal that is significantly weaker than the noise level.

In this paper, we optimize the Bayesian search method presented in Ref.~\cite{Ren:2021xbe} by considering a more realistic treatment of the QNM phase. We find that a significant portion of the phase information can be retained without introducing additional parameters. As a result, we develop a new phase-marginalized likelihood that can significantly improve the search sensitivity for QNMs that are well resolved in the frequency space. 
To comprehensively address theoretical uncertainties, we present a generic description of the echo waveform in Sec.~\ref{sec:theory}, with a particular focus on formulating the waveform as a superposition of the characteristic QNMs of UCOs.
We then derive the new phase-marginalized likelihood, and compare with the old one introduced in~\cite{Ren:2021xbe}. In Sec.~\ref{sec:validate1}, we first define a simple search template, dubbed $\uew$, with a small number of parameters. We then validate the Bayesian search algorithm by conducting a consistent injection-recovery study of $\uew$ using the two likelihoods.
In Sec.~\ref{sec:validate}, we construct a few complementary benchmarks for the echo waveforms. We then define a more generic search procedure and validate the algorithm with injections of the benchmarks. We summarize in Sec.~\ref{sec:summary}. 

\section{Formalism}
\label{sec:theory}

\subsection{Generic construction of echo waveforms}
\label{sec:generic}

A sufficiently compact UCO can be visualized as a leaky cavity bounded by the light-ring potential barrier and an effective inner boundary, with the average cavity length $x_0$ and round-trip time $t_d\approx 2x_0$. The light-ring potential barrier has been well constrained through current observations of black hole ringdown signals (e.g., \cite{LIGOScientific:2020tif}), as well as radio imaging with the Event Horizon Telescope (e.g., \cite{EventHorizonTelescope:2019dse}). The interior boundary encodes the essential information for the near-horizon corrections. If linear perturbations are present inside or outside the cavity, echoes are produced by the initial pulse, which undergoes repeated and damped reflections, gradually leaking out of the light-ring potential barrier.

The merger product typically has a significant spin. For simplicity, we represent the spinning UCO as a truncated Kerr black hole, where the inner boundary is located at a radius of $r_0$ outside the would-be horizon radius of $r_+$~\cite{Conklin:2017lwb, Nakano:2017fvh, Wang:2019rcf, Maggio:2019zyv, Xin:2021zir, Conklin:2021cbc, Srivastava:2021uku}. When $r_0$ is very close to $r_+$, corresponding to an extremely large compactness, the time decay is roughly twice the length of the cavity, with~\cite{Cardoso:2019rvt} 
\begin{eqnarray}\label{eq:td}
\frac{t_d}{M}\approx 2\left[1+(1-\chi^2)^{-1/2}\right] \ln\left(\frac{1}{\epsilon}\right)\,,
\end{eqnarray}
where $M$ and $\chi$ are the mass and dimensionless spin of the merger product. The parameter $\epsilon=(r_0-r_+)/r_+$ quantifies the compactness of the UCO, or the distance from the surface $r_0$ to the would-be horizon $r_+=M(1+\sqrt{1-\chi^2})$. 
For the coordinate and proper Planck length distance, $\ln\epsilon\approx -\eta \ln (M/\lp) = -\left[87.4 +\ln(M/M_\odot)\right]\eta$, for $\eta=1$ and $2$, respectively.
It is important to note that in the case of a realistic UCO, the time delay in Eq.~(\ref{eq:td}) that scales with $\ln(1/\epsilon)$ might be underestimated if the interior of the UCO manifests a considerably deep gravitational potential~\cite{Zimmerman:2023hua}.

For a sufficiently small $\epsilon$, the linear gravitational wave (GW) perturbations of UCOs are governed by the Teukolsky equation with a modified boundary condition at $r_0$. For a given source, the response of UCOs at infinity can be related to the responses of the corresponding Kerr black holes (BHs) at infinity and horizon. In the absence of mode mixing, the GW strain is related to the Teukolsky variables in a simple way. The observation for UCOs then manifests as a sum of BH ringdown and echoes, i.e. $h_{\rm UCO}(\omega)=h_{\rm RD}(\omega)+h_{\rm echo}(\omega)$, with~\cite{Xin:2021zir} (see Appendix.~\ref{sec:formalism} for the derivations)
\begin{eqnarray}\label{eq:hecho}
h_{\rm echo}(\omega)=\mathcal{P}(\omega) h_{\rm eff}(\omega),\quad
\mathcal{P}(\omega)=\frac{R_{\rm BH }(\omega)R_{\rm wall }(\omega)}{1-R_{\rm BH }(\omega)R_{\rm wall }(\omega)}\,.
\end{eqnarray}
The waveform is essentially determined by two ingredients: the BH response at the horizon $h_{\rm eff}(\omega)$ and the processing function $\mathcal{P}(\omega)$, which relies on the reflection coefficients of the light-ring potential barrier for waves coming from the left $R_{\rm BH }(\omega)$ and the reflection coefficient for the interior boundary $R_{\rm wall}(\omega)$. 
As expected, $h_{\rm UCO}$ reduces to $h_{\rm RD}$ with zero interior reflection, i.e. $R_{\rm wall}=0$. 
In the following, we discuss these two ingredients in more detail. 

The BH response at the horizon, $h_{\rm eff}(\omega)$, contains information about the source and determines the initial pulse profile. In certain cases, this response is closely related to the black hole ringdown waveform, $h_{\rm RD}(\omega)$. Specifically, as demonstrated in Appendix~\ref{sec:formalism}, if the source is an outgoing pulse originating from within the potential barrier or an incoming pulse originating from outside the potential barrier, then we find $h_{\rm eff}(\omega)=h_{\rm RD}(\omega)$ and $h_{\rm eff}(\omega)=-(\mathcal{T}_{\rm BH}^2/\mathcal{R}_{\rm BH}^2)h_{\rm RD}(\omega)$, respectively. $\mathcal{T}_{\rm BH}^2$ and $\mathcal{R}_{\rm BH}^2$ denote the transmission and reflection of energy flux for the light-ring potential barrier. This yields the ``inside'' and ``outside'' prescriptions of the echo waveform in the geometric ``optics'' picture~\cite{Wang:2019rcf}, i.e.
\begin{eqnarray}\label{eq:hechogeo}
h_{\rm echo}(\omega) \approx \mathcal{P}(\omega)
\left\{
\begin{array}{ll}
h_{\rm RD}(\omega)\,, & \textrm{inside}\\
-\frac{\mathcal{T}_{\rm BH}^2}{\mathcal{R}_{\rm BH}^2}h_{\rm RD}(\omega), & \textrm{outside}
\end{array}\right.\,.
\end{eqnarray}
The ringdown waveform can be modeled as the excitation of the dominant $l=m=2$ fundamental mode~\cite{Maggio:2019zyv}, i.e., $h_{\rm RD}(\omega) 
=A_{\rm RD}e^{i \delta_{\rm RD}}/(\omega-\omega_{\rm RD}+i/\tau_{\rm RD})+A_{\rm RD}e^{-i \delta_{\rm RD}}/(\omega+\omega_{\rm RD}-i/\tau_{\rm RD})$ \footnote{We only consider the plus polarization in our waveform model. See Eq. \eqref{eq:ZBH} in Appendix~\ref{sec:formalism} for the case of including both polarizations.}, which peaks around the ringdown frequency $\pm \omega_{\rm RD}$.
Considering that the coefficient $\mathcal{R}_{\rm BH}$ approaches unity for frequencies below $\omega_{\rm RD}$ and that the ratio $\mathcal{T}_{\rm BH}^2/\mathcal{R}_{\rm BH}^2$ is suppressed in this frequency regime, the signal obtained using the outside prescription would be significantly smaller than the inside  one.

More generally, $h_{\rm eff}(\omega)$ may not be closely related to the BH ringdown signal due to the additional contribution induced by the source. 
For example, a numerical study of echo waveforms for infalling particles~\cite{Xin:2021zir} reveals that $h_{\rm eff}$ could lie somewhere between the two prescriptions presented in Eq.~(\ref{eq:hechogeo}). 
Moreover, in the case of comparable-mass UCOs, the merger could generate additional perturbations that are independent of those originating from the light-ring potential barrier, owing to the significant time dilation associated with the inner boundary. As a result, the echo waveform could differ significantly from the predictions outlined in Eq.~(\ref{eq:hechogeo}). These findings highlight the considerable uncertainties associated with the source and underscore the limitations of the geometric optics picture.

The processing function, $\mathcal{P}(\omega)$, is intimately dependent on the combined reflectivity, $R_{\rm BH }(\omega)$ $R_{\rm wall}(\omega)$, of the cavity. For later discussion purposes, it is convenient to separate the amplitude and phase of the product so that $R_{\rm BH }(\omega)R_{\rm wall}(\omega)\equiv \mathcal{R}_{\rm eff}(\omega)e^{i \delta(\omega)}$. 
The QNMs of UCOs correspond to the zeros of the denominator of $\mathcal{P}(\omega)$ in the complex plane, i.e.
\begin{eqnarray}\label{eq:QNMeq}
1-\mathcal{R}_{\rm eff}(\omega)e^{i \delta(\omega)}=0,\; \textrm{for } \omega=\omega_R+i \omega_I\,.
\end{eqnarray}  
The square of amplitude $\mathcal{R}^2_{\rm eff}(\omega)$ is the product of the reflectivities of the energy flux on the two surfaces. To avoid ergoregion instability, a stable UCO must satisfy $\mathcal{R}_{\rm eff}\leq 1$. 
The phase, $\delta(\omega)$, encodes the dependence of the propagating distance.  For extremely compact UCOs, where the time delay $t_d$ is much longer than the typical timescale of the system $\sim M$, it is a reasonable approximation to take $\delta(\omega)\approx t_d \, \tilde\omega$, where $\tilde\omega=\omega-\omega_H$ denotes the wave frequency close to the inner boundary, with $\omega_H=m \Omega_H$ and $\Omega_H=\chi/(2r_+)$ the angular frequency of the horizon.
Assuming $|\omega_I|\ll|\omega_R|$ and $|d\mathcal{R}_{\rm eff}/d\omega|\ll t_d$, Eq.~(\ref{eq:QNMeq}) can be solved analytically with $(\omega_R,\omega_I)=(\omega_n, -1/\tau_n)$ for a series of integer $n$
\begin{eqnarray}\label{eq:QNMn}
t_d(\omega_n-\omega_H)\approx 2\pi n,\quad
t_d/\tau_n\approx -\ln \mathcal{R}_{\rm eff}(\omega_n)\,.
\end{eqnarray}
These are exactly the trapped modes of a long cavity with sufficiently strong reflection. 
The oscillation frequency is quasiperiodic, with the spacing between two modes roughly equal to the inverse of time delay, that is, $\omega_{n+1}-\omega_n\approx 2\pi\Delta f$, where $\Delta f=1/t_d$. The ratio of mode width, $1/\tau_n$, to spacing, $\Delta f$, is determined by the rate of energy dissipation. When $\mathcal{R}_{\rm eff}(\omega_n)$ is close to one,  $1/\tau_n$ is considerably smaller than $\Delta f$. Consequently, the echo waveform in the frequency space is characterized by a quasiperiodic and well-separated resonance pattern. 
Under the same approximation, the processing function around the simple pole $\omega_n-i/\tau_n$ can be expanded as 
\begin{eqnarray}\label{eq:Kexp}
 \mathcal{P}(\omega)\approx \frac{  \mathcal{R}_{\rm eff}(\omega_n)}{-i t_d}\frac{e^{i t_d \omega}}{\omega-\omega_n+i/\tau_n}+...\,,
\end{eqnarray}
and the amplitude at $\omega=\omega_n$ is given by $| \mathcal{P}(\omega_n)|= \mathcal{R}_{\rm eff}(\omega_n)/|\ln\mathcal{R}_{\rm eff}(\omega_n)|$.

In the absence of ergoregion instability, $\mathcal{P}(\omega)$ can be expressed as an infinite sum, i.e. $\mathcal{P}(\omega)=\sum_{k=1}^\infty R^k_{\rm BH }(\omega)R^k_{\rm wall }(\omega)$. The echo waveform is then given by 
\begin{eqnarray}\label{eq:hechosum}
h_{\rm echo}(\omega)=\sum_{k=1}^\infty h_{\rm eff}(\omega)\mathcal{R}^k_{\rm eff}(\omega)e^{i k t_d \tilde\omega}\,.
\end{eqnarray} 
For early pulses with distinct shapes, the $n$-th term in the infinite sum corresponds to the $n$-th pulse in the echo waveform, with, for example, $h_{\rm eff}(\omega)\mathcal{R}_{\rm eff}(\omega)e^{i t_d \tilde\omega}$ being the frequency content of the first pulse.

In practice, we typically analyze a finite segment of strain data. The corresponding waveform in the frequency domain is then described by a finite sum in Eq.~(\ref{eq:hechosum}), which is truncated at the number of pulses, $N_E=T/t_d$, i.e. 
\begin{eqnarray}\label{eq:hechosumT}
h_{\rm echo}^{(T)}(f)&=&h_{\rm eff}(f)\sum_{k=1}^{N_E} \mathcal{R}^k_{\rm eff}(f)e^{i k t_d \tilde\omega}
=h_{\rm eff}(f)\mathcal{R}_{\rm eff}(f)e^{i t_d \tilde\omega}\frac{1-\mathcal{R}^{N_E}_{\rm eff}(f)e^{i N_E t_d \tilde\omega}}{1-\mathcal{R}_{\rm eff}(f)e^{i t_d \tilde\omega}}\nonumber\\
&\approx& 
A_n e^{i\delta_n}e^{i 2\pi f t_d }
\frac{1-e^{-T/\tau_n}e^{i 2\pi (f-f_n) T}}{2\pi(f-f_n)+i/\tau_n},
\quad \textrm{for\;} f\sim f_n\,,
\end{eqnarray} 
where $f=\omega/(2\pi)$ and $f_n=\omega_n/(2\pi)$. We expand around the $n$-th mode in the second line, using the analytical approximation in Eq.~(\ref{eq:QNMn}) and $T\approx N_E t_d$. The amplitude and overall phase are given by $A_n\approx|h_{\rm eff}(f_n)|\mathcal{R}_{\rm eff}(f_n)/t_d$ and
$\delta_n\approx \arg(h_{\rm eff}(f_n))-2\pi f_n t_d+\pi/2$, respectively. The factor $1-e^{-T/\tau_n}e^{i 2\pi (f-f_n) T}$ represents the finite duration correction.\footnote{As demonstrated in Appendix~\ref{eq:echoQNMtime}, this exactly matches the finite time duration effects for the Fourier transform in Eq.~(\ref{eq:hfQNMT}).} The correction is negligible for the fully resolved mode (i.e., $T\gg \tau_n$), while it smears out the resonance when the sampling resolution is insufficient (i.e., $t_d\ll T\lesssim \tau_n$). 
The amplitude of resonance peak at $f=f_n$ is given by
\begin{eqnarray}\label{eq:hTuplimit}
|h_{\rm echo}^{(T)}(f_n)|\approx  |h_{\rm echo}^{(1)}(f_n)|\frac{1-e^{-T/\tau_n}}{|\ln \mathcal{R}_{\rm eff}(f_n)|}\,,
\end{eqnarray}
where $|h_{\rm echo}^{(1)}(f_n)|=|h_{\rm eff}(f_n)|\mathcal{R}_{\rm eff}(f_n)$ denotes the first pulse profile of echoes. 
Since the discrete Fourier transform may not sample exactly at the peak position, Eq.~(\ref{eq:hTuplimit}) depicts the envelope that bounds all the reconstructed resonances from above. Compared to the amplitude of the first pulse, the envelope of resonances is significantly enhanced.  For well-resolved modes, i.e., $T\gg \tau_n$, $|h_{\rm echo}^{(T)}(f_n)|$ is enhanced by a factor of $1/|\ln \mathcal{R}_{\rm eff}(f_n)|\approx \tau_n/t_d$. In the low-resolution limit, i.e., $t_d\ll T\ll \tau_n$,  the enhancement factor becomes $T/t_d$. Thus, as a good approximation, we have $|h_{\rm echo}^{(T)}(f_n)|/|h_{\rm echo}^{(1)}(f_n)| \approx \min\{\tau_n, T\}/t_d$  for a generic $T$. 
This also determines the improvement of the optimal signal-to-noise ratio (SNR) of $h_{\rm echo}^{(T)}$ relative to that of $h_{\rm echo}^{(1)}$, as given in Eq.~(\ref{eq:SNRopt1}). 
Note that the above estimates are no longer precise when $\mathcal{R}_{\rm eff}$ is considerably below one. For the wider modes, the envelope can be approximated by adding a factor of $ |h_{\rm eff}(f_n)|\mathcal{R}_{\rm eff}(f_n)$ on top of  Eq.~(\ref{eq:hTuplimit})~\cite{Conklin:2017lwb}.

Now, let us examine the expansion around the $n$-th mode in more detail. For $f_n-\mathcal{O}(1)/\tau_n\lesssim f\lesssim f_n+\mathcal{O}(1)/\tau_n$, the amplitude and phase are given by
\begin{eqnarray}
	\abs{h_{\rm echo}^{(T)}(f)} & \approx &  \frac{A_n}{\sqrt{4\pi^2(f-f_n)^2+1/\tau^2_n}} \abs{1-\ee^{-T/\tau_n}\ee^{\ii 2\pi (f-f_n)T}} \label{comb_abs} \label{eq:abshf}\\
	\arg \qty(h_{\rm echo}^{(T)}(f) ) & \approx &\arg \qty (\frac{1-\ee^{-T/\tau_n}\ee^{\ii 2\pi (f-f_n)T}}{\ii 2\pi(f-f_n)+ 1/\tau_n})
	+  \left(\delta_n+ 2\pi f t_d\right) \,. \label{eq:arghf}
\end{eqnarray}
The amplitude takes on a Lorentzian shape with a smearing factor and is described by four parameters $\{A_n,f_n,\tau_n,T\}$, which encode the essential information about $t_d$, $\mathcal{R}_{\rm eff}(f)$ and $h_{\rm eff}(f)$.
The phase has been separated into two parts. The first term comes exactly from the Lorentzian shape and is determined by the same parameters as the amplitude. The second term depends on the overall phase $\delta_n$, as well as the start time $t_d$. As a result, it is more affected by theoretical uncertainties.

Therefore, in order to improve the search for QNMs, we retain the first part of the phase to enhance the detection probability. As for the additional contributions, the term $2\pi f t_d$ varies more slowly with frequency compared to the first part and can be treated as a constant.\footnote{For $f-f_n\in [-1/\tau_n,1/\tau_n]$, $2\pi (f-f_n) t_d\sim t_d/\tau_n\ll 1$ for the long-lived QNMs, while the first phase term varies by $\sim \pi$.} 
In a more general scenario, there might be corrections associated with the interference of QNMs, leading to additional variations on top of Eq.~(\ref{eq:arghf}). To simplify matters, we choose to retain only the Lorentzian shape contribution and treat the rest as a constant.  
This motivates the development of our new phase-marginalized likelihood below, which accounts for the marginalization of a single constant phase for each mode. 
In Sec.~\ref{sec:BayesianUEW} and Appendix~\ref{eq:echoQNMtime}, we verify the assumption of a constant phase and find that the algorithm's performance is minimally affected by these intricacies.


\subsection{Two phase-marginalized likelihoods}
\label{sec:likelihood}

To derive the explicit forms of phase-marginalized likelihoods, we start from the Gaussian likelihood in polar coordinate
\begin{equation} 
\mathcal{L}(d_j | h_j)=\frac{|d_j|}{2\pi P_j}\exp\left(-\frac{1}{2}\frac{|d_j|^2+|h_j|^2}{\tilde{P}_j}\right)
\exp\left(\frac{|d_j||h_j|}{\tilde{P}_j} \cos(\phi_{j}-\psi_{j}) \right)\,,
\end{equation}
where $d_j$ and $h_j$ denote the data and signal model in the $j$-th frequency bin, and $\phi _{j}=\arg(d_j)$, $\psi _{j}=\arg(h_j)$ denote their phases respectively. $\tilde{P}_j = P_j/4\delta f$ is the normalized one-sided power spectral density (PSD), with $\delta f=1/T$ as the frequency resolution. 
In Ref.~\cite{Ren:2021xbe}, we marginalize the signal model phase in each bin separately with a flat prior $\pi(\psi_j)=1/(2\pi)$, and the resulting marginalized likelihood in each bin is given by
\begin{eqnarray}	
	\mathcal{L}(d_j | |h_j|) &\equiv& \int _0^{2\pi} \mathcal{L}(d_j | h_j) \pi(\psi_j) \dd \psi_j \nonumber\\
	&=& \frac{|d_j|}{2\pi P_j}\exp\left(-\frac{1}{2}\frac{|d_j|^2+|h_j|^2}{\tilde{P}_j}\right)
\int _0^{2\pi} \exp\left(\frac{|d_j||h_j|}{\tilde{P}_j} \cos(\phi_{j}-\psi_{j}) \right) \pi(\psi_j) \dd \psi_j \nonumber\\
	&=& \frac{|d_j|}{2\pi P_j}\exp \qty(-\frac{1}{2} \frac{|d_j|^2+|h_j|^2}{\tilde{P}_j}) I_0 \qty(\frac{|d_j||h_j|}{\tilde{P}_j})\,,
\end{eqnarray}
where $I_0$ is the zeroth-order modified Bessel function of the first kind. By combining all frequency bins, the marginalized log-likelihood normalized by the noise contribution is obtained as follows~\cite{Ren:2021xbe}
\begin{eqnarray}\label{eq:oldlnL}
\ln \mathcal{L}_\textrm{old}&\equiv&\sum_{j} \left[\ln \mathcal{L}(d_j| |h_j|)- \ln \mathcal{L}(d_j| 0)\right] \nonumber\\
&=&\sum_{j} \ln I_0 \qty(\frac{|d_j||h_j|}{\tilde{P}_j}) -\frac{1}{2} \sum_{j} \frac{|h_j|^2}{\tilde{P}_j}\,.
\end{eqnarray}
We denote this as the \emph{old} likelihood for later comparison. The second term is the conventional optimal signal-to-noise ratio (SNR) term: $\snr^2=\sum_{j} |h_j|^2/\tilde{P}_j$. The first term reflects the main effect of phase marginalization. Without phase information, a log-Bessel function acts on the overlap of the absolute values of signal and data. As a result, the sensitivity of the old likelihood depends mainly on $\snr^2$ per frequency bin, that is, $\snr^2_{\rm bin}$. The overlap term $|d_j||h_j|/\tilde{P}_j\approx |h_j|^2/\tilde{P}_j$ when $h_j$ dominates over the noise. However, the signal sensitivity is lost when $|h_j|^2/\tilde{P}_j\lesssim 1$. 

Next, we move to the more refined treatment of the phase in Eq.~(\ref{eq:arghf}). To deal with the unwanted phase for the $n$ th QNM, we first combine the likelihood of all frequency bins within the range $f_n-\Delta f/2\lesssim f_j\lesssim f_n+\Delta f/2$, that is, for $j\in \mathcal{J}_n\equiv [\lceil\frac{1}{\delta f}(f_n-\Delta f/2)\rceil, \lfloor\frac{1}{\delta f}(f_n+\Delta f/2)\rfloor]$. We then marginalize over the second phase term $\delta'_n\equiv \delta_n+2\pi f_n t_d $ by treating it as a constant. The combined likelihood for the $n$-th mode is given by
\begin{equation}
	\begin{aligned}
		 \mathcal{L}_n(d| h) & \equiv \int \qty(\prod_{j \in \mathcal{J}_n} \mathcal{L}(d_j | h_j)) \pi(\delta'_n) \dd \delta'_n\\
		& =\qty( \prod_{j \in \mathcal{J}_n} \frac{|d_j|}{2\pi P_j}) 
		\exp \qty(\sum_{j \in \mathcal{J}_n}-\frac{1}{2} \frac{|d_j|^2+|h_j|^2}{\tilde{P}_j}) 
		\int_0^{2\pi} \exp \qty(\sum_{j \in \mathcal{J}_n} \frac{|d_j||h_j|}{\tilde{P}_j} \cos(\varphi_{j}-\delta'_{n}) ) \pi(\delta'_n) \dd \delta'_n \\
		& = \qty( \prod_{j \in \mathcal{J}_n} \frac{|d_j|}{2\pi P_j}) 
		\exp \qty(\sum_{j \in \mathcal{J}_n}-\frac{1}{2} \frac{|d_j|^2+|h_j|^2}{\tilde{P}_j}) 
		I_0 \qty(\abs {\sum_{j\in \mathcal{J}_n}\frac{|d_j||h_j|}{\tilde{P}_j} \ee ^{-\ii \varphi_j}} ) \\
		& = \qty( \prod_{j \in \mathcal{J}_n} \frac{|d_j|}{2\pi P_j}) 
		\exp \qty(\sum_{j \in \mathcal{J}_n}-\frac{1}{2} \frac{|d_j|^2+|h_j|^2}{\tilde{P}_j}) 
		I_0 \qty(\abs {\sum_{j\in \mathcal{J}_n}\frac{d_j h_j^*}{\tilde{P}_j} } )
	\end{aligned}
\end{equation}
where $\varphi_j =\phi_j -(\psi_j-\delta'_n)$, $\psi_j-\delta'_n$ is the relative phase of $h$ and $h^*_j$ is the complex conjugate of $h_j$.
The integral is simplified with $\int_0^{2\pi} \exp \qty(\sum_i a_i \cos(x+b_i)) \dd x 		
= 2\pi I_0 \qty(\abs{\sum_{i}a_i \ee^{i b_i}})$ for $a_i,b_i \in \mathbb{R}$. 
Thus, the \emph{new} marginalized log-likelihood normalized by the noise contribution is given by, 
\begin{eqnarray}\label{eq:newlnL}
\ln \mathcal{L}_\textrm{new}&\equiv &\sum_{n} \left[\ln \mathcal{L}_n(d| h)- \ln \mathcal{L}_n(d|0)\right] \nonumber\\
&=&\sum_{n} \ln I_0 \qty(\abs {\sum_{j\in \mathcal{J}_n}\frac{d_j h_j^*}{\tilde{P}_j} } ) -\frac{1}{2} \sum_{j} \frac{|h_j|^2}{\tilde{P}_j}\,.
\end{eqnarray}
Compared to the old likelihood in Eq.~(\ref{eq:oldlnL}) used in Ref.~\cite{Ren:2021xbe}, the new likelihood in Eq.~(\ref{eq:newlnL}) coherently combines different frequency bins belonging to one QNM, and the log-Bessel function acts on the coherent sum. With the information of relative phases maintained, the new likelihood depends mainly on $\snr^2$ per mode, that is, $\snr_{\rm mode}^2$. For the $n$-th mode, the overlap term is $\sum_{j\in \mathcal{J}_n} d_j h_j^*/\tilde{P}_j\approx \sum_{j\in \mathcal{J}_n} |h_j|^2/\tilde{P}_j=\snr_n^2$. When $T\gg \tau_n$, $\snr_n^2$ could be considerably larger than $\snr_j^2$ for a well-resolved QNM. Thus, the new likelihood allows for the detection of these modes at the large $T$ limit.


\section{Bayesian search for $\uew$ injections}
\label{sec:validate1}

Very compact UCOs may have a large number of long-lived QNMs that can be excited in the postmerger stage from the master equation in Eq.~(\ref{eq:hecho}).
Under the approximation of one-mode dominance, the echo waveform around each peak resonance is described by at least three parameters, i.e. the height, position, and width, even if the unwanted phases are discarded. 
Thus, a detailed modeling of the full spectrum would still include many search parameters and may suffer strongly from the Occam penalty for Bayesian searches.
However, it may suffice to capture the dominant modes within a narrower frequency range for an efficient and model-independent search.
For this purpose, we consider in Sec.~\ref{sec:uniformmodel} a simple periodic and uniform model of the echo waveform, dubbed $\uew$, as the search templates. We then validate the search algorithm with the $\uew$ injections in Gaussian noise by considering $\mathcal{N}=100$ different noise realizations to account for noise uncertainties. Sec.~\ref{sec:simpleUEW} presents a simple comparison between the two likelihoods. Sec.~\ref{sec:BayesianUEW} examines various factors that may influence the performance of the Bayesian search.  

\subsection{Uniform echo waveform ($\uew$)} 
\label{sec:uniformmodel}

Under the approximation of one-mode dominance (i.e., distinct and well-separated QNMs), we define a simple periodic and uniform model of echo waveform ($\uew$) within a frequency band $[f_{\rm min}, f_{\rm max}]$ as 
\begin{eqnarray}\label{eq:uniformQNM}
h_{\uew}(f) = \sum_{n=N_{\rm min}}^{N_{\rm max}}
\left\{\begin{array}{ll}
 A \frac{1-\ee^{-T/\tau}\ee^{\ii 2\pi (f-f_n)T}}{2\pi(f-f_n)+i/\tau},& |f-f_n|\leq f_\textrm{cut}\\
 0, & |f-f_n|> f_\textrm{cut} \\
 \end{array}\right.\,,
\end{eqnarray}
where $f_n = \Delta f(n +\rshift)$, $N_{\rm min}=\lceil f_{\rm min}/\Delta f\rceil$ and $N_{\rm max}=\lfloor f_{\rm max}/\Delta f\rfloor$.\footnote{Specifically, we define $f_{\rm cut}=\min \left \{\max\left [ \sqrt{6} / (\pi \tau), 3/T \right ], \Delta f/2 \right \}$, which is upper bounded by $\Delta f/2$. In the high-frequency resolution limit, it corresponds to cutting at 20\% of the peak height. In the low resolution limit, we retain six frequency bins.} $N\equiv N_{\rm max}-N_{\rm min}+1$ denotes the number of QNMs within the frequency band. For frequency bins sufficiently close to the peak, this accounts for the amplitude in Eq.~(\ref{eq:abshf}) and the first phase term in Eq.~(\ref{eq:arghf}). 
The second part $\delta'_n\equiv\delta_n+2\pi f_n t_d$ is approximated as a constant and marginalized over in the new likelihood Eq.~(\ref{eq:newlnL}).  
Data points away from the peak are set to zero to guarantee the one-mode dominance in the simple model.
For UCOs with strong reflection and large compactness, the $\uew$ model provides a leading-order approximation to the echo waveform if the amplitude and width of QNMs vary slowly with frequency. This is possible if the initial frequency content and interior reflection vary more slowly than that of the Lorentzian shape around the pole.

The simple model $\uew$ is fully specified by seven parameters: 
\begin{eqnarray}\label{eq:parameters}
\Delta f,\;
\rshift,\;
A,\;
\tau,\;
T,\;
f_{\rm min},\;
f_{\rm max}\,,
\end{eqnarray}
denoting the spacing, relative shift, amplitude, damping time, time duration, and minimum and maximum frequency, respectively. The mode number is given by $N=\lceil(f_{\rm max}-f_{\rm min})/\Delta f \rceil$. 
These parameters capture the essential features of QNMs associated with echoes, and provide an excellent estimate of the average spacing, height, width, and frequency range of the dominant QNMs. 

The preferred values of the parameters in Eq.~(\ref{eq:parameters}) are well motivated by the physics behind the UCOs. 
The highest frequency scale is set by the black hole ringdown frequency ($l=m=2$)
\begin{eqnarray}\label{eq:fRD}
M f_{\rm RD}=0.243-0.184(1-\chi)^{0.129}\,.
\end{eqnarray}
Above $f_{\rm RD}$, the reflectivity of light-ring potential barrier $\mathcal{R}_{\rm BH}$ is strongly suppressed, and the QNMs are no longer long-lived. 
The average spacing $\Delta f$ of the quasiperiodic pattern is roughly the inverse of time delay $t_d$ in Eq.~(\ref{eq:td}), with
\begin{eqnarray}\label{eq:Deltaf}
M \Delta f \approx \frac{\bar{R}}{\eta},\quad
\bar{R}= \frac{0.00572}{1+(1-\chi^2)^{-1/2}}\,.
\end{eqnarray}
With $\eta\sim \mathcal{O}(1)$, there is a hierarchy between $f_{\rm RD}$ and $\Delta f$ due to the large redshift close to the would-be horizon. Their radio then indicates the maximal number of long-lived QNMs to search for, i.e. $N\lesssim f_{\rm RD}/\Delta f\sim 30\eta-60\eta$, which ranges from a few tens to hundreds. 
The narrow width of QNMs constitutes another important feature of our search target. 
At $\omega\lesssim \omega_H$, the combined reflectivity $\mathcal{R}_{\rm eff}$ could approach one sufficiently close, and the width-to-spacing ratio $1/(\tau_n\Delta f)$ could be considerably smaller than one from Eq.~(\ref{eq:QNMn}). This then produces another hierarchy. 
Since the new likelihood Eq.~(\ref{eq:newlnL}) would not be deteriorated by the increasing time duration $T$, it is good to choose a sufficiently large $T$ to increase the possibility of probing a strong reflecting interior surface. Considering the limitation of computational resources, we manually scan over a list of $T$  for a practical Bayesian search. In this paper, we take $\tau_n \Delta f$ and $T \Delta f$ to be a few tens to hundreds below for demonstration purposes.

\subsection{Simple comparison of the two likelihoods} 
\label{sec:simpleUEW}

To get some insight, we first compare the two likelihoods by considering a simple search of the maximum log-likelihood by varying the amplitude $A$, where all other parameters are fixed as the injected values. More explicitly, we search for $A_\textrm{max}$ that maximizes the log-likelihood in Eqs.~(\ref{eq:oldlnL}) and (\ref{eq:newlnL}) for each noise realization, respectively, and then produce a distribution of the maximum log-likelihood $\ln\mathcal{L}_\textrm{max}$ and the corresponding amplitude $A_\textrm{max}$ for different noise realizations. The distributions peak around zero for small signals, and are approximately Gaussian for sufficiently large signals. 
Below we compare the two likelihoods from different perspectives.

\begin{figure}[!h]
	\centering
	\includegraphics[width=7.5cm]{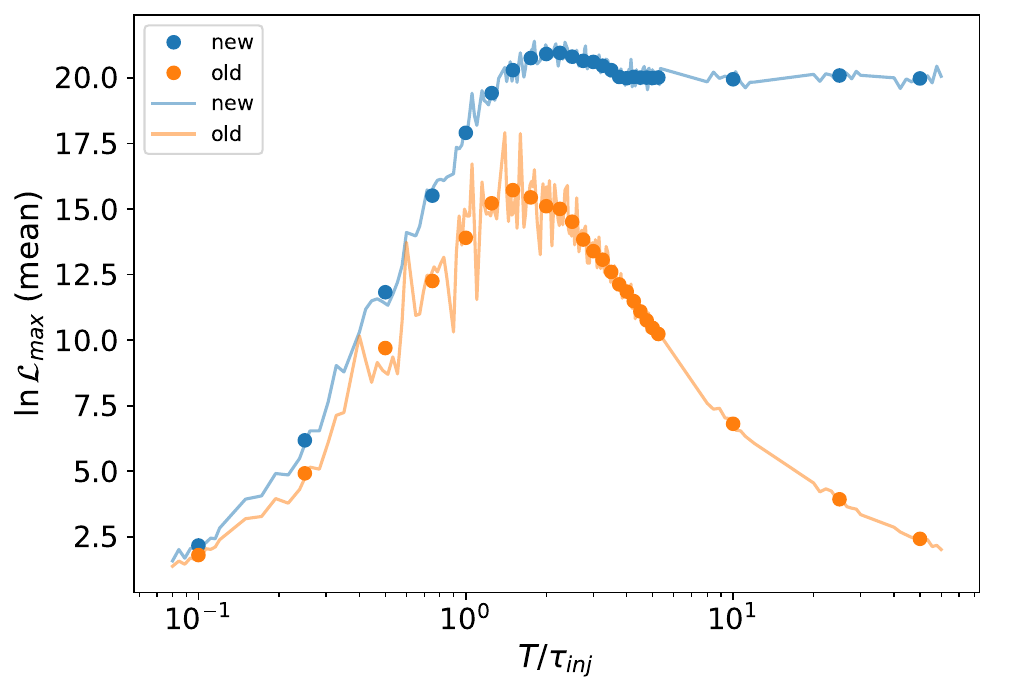}
	\includegraphics[width=7.5cm]{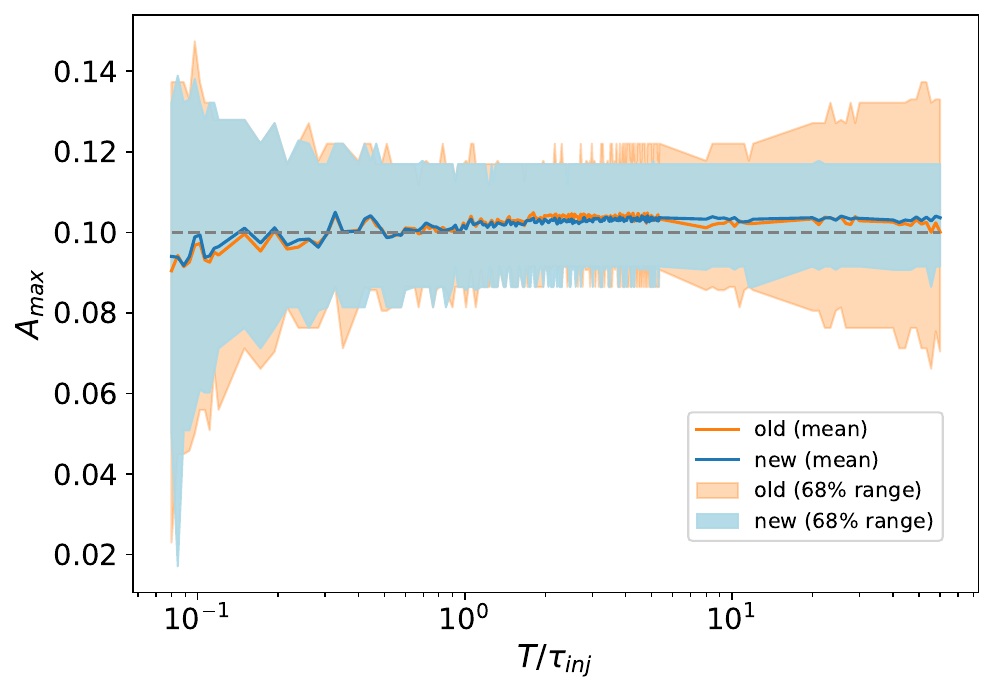}	
	\caption{\label{fig:TLoglikelihood} {The time duration $T$ dependence of the old likelihood (orange) and the new likelihood (blue) for the injected toy signal with $\snr=7.3$. Left: the mean value of the maximum log-likelihood. Right: the mean value (solid) and  the range of 68\% (shaded) of the search amplitude $A_\textrm{max}$ compared with $A_\textrm{inj}$ (dashed). }}
\end{figure}

Figure~\ref{fig:TLoglikelihood} compares the time duration dependence for a given injected signal amplitude. The top row shows the dependence for the maximum log-likelihood $\ln\mathcal{L}_\textrm{max}$. The performances of the two likelihoods are similar at low-frequency resolution, i.e. $T/\tau_{\rm inj}\lesssim 1$, while they differ significantly when the resonance gets well resolved, i.e. $T/\tau_{\rm inj}\gtrsim 1$. This is expected since the difference of Eqs.~(\ref{eq:oldlnL}) and (\ref{eq:newlnL}) comes down to the number of frequency bins contributing to one QNM. In the low-frequency resolution limit, the signal $\snr\propto T^{1/2}$ due to the exponential factor in Eq.~(\ref{eq:hfQNMT}) and so the mean of $\ln\mathcal{L}_\textrm{max}$ is proportional to $T$ approximately. 
When $T/\tau_{\rm inj}\gtrsim 1$, $\snr$ per mode approaches the continuous limit, and the new likelihood becomes quite insensitive to $T$. On the other hand, SNR per bin scales as $1/\sqrt{T}$, and the old likelihood drops quickly with increasing $T$. 
The relative error of $\ln\mathcal{L}_\textrm{max}$ is inversely related to the mean, which reaches the minimum around $T/\tau_{\rm inj}\sim 1$ for the old likelihood and  approaches a constant when $T/\tau_{\rm inj}\gg 1$ for the new likelihood.  
The distributions for the search amplitudes $A_\textrm{max}$ are more similar for the two likelihoods. The mean is around the injected value, and the error does not vary much.

\begin{figure}[!h]
	\centering
	\includegraphics[width=7.5cm]{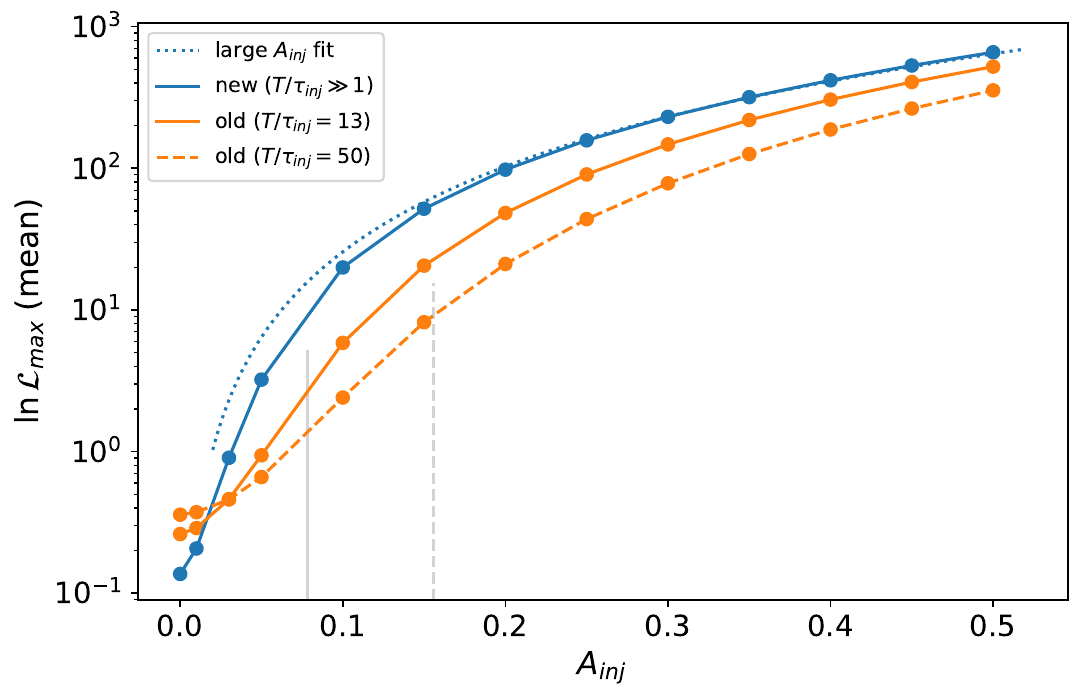}
	\includegraphics[width=7.5cm]{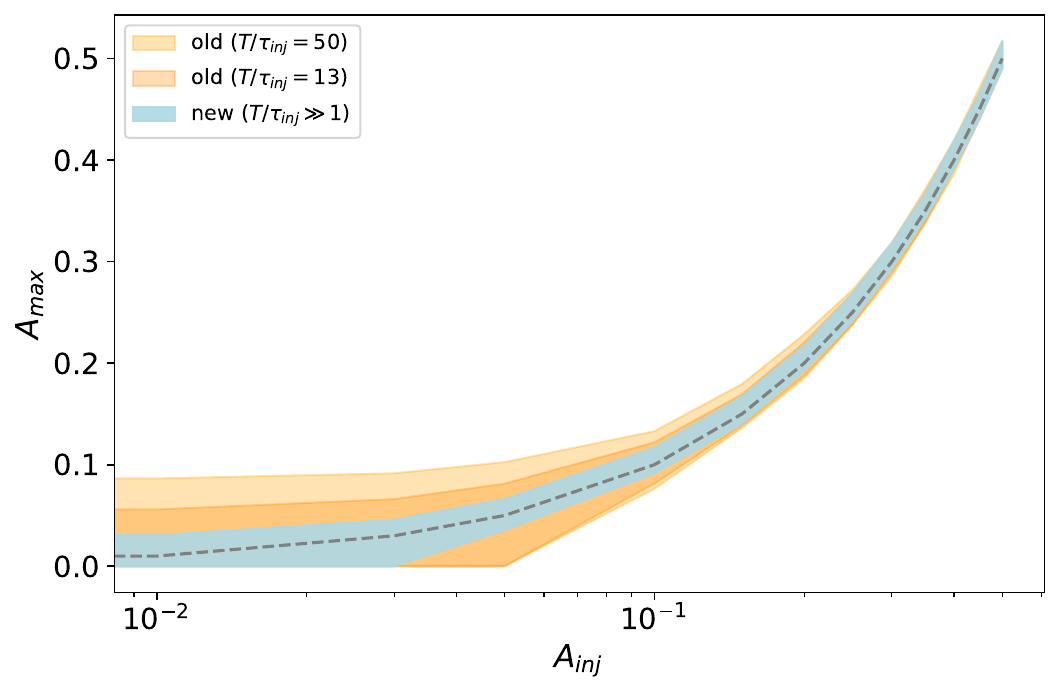}	
	\caption{\label{fig:ALoglikelihood} {The injected amplitude  $A_\textrm{inj}$ dependence of the new likelihood (blue)  and the old likelihood (orange) for two duration $T$ in high frequency resolution limit. Left: the mean value of the maximum log-likelihood. The blue dotted lines show a large signal fit for the new likelihood. The gray vertical lines denote when the normalized value of  $A_\textrm{inj}$ is 1. Right: the $1\sigma$ range of searched amplitude $A_\textrm{max}$. The gray dashed line denotes $A_\textrm{max}=A_\textrm{inj}$.}}
\end{figure}

Figure~\ref{fig:ALoglikelihood} compares the injected amplitude dependence. 
For the new likelihood, the mean value and the relative error can be nicely fitted by $a\, A_\textrm{inj}^2 $ and $a'/A_\textrm{inj}$ (blue dotted line), respectively, in the large signal limit. The old likelihood, on the other hand, is sensitive to the normalized amplitude, and so exhibits a more complicated dependence on $A_\textrm{inj}$.
For the searched amplitude, it is useful to compare the $1\sigma$ range of $A_\textrm{max}$ distribution to the injected value.  The $1\sigma$ range is symmetric around $A_\textrm{inj}$ for a large signal, while it  broadens and becomes asymmetric for decreasing signal amplitude. For the old likelihood, the upper boundary is driven large mainly by the noise and thus is quite insensitive to $A_\textrm{inj}$. The value becomes larger for increasing $T$, corresponding to a constant normalized value $\sim 0.6$.

In addition, in the high-resolution limit, the old likelihood exhibits a strong dependence on the QNM shape. Considering two models with the same SNR per mode but different $\tau_{\rm inj}$, we find distributions not much changed for the new likelihood, while those for the old likelihood are drastically different due to varying SNR per bin. 

In summary, the new likelihood can significantly improve the search sensitivity over the old one when the QNMs can be fully resolved. 
This comes directly from the coherent combination of multiple frequency bins belonging to one QNM. As a result, the performance of the new likelihood is more directly related to the signal SNR, while previous searches with the old likelihood~\cite{Ren:2021xbe} are subject to strong dependence on the time duration and the QNM shape. This is important for realistic echo searches, where the QNMs are nonuniform and the width can vary a lot with frequency. 
Therefore, the new likelihood not only mitigates intrinsic uncertainties associated with echo searches, but also enables a more physical interpretation of the search results.

\subsection{Bayesian search for the $\uew$ injections}
\label{sec:BayesianUEW}

Now we verify the reliability of the Bayesian search algorithm for the $\uew$ injections, and examine the influence of various factors on the search performance. 
Since $\uew$ is uniform, we set $\{\Delta f, q_0, A, 1/\tau\}$ as the search parameters and fix $f_{\rm min}, f_{\rm max}$ as their injected values. Tab.~\ref{tab:parameter1} summarizes the priors for the four search parameters. Compared to our previous studies in Ref.~\cite{Ren:2021xbe}, we add the width $1/\tau$ as the fourth search parameter, with the ratio  $1/(\tau\Delta f)$ ranging from $1/(T\Delta f)$ to $\mathcal{O}(1)$, where $T\Delta f \sim\mathcal{O}(100)$. This enables a more efficient and model-independent search of the long-lived QNMs with vastly different $\tau$, while the search performance is not significantly affected, as we will show below. The influence of the frequency band is manifested as the number of QNMs $N$. We examine this by considering $\uew$  injections with different $N$.

\begin{table}[h]
\begin{center}
\begin{tabular}{l||l}
\hline\hline
&
\\[-3mm]
Parameters & Priors 
\\
&
\\[-3.5mm]
\hline
$\Delta f$ & Uniform in $[0.2, 2]\Delta f_{\rm inj}$
\\
$\rshift$ & Uniform in $[0,1]$
\\
$A$ & Uniform in $[10^{-2}, 10]\langle\tilde{P}\rangle^{1/2}$
\\
$1/\tau$ & Log-uniform in $[1/T, \max\{2/\tau_{\rm inj}, 2/T\}]$
\\
\hline
$T$ & $\{1/4, 1/2, 3/4, 1, 2, 3, 6, 13\}\,\tau_{\rm inj}$\\
\hline\hline

\end{tabular}
\caption{The parameter settings for the four-parameter Bayesian search for the $\uew$ injections.  
The range of $A$ is specified in terms of the normalized PSD  $\langle\tilde{P}\rangle$  for noise. The last line denotes the $T$ range over which we scan.}
\label{tab:parameter1}
\end{center}
\end{table}

We use BILBY~\cite{Ashton:2018jfp} to analyze data, which utilizes the nested sampling algorithm (DYNESTY sampler~\cite{Speagle:2019ivv}) to explore the complicated likelihood distributions, as in Ref.~\cite{Ren:2021xbe}. A proper choice of the sampler settings is essential for the detection of narrow resonances. Following Ref.~\cite{Ren:2021xbe}, we choose $n_{\rm live}=1000$, ${\rm walks}=100$, $n_{\rm act}=10$ and ${\rm maxmcmc}=10000$ as the default setting for the following search. At the end of this section, we will comment on the influence of the sampler settings.
Since the background search results are relatively insensitive to various parameter settings, we focus on the search results for the injected signals below. The details of the background search are presented in Appendix~\ref{sec:moreresults}.


\begin{figure}[!h]
	\centering
	\includegraphics[width=7.2cm]{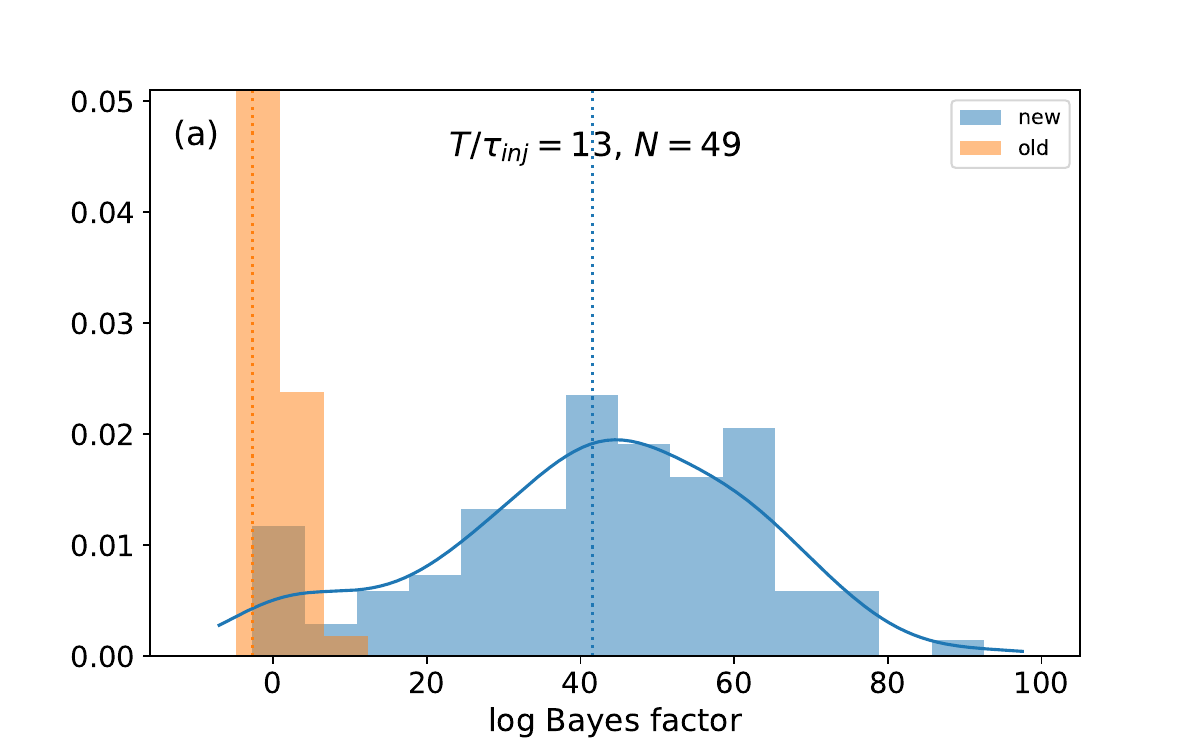}\;\;
	\includegraphics[width=7.2cm]{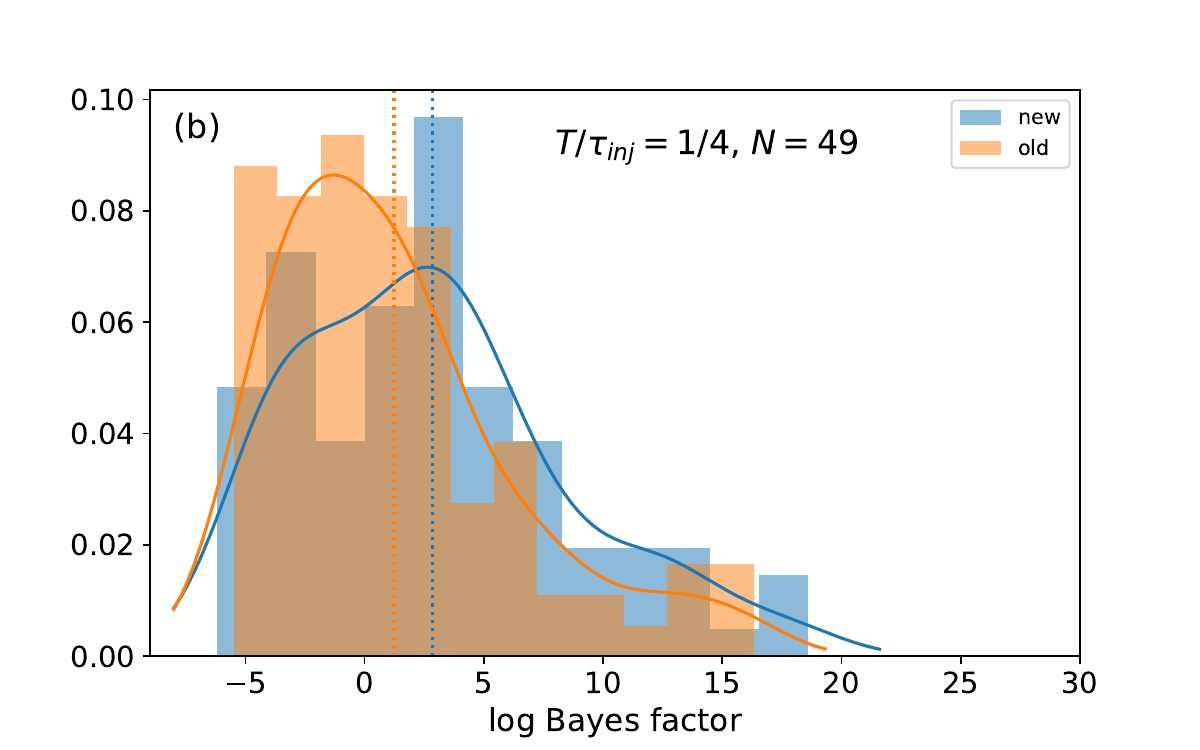}\\
	\includegraphics[width=7.2cm]{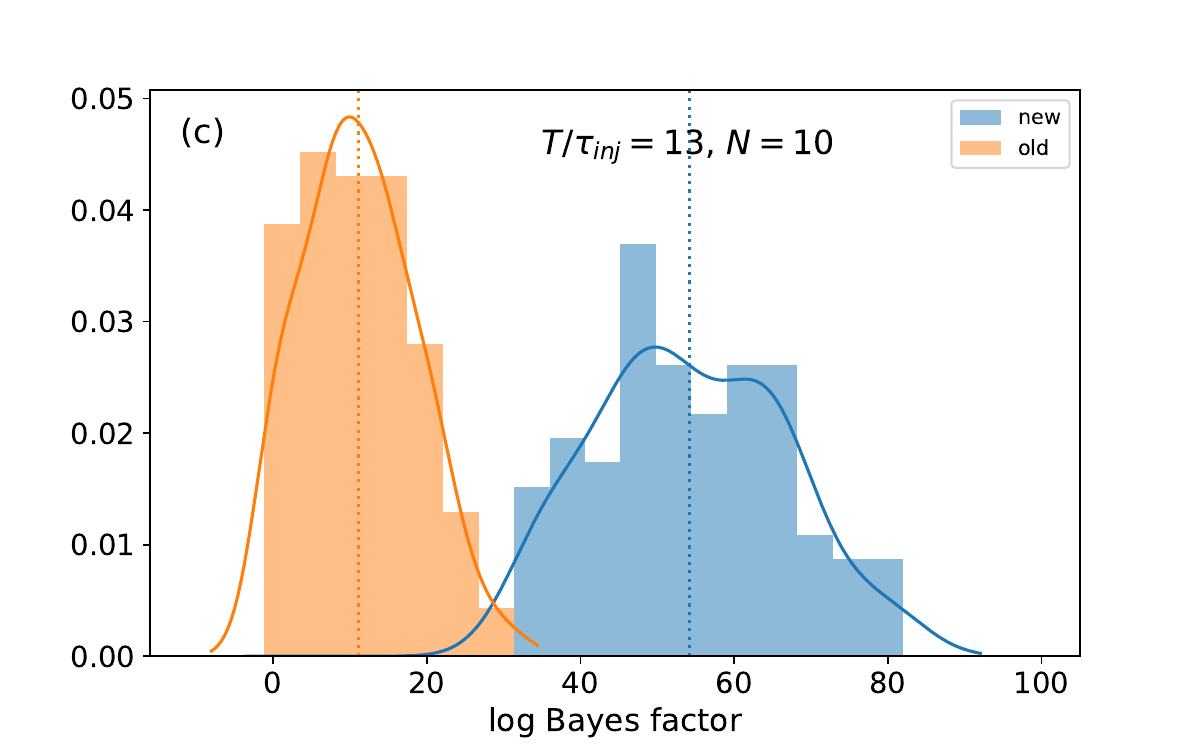}\;\;
	\includegraphics[width=7.2cm]{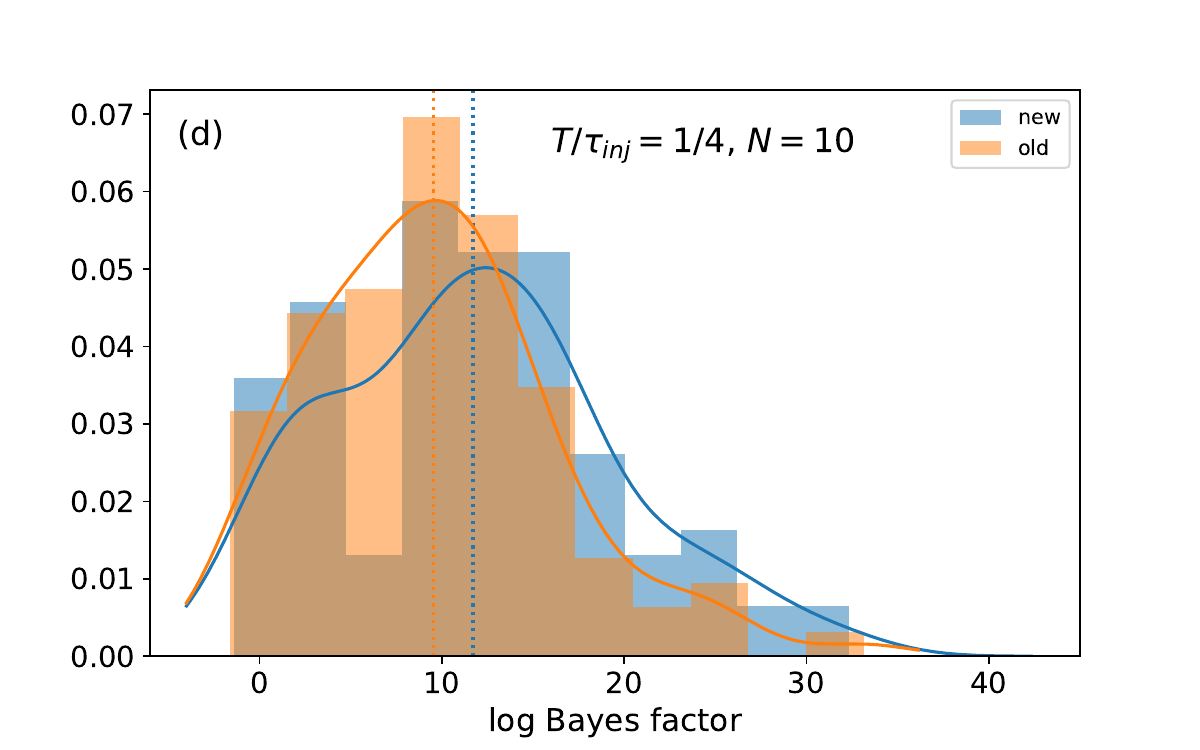}\;\;
	\caption{\label{fig:uewresult1} {The log Bayes factor distributions with the new (blue) and old (orange) likelihoods for the $\uew$ injection with $\tau_{\rm inj}\Delta f_{\rm inj}=23$  in $\mathcal{N}=100$ noise realizations. The panels (a) and (b) represent the $N=49$ case with $\textrm{SNR}\approx 16$ in the continuous limit, while (c) and (d) represent the $N=10$ case with $\textrm{SNR}\approx 13$ in the continuous limit. Panels (a) and (c) are for high-frequency resolution limit, and panels (b) and (d) are for the low-frequency resolution limit.}}
\end{figure}

\begin{figure}[!h]
	\centering
	\includegraphics[width=7.2cm]{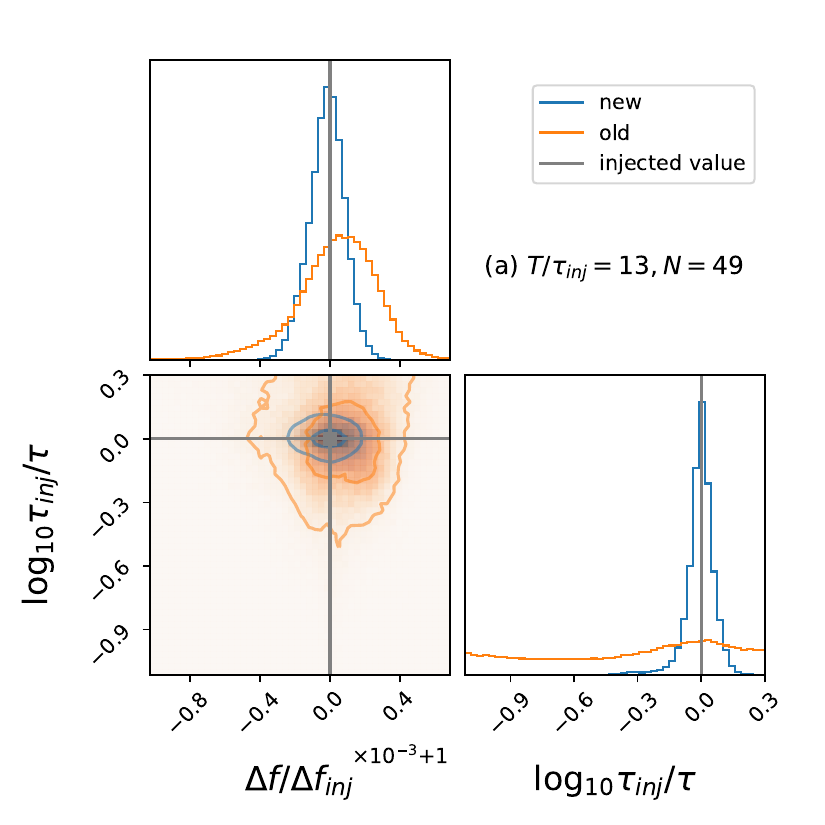}\;\;
	\includegraphics[width=7.2cm]{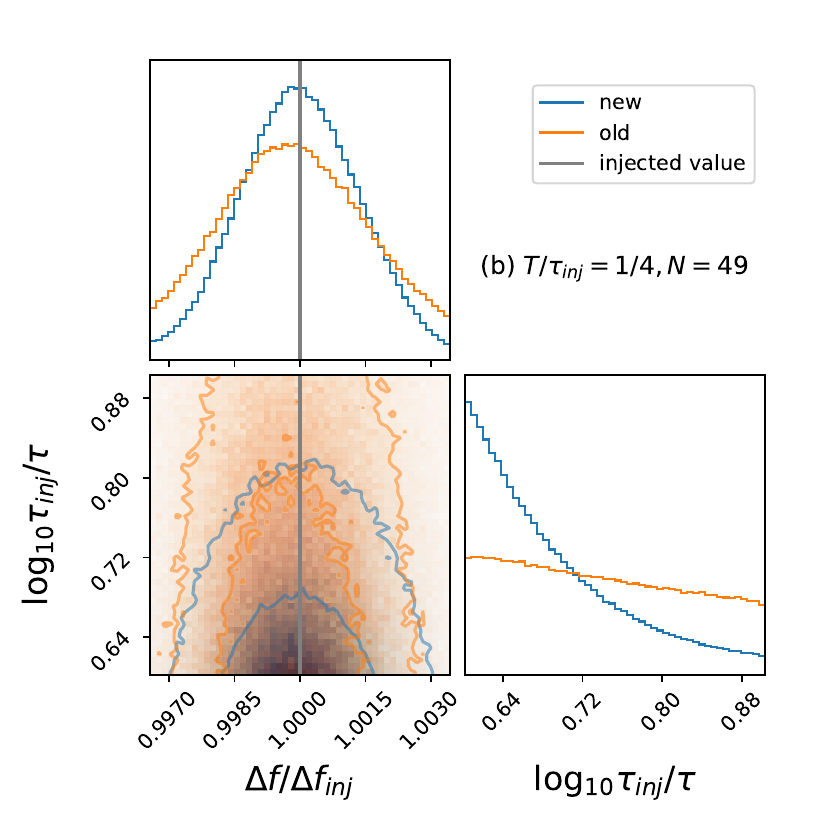}	\\
	\includegraphics[width=7.2cm]{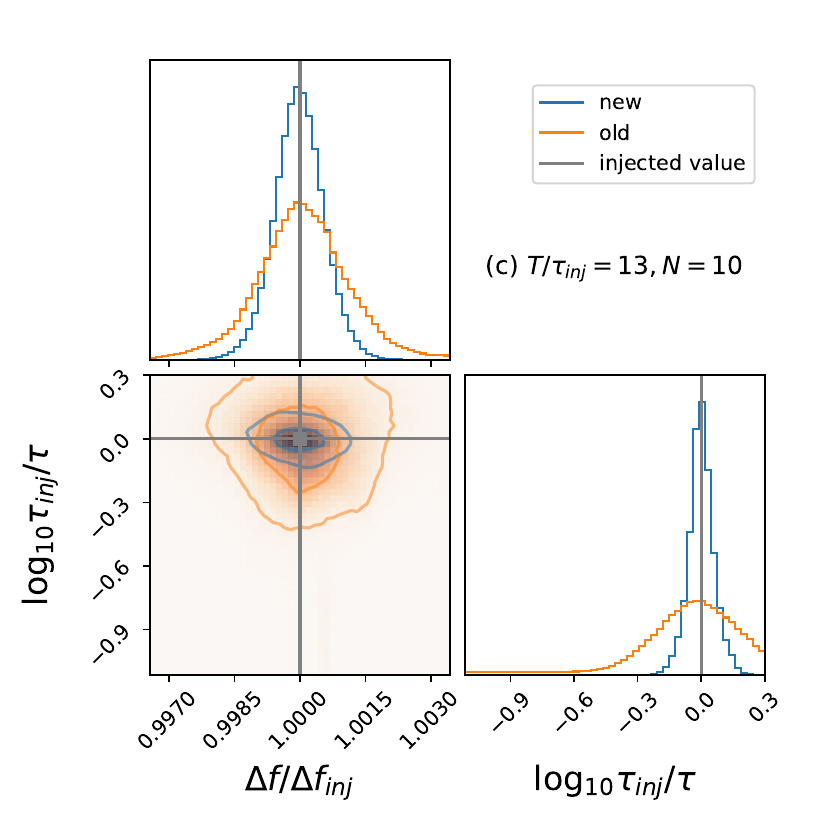}\;\;
	\includegraphics[width=7.2cm]{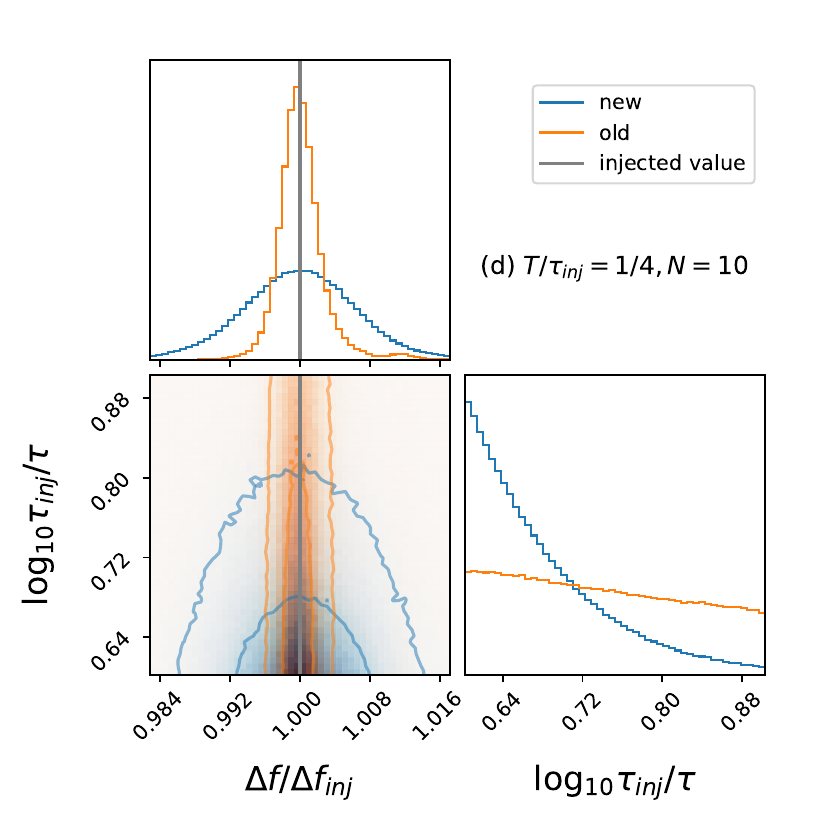}	
	\caption{\label{fig:posterior1}{Corner plots for the overall posterior distributions for the spacing $\Delta f$ and the width $1/\tau$ for the $\uew$ injection with $\tau_{\rm inj}\Delta f_{\rm inj}=23$. The blue and orange are for new and old likelihoods, respectively. The contours in the 2D posteriors for the spacing and width represent the $1\sigma$ and $2\sigma$ ranges. Similar to Fig.~\ref{fig:uewresult1}, (a) and (b) are for $N=49$, while (c) and (d) are for $N=10$. (a) and (c) represent high-frequency resolution, and (b) and (d) represent low-frequency resolution.}}
\end{figure}

We begin by examining the influences of the time duration $T$ and the QNMs number $N$ for examples with a relatively small spacing-to-width ratio $\tau_{\rm inj} \Delta f_{\rm inj}=23$.  For the time duration, we consider two cases with $T/\tau_{\rm inj}\gg 1$ and $T/\tau_{\rm inj}\ll 1$ for the high and low frequency resolution limits, respectively. For the number of QNMs, we choose $N\sim 50$ to account for excitation of a large number of modes and $N=10$ for the case where only a subset of QNMs dominates the SNR. The former is bounded by the maximum total number of modes $\sim f_{\rm RD}/\Delta f$. The latter corresponds to a lower bound set previously for the real data analysis~\cite{Ren:2021xbe}, which helps to avoid contamination from large spectral lines in detector noise.  
Figs.~\ref{fig:uewresult1} and \ref{fig:posterior1} display the Bayesian search results for the four cases: (a) $T/\tau_{\rm inj}=13$, $N=49$; (b) $T/\tau_{\rm inj}=1/4$, $N=49$; (c) $T/\tau_{\rm inj}=13$, $N=10$; (d) $T/\tau_{\rm inj}=1/4$, $N=10$.

As expected, the new likelihood brings in significant improvements in the high frequency resolution limit, while not much difference can be seen when the QNMs are not well resolved. Specifically,  for the $T/\tau_{\rm inj}=13$ case, the log Bayes factor distributions for new likelihoods in Fig.~\ref{fig:uewresult1} are well separated from noise and approximately Gaussian. The old likelihood ones are much worse, in particular for the $N=49$ case, where the $\textrm{SNR}$ per bin is significantly smaller than 1. This case is very similar to the example considered in Ref.~\cite{Ren:2021xbe}. 
Although the performance degrades slightly with the additional search parameter,  the new likelihood brings in much more significant improvement, demonstrating the efficiency of  this search method.  
In the low resolution limit, i.e. $T/\tau_{\rm inj}=1/4$, the overall shape of log Bayes factor distributions for two likelihoods is quite similar, with a slightly longer tail for the new likelihood only. This is expected because the SNR per mode is dominated by only one bin.

Regarding the numbers of QNM dependence, we choose the SNRs such that the two $N$ cases have similar maximum log-likelihood distributions for the new likelihood in high frequency resolution limit.
The $\textrm{SNR}$ per mode is then larger for the $N=10$ case, although its total $\textrm{SNR}$ is smaller.  As a result, the old likelihood performs better for the $N=10$ case in the high frequency limit, with the $\textrm{SNR}$ per bin $\sim 1$. The performance at low frequency limit is also better than the $N=49$ case for both likelihoods.

To account for noise uncertainties, the posteriors for $\mathcal{N}=100$ different noise realizations are averaged with equal weights to obtain the overall distributions~\cite{Breschi:2022ens},
\begin{eqnarray}
p(\mathbf{\theta}|d)= \sum_{k=1}^{\mathcal{N}}p(\mathbf{\theta}|d_k)p(d_k),\quad \textrm{with } p(d_k)=1/\mathcal{N}\,.
\end{eqnarray} 
Fig.~\ref{fig:posterior1} displays the overall posterior distributions of the spacing $\Delta f$ and the width $1/\tau$ with respect to the injected values. Specifically, we list the inferred parameters with symmetric 90\% credible intervals for the new likelihood as follows,
\begin{eqnarray}\label{eq:uew90CI} 
&&\textrm{(a)}:\; \Delta f/\Delta f_{\rm inj}\approx 1.0000^{+0.0003}_{-0.6665},\;
\log_{10} \tau_{\rm inj}/\tau\approx 0.00^{+0.09}_{-0.11}\nonumber\\
&&\textrm{(b)}:\; \Delta f/\Delta f_{\rm inj}\approx 1.0^{+0.4}_{-0.3},\;
\log_{10} \tau_{\rm inj}/\tau\approx 0.67^{+0.17}_{-0.06}\nonumber\\
&&\textrm{(c)}:\; \Delta f/\Delta f_{\rm inj}\approx 1.0000^{+0.0009}_{-0.0009},\;
\log_{10} \tau_{\rm inj}/\tau\approx 0.00^{+0.10}_{-0.10}\nonumber\\
&&\textrm{(d)}:\; \Delta f/\Delta f_{\rm inj}\approx 1.00^{+0.01}_{-0.01},\;
\log_{10} \tau_{\rm inj}/\tau\approx 0.66^{+0.16}_{-0.05}\,.
\end{eqnarray}
A key feature of the search is the ability to estimate the spacing $\Delta f$ with high precision, particularly when $N$ is large. This is due to the periodic nature of the $\uew$ model, where a small change in $\Delta f$ can cause a significant mismatch of the entire template, resulting in a relative error of $\Delta f$ that is suppressed by a factor of  $1/N$. This can be observed by comparing the cases $N=49$ and $N=10$ with the new likelihood. In the high frequency resolution limit, the relative error of $\Delta f$ for the case $N=49$ is reduced by a factor of a few compared to the case $N=10$, reaching an order of magnitude of 0.01\%.\footnote{The large lower error for case (a) corresponds to non-negligible probability of the average spacing at half of the best-fit value due to the partial overlap between the template and the injected signal~\cite{Conklin:2017lwb,Ren:2021xbe}.} For the old likelihood, the difference in spacing errors is mainly determined by the log Bayes factor. 
For the inferred width, the distributions peak well around the injected values in the high-resolution limit. Since the width error is insensitive to $N$, we observe very similar width errors for both $N$ cases with the new likelihood.  
In the low-resolution limit, the width distributions peak around the lower end $\sim 1/T$ for both cases. This is expected because the mode has been smeared out and the SNR is dominated effectively by one bin with the width $1/T$.\footnote{As for a test, we  consider a different prior $1/\tau \in[0.1/T,1/T]$ (log-uniform) for this case, with the injected value included. It turns out that the likelihood is quite insensitive to the variation of small $\tau$ in this range, and the sensitivity worsens for both likelihoods.}

\begin{figure}[!h]
	\centering
	\includegraphics[width=14cm]{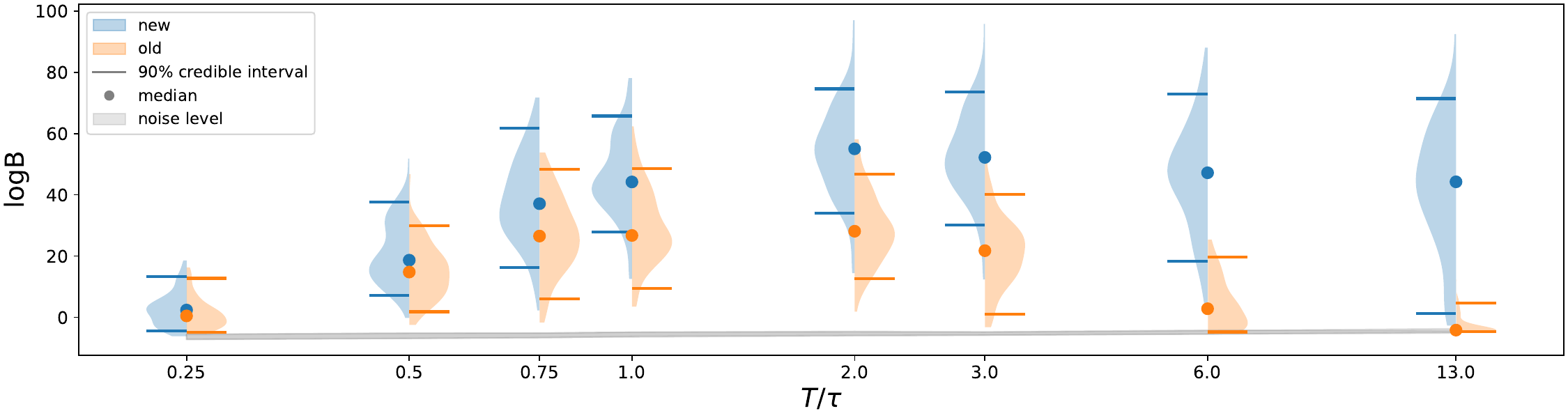}
	\includegraphics[width=14cm]{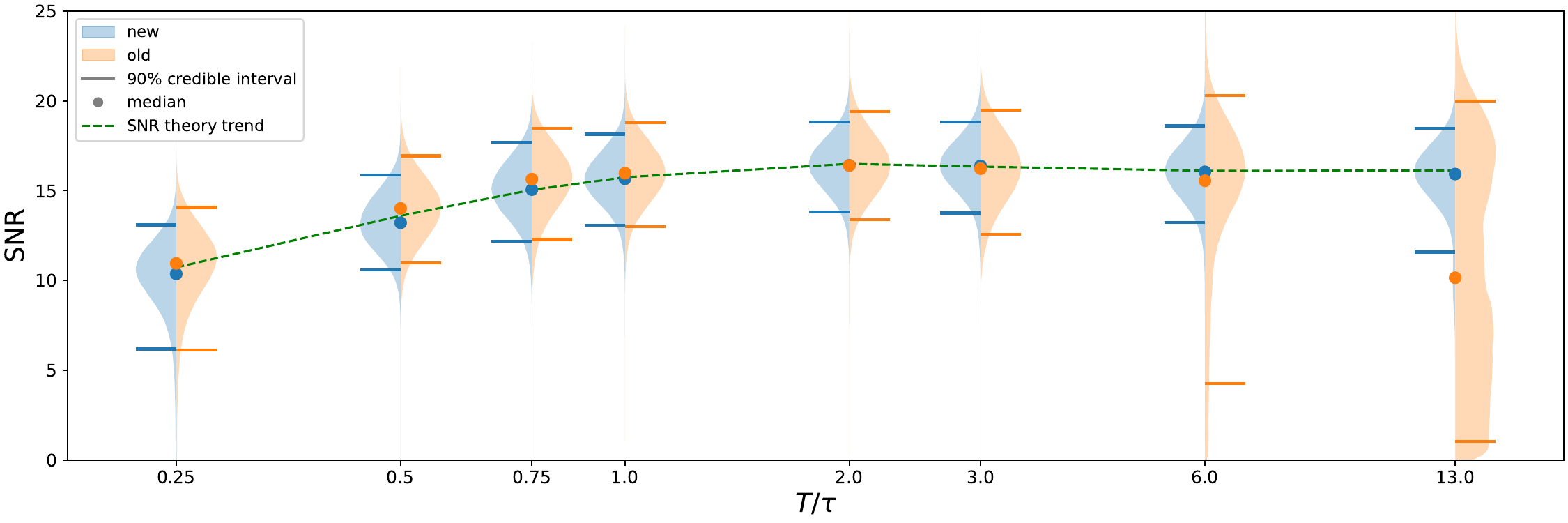}
        \includegraphics[width=14cm]{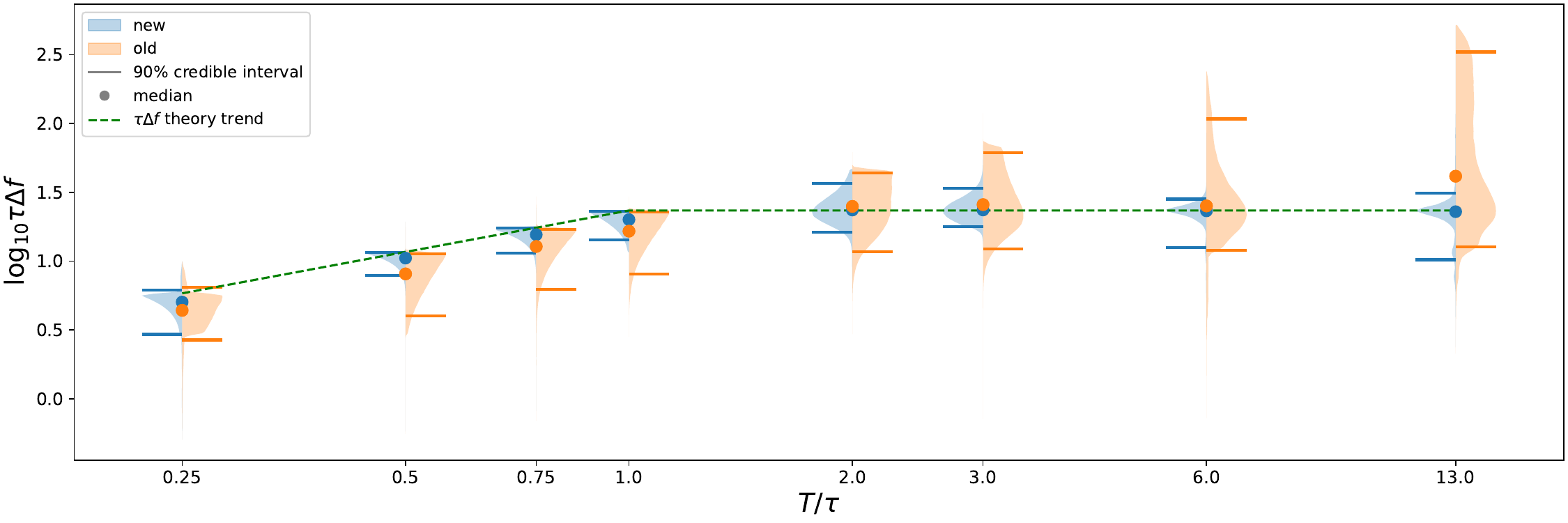}
	\caption{\label{fig:violin1}{The log Bayes factor distributions (top), the overall posterior distributions of the inferred SNR (middle) and the combination $\tau \Delta f$ (bottom) for the $\uew$ injection with $\tau_{\rm inj} \Delta f_{\rm inj}=23$ and $N=49$, as a function of the time duration $T/\tau_{\rm inj}$. For all panels, the blue and orange colors represent the results obtained using the new and old likelihoods, respectively. The upper and lower bars represent the symmetric 90\% credible intervals, and the dots denote the median values. In the top panel, the gray band represents the 90\% credible interval of the noise distributions. For the middle and bottom ones, the green dashed lines denote the theoretical prediction of the $\snr$ and the combination $\{\tau_{\rm inj}, T\}\Delta f_{\rm inj}$, respectively.}}
\end{figure}

For a generic search, we scan over a list of time durations $\{T_i\}$ for a given $\uew$ injection. For each $T_i$, we perform the Bayesian search with the two phase-marginalized likelihoods and the priors listed in Tab.~\ref{tab:parameter1}. 
The search results for the previous example with $\tau_{\rm inj}\Delta f_{\rm inj}=23$ and $N=49$ are presented in Fig.~\ref{fig:violin1}. The two likelihoods are comparable at small $T< \tau_{\rm inj}$, and their performance improves as $T$ increases.  When $T$ exceeds $\tau_{\rm inj}$, the performance of the new likelihood saturates and varies less with $T$, while the performance of the old likelihood degrades significantly as $T$ increases. This change in behavior also provides an estimate of the injected value of $\tau$. In the high resolution limit, i.e. $T\gg \tau_{\rm inj}$, the  sampling uncertainties increase, yielding a larger error of log Bayes factor distribution. 
Regarding the inferred SNR, the median value of the posterior distributions agrees well with the injected value, except for the large $T$ case for the old likelihood with very poor sensitivity. The relative error also remains stable with respect to $T$. The combination $\tau \Delta f$ is useful because it provides an estimate of the combined reflectivity $\mathcal{R}_{\rm eff}$ for long-lived QNMs through Eq.~(\ref{eq:QNMn}), i.e. $\tau_n \Delta f\approx 1/|\ln\mathcal{R}_{\rm eff}(f_n)|$. Specifically, the median value of this quantity traces well the combination $\{\tau_{\rm inj}, T\}\Delta f_{\rm inj}$. When the frequency resolution is insufficient, i.e., $T<\tau_{\rm inj}$, its posterior distribution is less informative and tends to concentrate around the upper boundary $T\Delta f_{\rm inj}$. Only when the mode becomes resolved at $T>\tau_{\rm inj}$, the posterior distribution becomes more symmetric and informative inferences are possible.


\begin{figure}[!h]
	\centering
	\includegraphics[width=7.2cm]{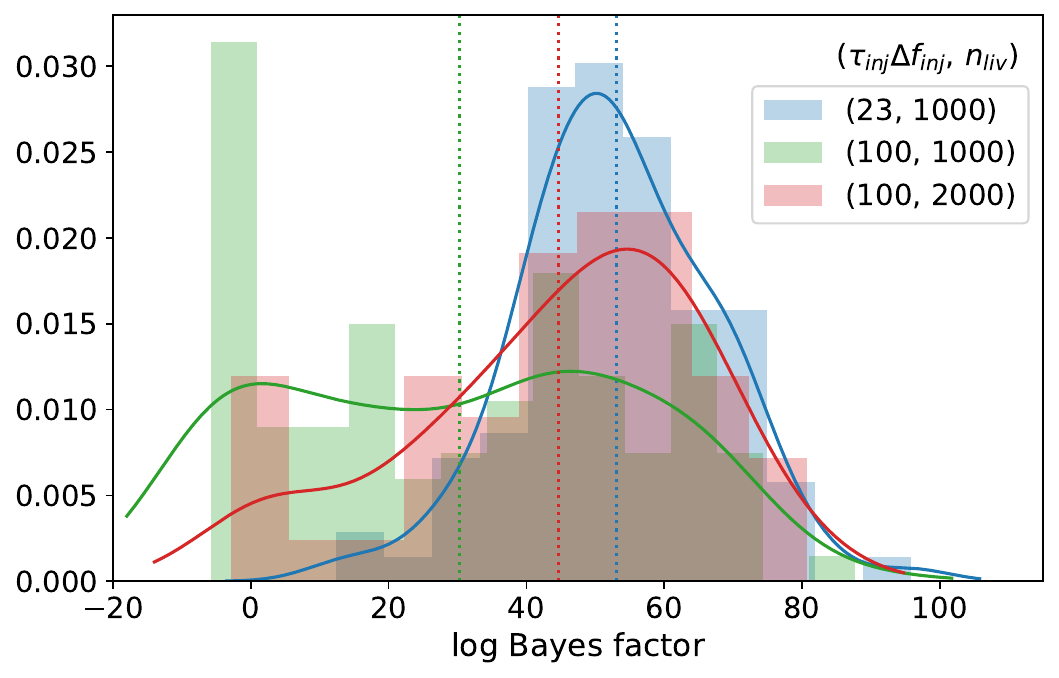}\;\;
	\includegraphics[width=7.2cm]{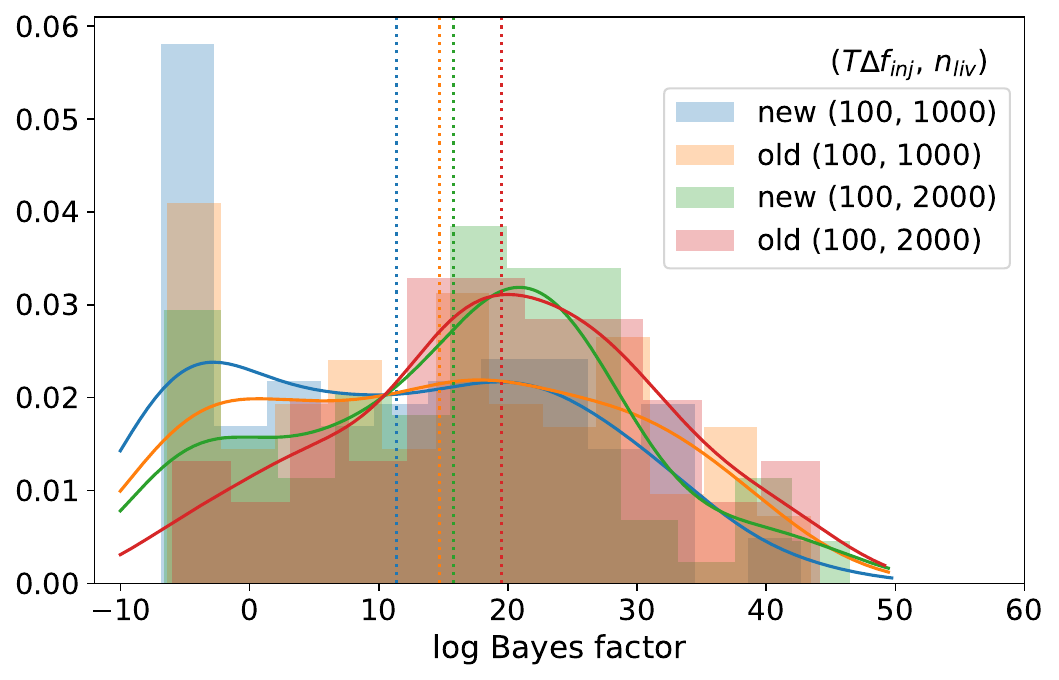}
	\caption{\label{fig:uewresultnliv} {Four-parameter Bayesian search results of the injected $\uew$ signal with $N=49$. Left: the log Bayes factor distribution for the new likelihood in the high-frequency limit, with $T/\tau_{\rm inj}=3$, $\textrm{SNR}\approx 16$,  and different choices of $\tau_{\rm inj}\Delta f_{\rm inj}$ and $n_{\rm live}$. Right: the log Bayes factor distributions for both new and old likelihoods in the low-frequency limit, with $T/\tau_{\rm inj}=1/2$, $\textrm{SNR}\approx 14$, $T\Delta  f_{\rm inj}=100$ and different choices of $n_{\rm live}$.}}
\end{figure}

The search performances are also sensitive to the injected value of spacing-to-width  ratio, i.e. $\tau_{\rm inj} \Delta f_{\rm inj}$. As $\tau_{\rm inj} \Delta f_{\rm inj}$ increases, the peak region takes a smaller fraction of the total number of frequency bins, making it numerically more challenging to find the signal. The sampler settings for the Bayesian search then have to be adjusted accordingly. The left panel of Fig.~\ref{fig:uewresultnliv} displays the correlation between $\tau_{\rm inj} \Delta f_{\rm inj}$ and the number of live points $n_{\rm live}$ of nested sampling in the high-resolution limit. Performance degrades significantly if we increase $\tau_{\rm inj} \Delta f_{\rm inj}$ from 23 to 100 but with the sampler setting unchanged, that is, $n_{\rm live}=1000$. The peak of the log Bayes factor distribution around zero corresponds to a large number of failed searches that miss the narrow resonances. The search is efficient only for a small number of cases where the injected signal is found. The situation is greatly improved if we increase $n_{\rm live}$ to 2000 for $\tau_{\rm inj} \Delta f_{\rm inj}=100$. The missed cases are then largely eliminated, making the search results more similar to that for  $\tau_{\rm inj} \Delta f_{\rm inj}=23$.  Since $\tau_{\rm inj} \Delta f_{\rm inj}$ is directly related to the combined reflectivity of the cavity $\mathcal{R}_{\rm eff}$, as shown in Eq.~(\ref{eq:QNMn}), this reveals that a sufficiently fine setting of the stochastic sampler is required to efficiently probe the strong reflectivity case.  
In the low resolution limit, the effective width of the resonance is limited by $1/T$, and thus the sampler setting is mainly based on $T\Delta f_{\rm inj}$. As shown in the right panel of Fig.~\ref{fig:uewresultnliv}, the old likelihood performs slightly better in the case of inefficient sampling, i.e. $T\Delta f_{\rm inj}=100$, $n_{\rm live}=1000$. This is reasonable because the new likelihood incorporates more information, allowing for more possibilities. Both likelihood performances improve as $n_{\rm live}$ increases to 2000. In particular, the performance of the old likelihood for this case becomes comparable to that for the low $T\Delta f_{\rm inj}$ case in Fig.~\ref{fig:violin1}. Therefore, choosing the right sampler settings is crucial for an efficient search. Given that the choice of $n_{\rm live}$ relies on the quantity $\min\{T, \tau_{\rm inj}\}\Delta f_{\rm inj}$, for a generic search that scans over a list of $T_i$, it is safe to choose $n_{\rm live}$ according to $T_i\Delta f_{\rm max}$, which represents the maximum  number of pulses included in the segment of strain data.\footnote{Another approach to address the challenge of finding narrow resonances within a wide frequency range is to manually scan over the peak location (e.g. $\Delta f$), similar to the continuous wave search~\cite{Riles:2022wwz}. However, this method requires additional computational resources, and we leave it for future investigation. }

As a final note, we perform injections of the $\uew$ model given by Eq.~(\ref{eq:uniformQNM}) in the frequency domain, assuming a constant phase $\delta'_n$ per mode that can be properly marginalized. 
To examine the impact of QNM interference on the Bayesian search performance, we discuss the superposition of QNMs in the time domain in Appendix~\ref{eq:echoQNMtime}. By considering a uniform and periodic model of QNMs in the waveform given by Eq.~(\ref{eq:htQNM}), we have verified that the search algorithm's performance remains robust as long as the spacing-to-width ratio is considerably larger than 1.

\section{Bayesian search for echo waveform injections}
\label{sec:validate}

\subsection{Benchmarks for echo waveforms}
\label{sec:benchmark}

With the generic construction of echo waveforms in Sec.~\ref{sec:generic}, the main uncertainties come down to the effective reflection from the interior boundary and the frequency content of the initial pulse. To validate our search algorithm, we consider a few benchmarks below to demonstrate its ability to recover echo waveforms of different shapes.

For interior reflection, the inputs are the energy flux reflection of the interior boundary $\mathcal{R}_{\rm wall}(\omega)$ and the phase $\delta(\omega)$. A toy model is utilized as a reference, with constant $\mathcal{R}_{\rm wall}$ and $\delta(\omega)=t_d\tilde\omega$. Subsequently, two more representative and complementary examples are explored, which are more physically realistic. 
The first comes from an explicit model of UCOs, the 2-2-hole in quadratic gravity~\cite{Holdom:2016nek, Holdom:2019ouz, Ren:2019afg, Holdom:2022zzo}. Since this object is extremely compact, with $\eta\sim2$ in Eq.~(\ref{eq:td}), the phase is dominated by the time delay contribution. In addition, a perfect reflecting boundary condition is naturally defined at the origin of 2-2-holes. Thus, the effective energy flux reflection for this case is fully determined by the energy loss experienced by gravitational waves traveling through the matter source inside 2-2-holes. The prediction of this model is found as~\cite{Holdom:2020onl}  
\begin{align}\label{eq:Rdamp}
 \mathcal{R}_{\mathrm{damp}}(\omega)=\exp \left(-4 \pi \mathcal{V}(\alpha) \zeta\left[1+\left(8 \pi \tilde{\omega} \frac{\zeta}{\alpha^{2}}\right)^{2}\right]^{-1} \int_{0}^{t_d/2M} \hat{s}(\tilde{x}) e(\tilde{\omega}, \tilde{x}) d \tilde{x}\right),\; \delta(\omega)=t_d \tilde\omega\,,
\end{align}
where $\alpha$ is the dimensionless coupling, $\mathcal{V}(\alpha)$ is the viscosity-to-entropy density ratio for the matter source and $\zeta=2$ is a parameter characterizing the entropy of the 2-2-hole.  $\hat{s}(\tilde{x}) $ is the entropy density at position $\tilde{x}$.  $e(\tilde{\omega}, \tilde{x})$ is the energy density profile of the gravitational wave. 
This model features strong energy absorption near the special frequency $\omega_H$, while the damping becomes negligible away from $\omega_H$.

The second model considers a reflection following the Boltzmann distribution, motivated by assuming that black holes are quantum systems that satisfy either thermodynamic detailed balance, CP-symmetry, or (a version of) the fluctuation-dissipation theorem~\cite{Oshita:2018fqu,Oshita:2019sat,Wang:2019rcf}. Specifically, using the fluctuation-dissipation theorem, the general form of the Boltzmann reflection is given by
\begin{align}\label{eq:Rbolz}
    \mathcal{R}_{\rm Boltz}(\omega) =\exp \left(-\frac{|\tilde{\omega}|}{2 T_{\rm QH}}\right),\; 
    \delta(\omega)=-\frac{\tilde\omega}{\pi T_{\rm H}}\ln(\gamma|\tilde\omega|)\,,
\end{align}
where  $T_{\rm QH}$ denotes the quantum horizon temperature and $T_{\rm H}$ is the Hawking temperature. $\gamma$ is a constant that determines the energy scale of exotic physics responsible for the reflection. In this paper, we assume $\gamma= 1 $, which is equivalent to assuming that the relevant energy scale at the would-be horizon is the Planckian energy. With no definite prediction for the latter, we leave $T_{\rm QH}$ as a free parameter here. 
The energy reflection reaches its maximum at the special frequency $\omega_H$, and is exponentially suppressed away from $\omega_H$.
The phase is not exactly linear in $\omega$ due to the log correction term, yielding a frequency-dependent spacing for the QNMs and a small deviation for the periodicity of $\omega_n$.

For both models, the ergoregion instability can be easily quenched by the absorption around $\omega_H$ with a proper choice of the parameters, i.e. $\mathcal{R}_{\rm eff}(\omega)\leq 1$ with a sufficiently large $\mathcal{V}\alpha^4$ in Eq.~(\ref{eq:Rdamp}) or a sufficiently small $T_{\rm QH}$ in Eq.~(\ref{eq:Rbolz}). 

\begin{figure}[H]
\centering
\includegraphics[width=12cm]{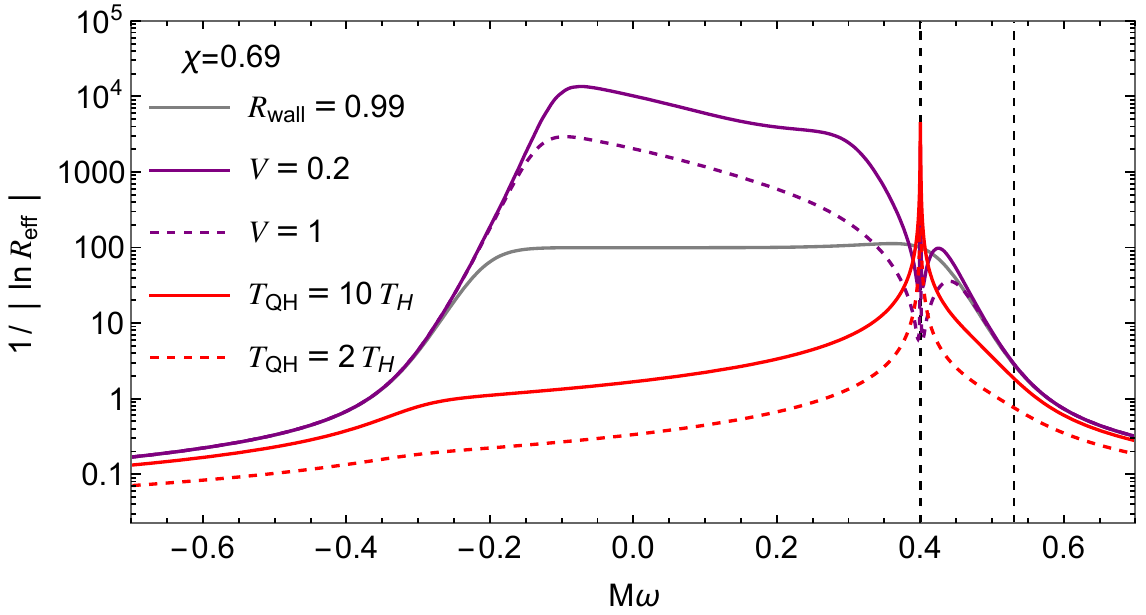} 
\caption{The frequency dependence of $1/|\ln\mathcal{R}_{\rm eff}(\omega)|$ for the combined reflectivity $\mathcal{R}_{\rm eff}=\mathcal{R}_{\rm BH}\mathcal{R}_{\rm wall}$ for the three representative models of $\mathcal{R}_{\rm wall}$ and  $\mathcal{R}_{\rm BH}$ with $\chi=0.69$. The gray solid curve is for the constant reflectivity $R_{\rm wall}=0.99$. The purple curves are for the damping model, with $\mathcal{V}=0.2$ (solid) and $\mathcal{V}=1$ (dashed) in Eq.~(\ref{eq:Rdamp}) ($\alpha^4=0.01$ for both cases). The red curves are for the Boltzmann reflection model, with $T_{\rm QH}=10T_H$ (solid) and $T_{\rm QH}=2T_{\rm H}$ (dashed) in Eq.~(\ref{eq:Rbolz}). The ergoregion instability is successfully quenched for all these cases.
The two vertical dashed lines are for $M\omega_H=0.40$ and $M\omega_{\rm RD}=0.53$.
}
\label{fig:Rwall}
\end{figure}

Figure~\ref{fig:Rwall} displays the quantity $1/|\ln\mathcal{R}_{\rm eff}(\omega)|$, which is approximately the spacing-to-width ratio $\tau \Delta f$ (or quality factor) for narrow QNMs as a function of their real frequency,  as given in Eq.~(\ref{eq:QNMn}). In addition to the strong absorption case for the Boltzmann reflection model, the main difference of these models appears at $\omega\lesssim \omega_H$. The constant reflection has roughly $\tau \Delta f \approx 1/(1-\mathcal{R}_{\rm wall})$ for $\omega\lesssim \omega_H$. The damping model, on the other hand, predicts an increase $\tau \Delta f$ as $\omega$ decreases. The Boltzmann model is the opposite, where narrow resonances with $\tau \Delta f\gg1$ only appear around $\omega_H$.

For the frequency content of the initial pulse, in addition to the inside and outside prescriptions in the geometric optics limit as given in Eq.~(\ref{eq:hechogeo}), we also consider the case of infalling particles with $h_{\rm eff}(\omega)$ extracted from the numerical results of a recent analysis~\cite{Xin:2021zir}. For the outside case, $h_{\rm eff}(\omega)$ is suppressed by $\mathcal{T}_{\rm BH}^2$ at low frequency and enhanced by $1/\mathcal{R}_{\rm BH}^2$ at high frequency, compared to that for the inside one. The resulting echoes are thus dominated by quickly damped modes at high frequencies, which are not the focus of our search. 
The prediction of the ``infalling particle'' case lies between the inside and outside prescriptions, with a relatively larger contribution from QNMs at higher frequencies, but not significantly so. This provides another example to illustrate the variability of initial conditions, in addition to the commonly used inside scenario.

\begin{figure}[h]
\centering
\begin{minipage}{.70\textwidth}  
  \centering
  \includegraphics[width=1\linewidth]{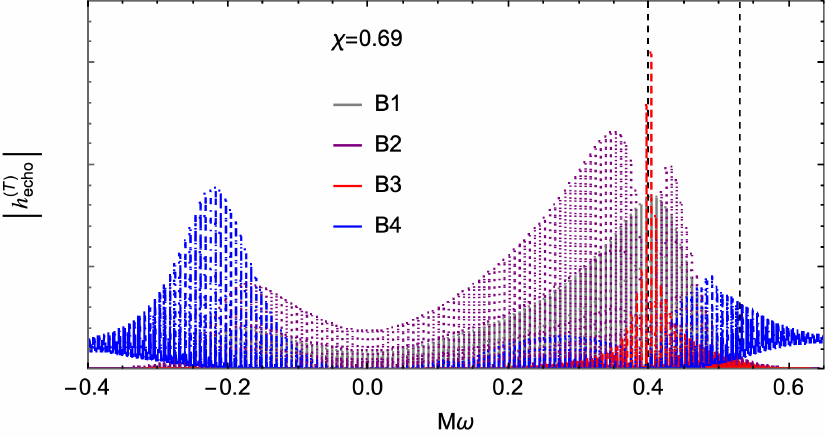}
\end{minipage}%
\;
\begin{minipage}{.28\textwidth}
  \centering
  \includegraphics[width=1\linewidth]{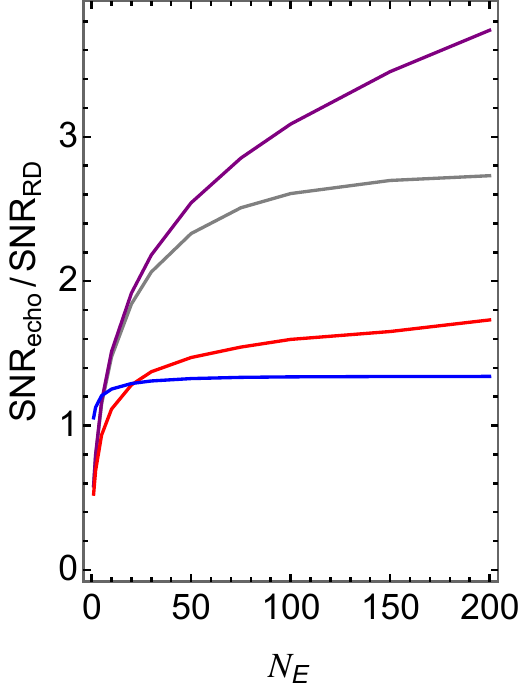}
\end{minipage}
\caption{Left: the relative echo amplitude $|h_{\rm echo}^{(T)}(\omega)|$ for the four benchmarks with the dimensionless spin $\chi=0.69$ and the number of pulses $N_E = 200$. The gray, purple, red and blue lines are for B1, B2, B3 and B4, respectively. The two vertical black dashed lines are for $M\omega_H=0.40$ and $ M\omega_{\rm RD}=0.53$. Right: the ratio of SNRs as a function of $N_E$ for the four benchmarks, with the same legends as the left. 
The symbol $\snr_{\rm echo}$ represents the total SNR of echoes within either the range $0<\omega<\omega_{\rm RD}$ or $-\omega_{\rm RD}<\omega<0$, depending on which range yields a higher SNR value.}
\label{fig:SpecBol}
\end{figure}

Given the above choices for the interior reflection and initial condition,  we then define four benchmarks for the echo waveforms,  
\begin{itemize}
    \item B1: Constant reflection $+$ inside: $\mathcal{R}_{\rm wall}(\omega)=0.99$, $h_{\rm eff}(\omega)=h_{\rm RD}(\omega)$
    \item B2: Damping 2-2-hole $+$ inside: $\mathcal{R}_{\rm wall}(\omega)=\mathcal{R}_{\mathrm{damp}}(\omega)$ in Eq.~(\ref{eq:Rdamp}) with $\mathcal{V}=0.2, \alpha^4=0.01$, $h_{\rm eff}(\omega)=h_{\rm RD}(\omega)$ 
    \item B3: Boltzmann reflection $+$ inside: $\mathcal{R}_{\rm wall}(\omega)=\mathcal{R}_{\mathrm{Boltz}}(\omega)$ in Eq.~(\ref{eq:Rbolz}) with $T_{\rm QH}=10 T_{\rm H}$, $h_{\rm eff}(\omega)=h_{\rm RD}(\omega)$
    \item B4: Constant reflection $+$ infalling particle: $\mathcal{R}_{\rm wall}(\omega)=0.99$, $h_{\rm eff}(\omega)$ given by Ref.~\cite{Xin:2021zir}
\end{itemize}
%
The properties of four benchmarks are displayed in Fig.~\ref{fig:SpecBol}. The relative echo amplitude in the left panel demonstrates a significant degree of complementarity. The spectrum is dominated by the positive frequency component for the first three benchmarks, while the negative frequency component is larger for the last one.
Compared to the constant reflection case B1, the damping model B2 exhibits a significantly broader spectrum, with the exception of a slight drop around $\omega_H$. Notably, there exists a larger number of narrow modes at lower frequencies that grow larger as time increases. 
In contrast, the spectrum with the Boltzmann reflection (B3) exhibits a peak around the special frequency $\omega_H$, which rapidly decreases as $\omega$ deviates from $\omega_H$. The spectrum associated with the infalling particle (B4) features strong modes at high frequencies. 
Since the current search algorithm is designed to detect either the positive or negative frequency component of the echo spectrum, as we explain below, we present the ratio of the total SNR of the dominant echo component to that of the ringdown as a function of $N_E$ in the right panel for the four benchmarks. The ratio for B2 grows the fastest with $T$ due to the dominance of low frequency modes with much longer lifetime. In contrast, the ratio for B4 saturates at very early time due to the dominance of quickly damped modes at high frequency.\footnote{To find this ratio for B4, we fit the surrogate model ``NRSur7dq4''~\cite{Varma:2019csw} used in Ref.~\cite{Xin:2021zir} with the ringdown waveform as dominated by the fundamental mode. A nice fit can be obtained with the ringdown initial time larger than $15/M$. We then use this earliest initial time to evaluate $\snr_{\rm RD}$ with $|M\omega|<1$. 
} Our method is most effective for models such as B1, and B2, which are dominated by a large number of long-lived QNMs with comparable heights. However, as we will demonstrate below, it is still capable of detecting models like B3 and B4, albeit with a lower probability.


\subsection{Search results for the benchmarks}
\label{sec:searchbenchmark}

Now we are ready to carry out  the model-independent search for the four complementary benchmarks designed above. To account for the detector response,  we consider a simple constant effective impulse response, and the waveform can be written as $h_{\rm det}(f)=\frac{1}{2}(h_{\rm echo}(f)+h_{\rm echo}^*(-f))$ with the impulse response absorbed in the amplitude. More generally, as long as the detector response varies slowly with frequency, our current search strategy for narrow QNMs would not be significantly affected. Also, considering the narrow widths of the QNMs targeted, it is unlikely that the positive and negative frequency components in $h_{\rm det}(f)$ are overlapping in general. The search algorithm will then look for the dominant component of the echoes with the signal-to-noise ratio $\snr_{\rm echo}$. 

\begin{table}[h]
\begin{center}
\begin{tabular}{l||l}
\hline\hline
&
\\[-3mm]
Parameters & Priors and scan values
\\
&
\\[-3.5mm]
\hline
$M\Delta f$ & Uniform in $[\bar{R}/\eta_{\textrm{max}}$, $\bar{R}/\eta_{\textrm{min}}]$
\\
$\rshift$ & Uniform in $[0,1]$
\\
$A$ & Uniform in $[10^{-2}, 10]\langle\tilde{P}\rangle^{1/2}$
\\
$1/\tau$ & Log-uniform in $[1/T, \Delta f_{\rm max}]$
\\
$f_\textrm{min}$, $f_\textrm{max}$ & Uniform in $[f_\textrm{cut} ,f_\textrm{RD}]$
\\
 & with $f_\textrm{max}-f_\textrm{min}>10\Delta f$
\\
\hline
$T\Delta f_{\rm max}$ & $\{20, 40, 100, 200, 300, 400\}$ \\
$n_{\rm live}$ & $\{1000, 1000, 1000, 2000, 2000, 2000\}$\\
\hline\hline

\end{tabular}
\caption{The parameter settings for the echo search with $\uew$. $\Delta f$, $\rshift$, $A$, $1/\tau$, $f_\textrm{min}$, $f_\textrm{max}$ are search parameters and the priors are given. $\bar{R}=0.0024$ is given by Eq.~(\ref{eq:Deltaf}) for $\chi=0.69$. $\eta_{\textrm{max}}=4$, and $\eta_{\textrm{min}}=1$.
The range of $A$ is specified in terms of the normalized PSD  $\langle\tilde{P}\rangle$  for noise.  $ f_\textrm{RD}$ is given in Eq.~(\ref{eq:fRD}) and $f_\textrm{cut}=0$ is used for this study with Gaussian noise. The frequency band satisfies an additional constraint $f_\textrm{max}-f_\textrm{min}>10\Delta f$. We scan over a list of $T/M$ with $T\Delta f_{\rm max}$ ranging from 20 to 400 and  varying $n_{\rm live}$.}
\label{tab:parameter2}
\end{center}
\end{table}

For a given benchmark, we  inject the waveform in $\mathcal{N}=100$ Gaussian noise realizations and analyze the resulting data with a series of time duration $\{T_i\}$. For each data sample, we conduct a six-parameter Bayesian search with the two likelihoods. 
The parameter settings for the searches are summarized in Table~\ref{tab:parameter2}. $\Delta f$ denotes the spacing with $\eta\sim\mathcal{O}(1)$. The duration is chosen such that the number of pulses included in the data segment, i.e. $N_E\approx T\Delta f$, ranges from $\mathcal{O}(10)$ to $\mathcal{O}(100)$. The prior range of width corresponds to setting the ratio $1/(\tau\Delta f)$ from $1/(T\Delta f)$ to $\mathcal{O}(1)$. The upper and lower ends of the frequency band are determined by $f_{\rm RD}$ and $f_{\rm cut}$, respectively. For studies with Gaussian noises, we set $f_{\rm cut}=0$, while a larger value of $f_{\rm cut}$ might be needed for real data analysis. Additionally, we  require the frequency band to include a sufficiently large number of QNMs to avoid contamination from spectral lines in detector noise~\cite{Ren:2021xbe}.

To enable a fair comparison of different models and their search efficiencies, we fix the amplitude of the injected signals by requiring $\snr_{\rm echo}\approx 16$ for $N_E\approx 100$ (i.e. $T\Delta f_{\rm max}\approx 200 $) for all four benchmarks.
It is worth noting that the expected value of $\textrm{SNR}_{\textrm{echo}}$ strongly depends on the models under consideration. Taking the four benchmarks as examples, for a typical binary black hole event with $\snr_{\textrm{RD}}\approx 8$, such as GW150914, we would anticipate $\textrm{SNR}_{\textrm{echo}}\approx 20$, 24, 13, 11 for $N_E\approx 100$ for models B1, B2, B3, B4, respectively, as indicated in the right panel of Fig.~\ref{fig:SpecBol}. Thus, for fast-damping models like B3 and B4, $\textrm{SNR}_{\textrm{echo}}$ is smaller than 16, while for models featuring long-lived QNMs like B1 and B2, it can be significantly larger. 
Given that our search method specifically targets the latter case, a $\textrm{SNR}_{\textrm{echo}}$ of approximately 16 is a reasonable choice.
This choice also allows for a direct comparison of the search results with those for $\uew$ injections of similar $\snr$ presented in Sec.~\ref{sec:BayesianUEW}.

\begin{figure}[!h]
	\centering
        \includegraphics[width=12cm]{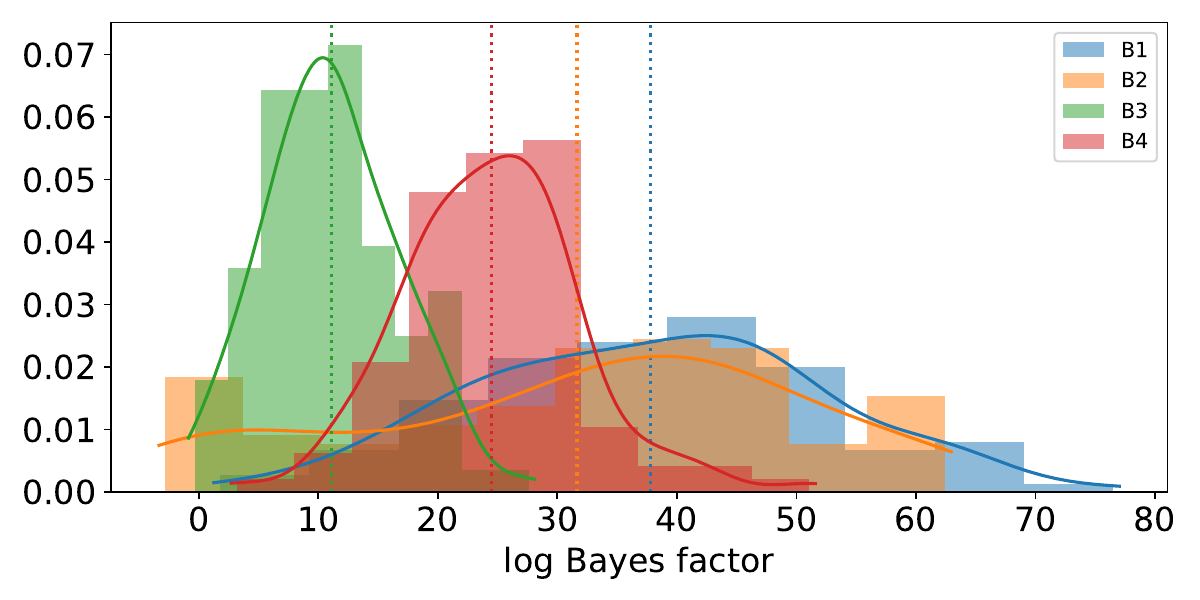} \\
	\caption{\label{fig:logBBmodel} {The log Bayes factor distribution with the new likelihood for the four benchmark injections in $\mathcal{N}=100$ Gaussian noise realizations, with $\snr_{\rm echo}\approx 16$ and $T\Delta f_{\rm max}\approx 200$.}}
\end{figure}

\begin{figure}[!h]
	\centering
    \includegraphics[width=10cm]{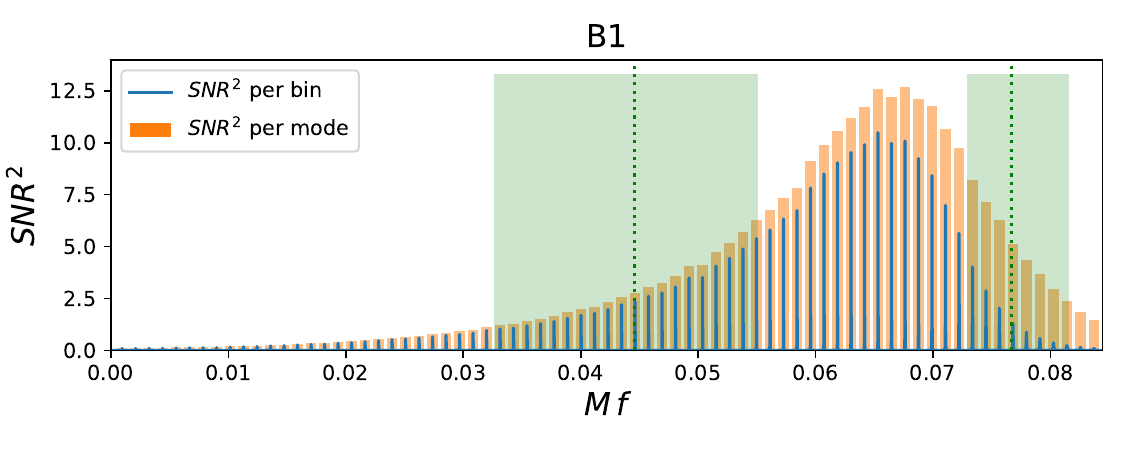} \,    \includegraphics[width=4.5cm]{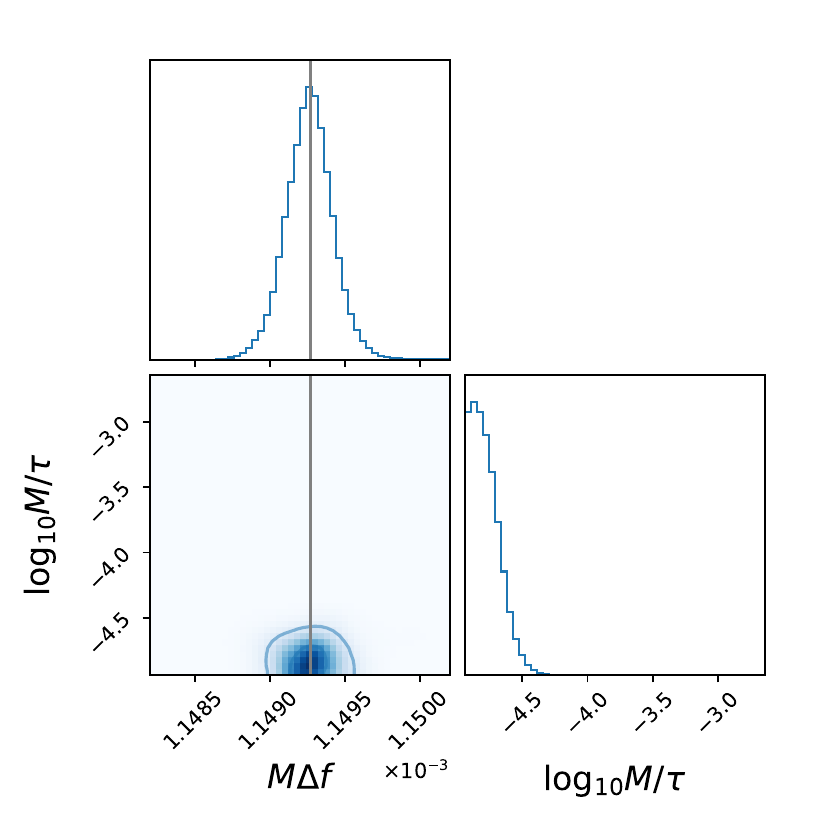} \\
    \includegraphics[width=10cm]{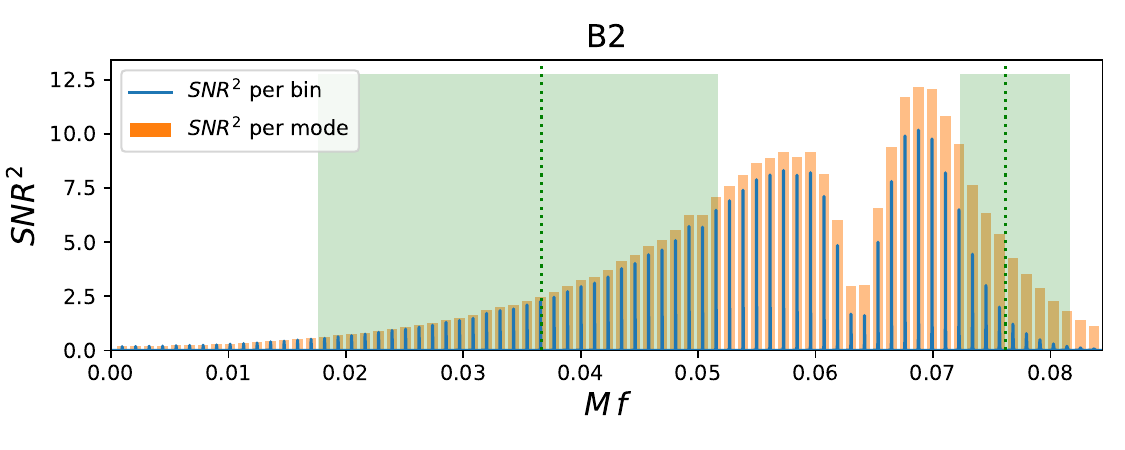} \,    \includegraphics[width=4.5cm]{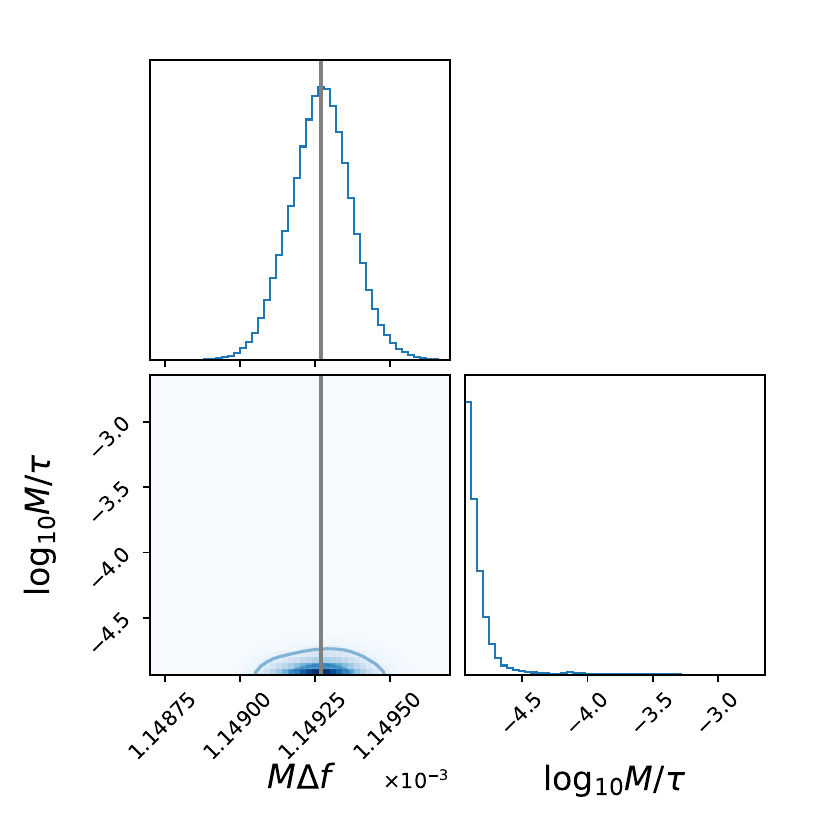} \\ 
    \includegraphics[width=10cm]{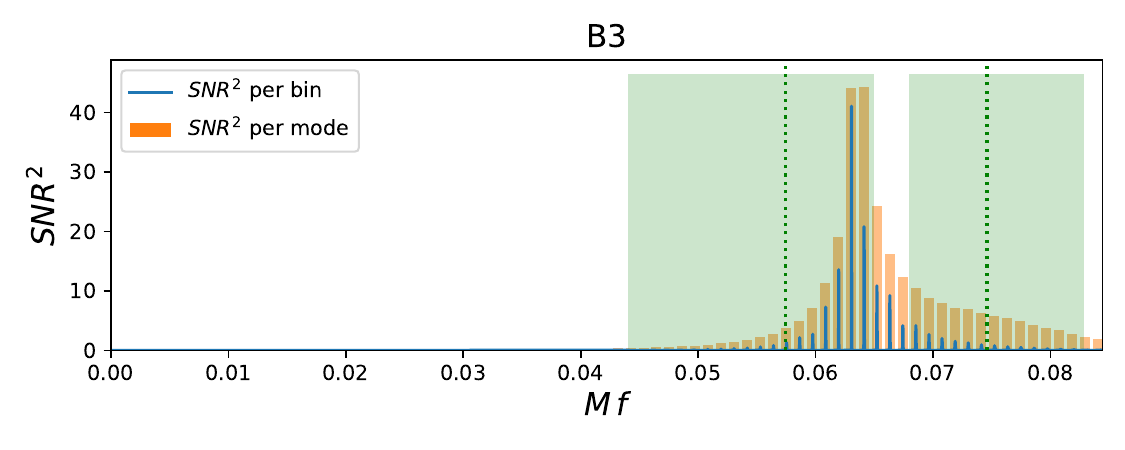} \,    \includegraphics[width=4.5cm]{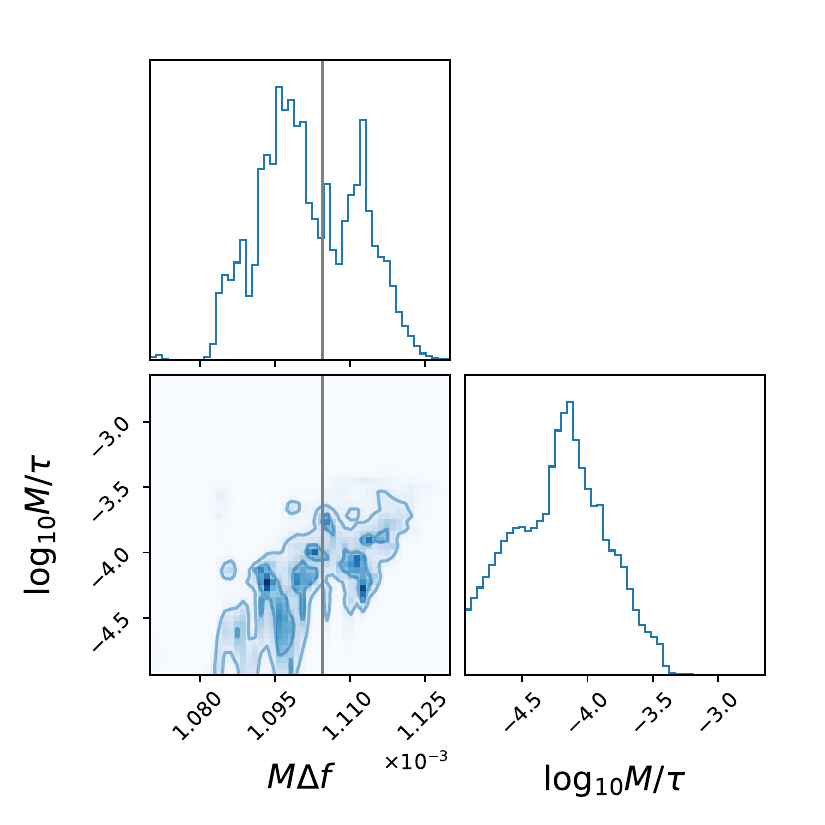} \\    
    \includegraphics[width=10cm]{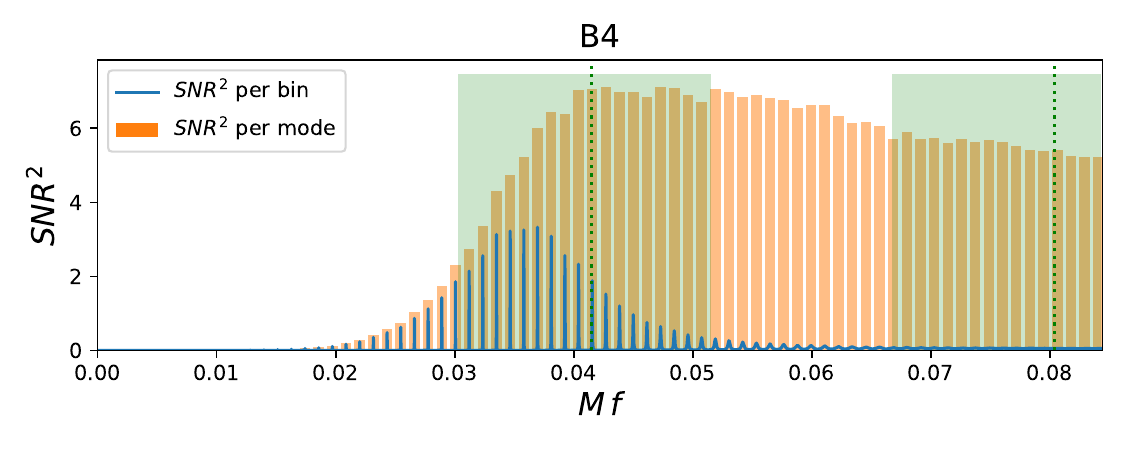} \,    \includegraphics[width=4.5cm]{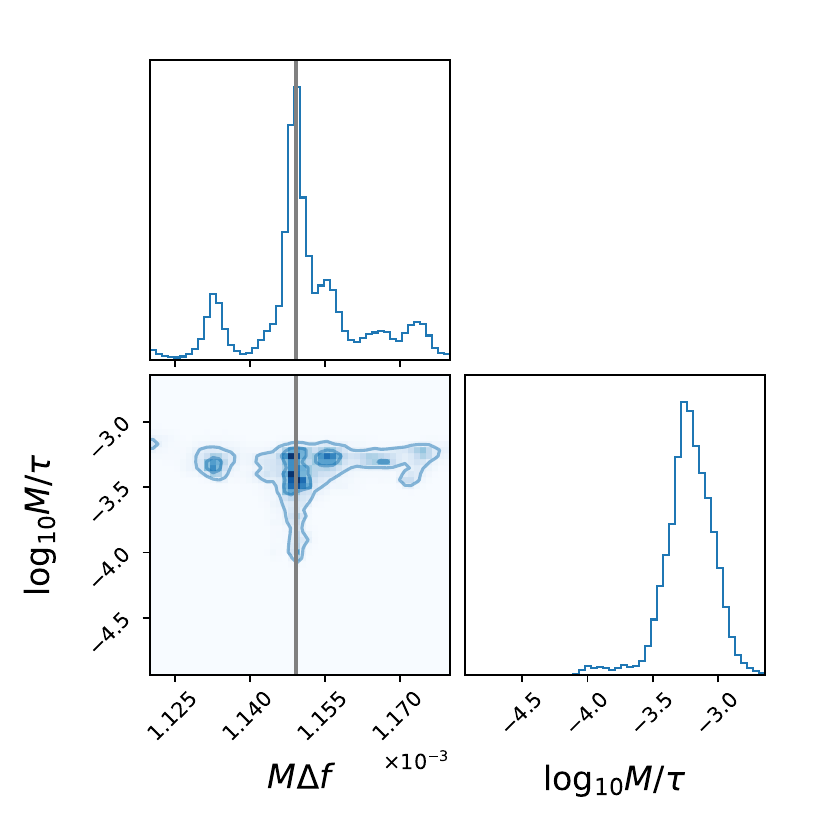}    
\caption{\label{fig:Bmodel} {The four benchmarks and search results for their injections with $\snr_{\rm echo}\approx 16$ and $T\Delta f_{\rm max}\approx 200$. Left column: $\snr^2$ per frequency bin with a bin resolution $M/T\approx10^{-5}$ (blue lines) and $\snr^2$ per QNM (orange bars). A well-resolved QNM has the orange bar much higher than the blue line. The two green bands (dashed vertical lines) denote the symmetric 90\%
credible intervals (median values) of $f_{\rm min}$ and $f_{\rm max}$ from the overall posterior distributions. 
Right column: the corner plots for the spacing $M \Delta f$ and the width $\log_{10}M/\tau$ with the overall posterior distributions. The vertical gray lines present the best-fit values of $M\Delta f$ from the injected spectrum within the inferred frequency ranges.}
}
\end{figure}

As an initial exploration, let us closely examine the search results for the four benchmarks at a specific time duration, i.e. $T\Delta f_{\rm max}\approx 200$. Figs.~\ref{fig:logBBmodel} and \ref{fig:Bmodel} present the results with the new likelihood (See Tab.~\ref{tab:BMresults_new} in Appendix.~\ref{sec:moreresults} for all results of the inferred parameters). In all cases, the log Bayes factor distributions are approximately Gaussian, with mean values considerably greater than zero. This indicates a high detection probability of the injected signals. 

The search performance for B1 injections outperforms the other three, but due to the mismatch with the search template and increasing number of search parameters, its log Bayes factor distribution is slightly worse than that of the $\uew$ injections with similar $\snr$ in Fig.~\ref{fig:violin1} at $T\gtrsim\tau_{\rm inj}$. Its overall posterior distributions of width and spacing closely resemble those in Fig.~\ref{fig:posterior1}. Specifically, the inferred parameters with the symmetric 90\% credible interval from the overall posterior distribution are found as 
\begin{eqnarray}
&&M \Delta f\approx 11492_{-5746}^{+3}\times 10^{-7},\quad
q_0=0.83^{+0.01}_{-0.16},\quad
A/\langle\tilde{P}\rangle^{1/2}\approx 3.3^{+1.3}_{-1.2},\nonumber \\
&&\log_{10} M/\tau\approx -4.8^{+0.2}_{-0.1},\quad
M f_{\rm min}\approx 0.046^{+0.010}_{-0.013},\quad
M f_{\rm max}\approx 0.076^{+0.005}_{-0.003}\,.
\end{eqnarray}
The average spacing $M \Delta f$ accurately peaks at the best-fit value of the injected spectrum within the inferred frequency range, suggesting a small variation in mode spacing within this range. This parameter is also determined with exceptional precision, demonstrating a tiny relative error of $\mathcal{O}(0.01\%)$ as in the case for $\uew$ injections in Eq.~(\ref{eq:uew90CI}).
The average width, height, and frequency range of the dominant QNMs are determined less accurately, with relative errors of $\sim\mathcal{O}(10\%)$. 
This information allows for the inference of a variety of interesting physical quantities. The number of pulses included in the data segment is $N_E\approx T\Delta f \approx 100$. The template SNR is found to be $\snr\approx 14.3^{+2.6}_{-2.8}$, which is only slightly below $\snr_{\rm echo}$ for the injected signal, indicating that the majority of QNMs have been captured by the template. The combined reflectivity within the searched frequency band can be inferred with $1/|\ln\mathcal{R}_{\rm eff}|\approx\tau \Delta f\approx 70_{-34}^{+26}$, which exactly recovers the theoretical prediction in Fig.~\ref{fig:Rwall}. 
These results suggest that the simple $\uew$ template can effectively capture B1-like echo signals and allow for extracting valuable information about the underlying physics.

\begin{figure}[!h]
	\centering
   \includegraphics[width=15cm]{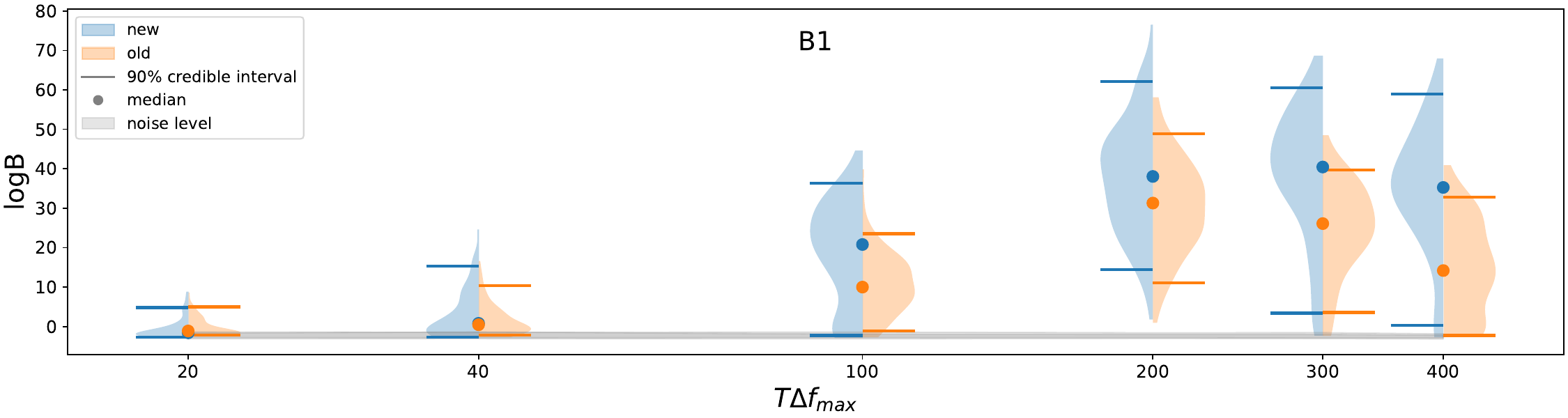} 
   \\
    \includegraphics[width=15cm]{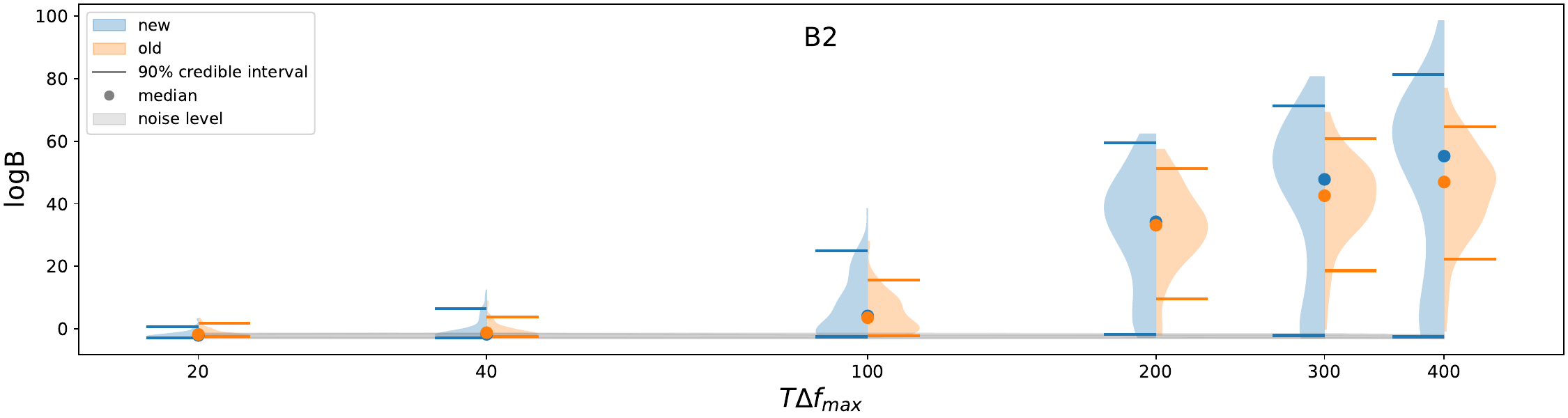} \\ 
    \includegraphics[width=15cm]{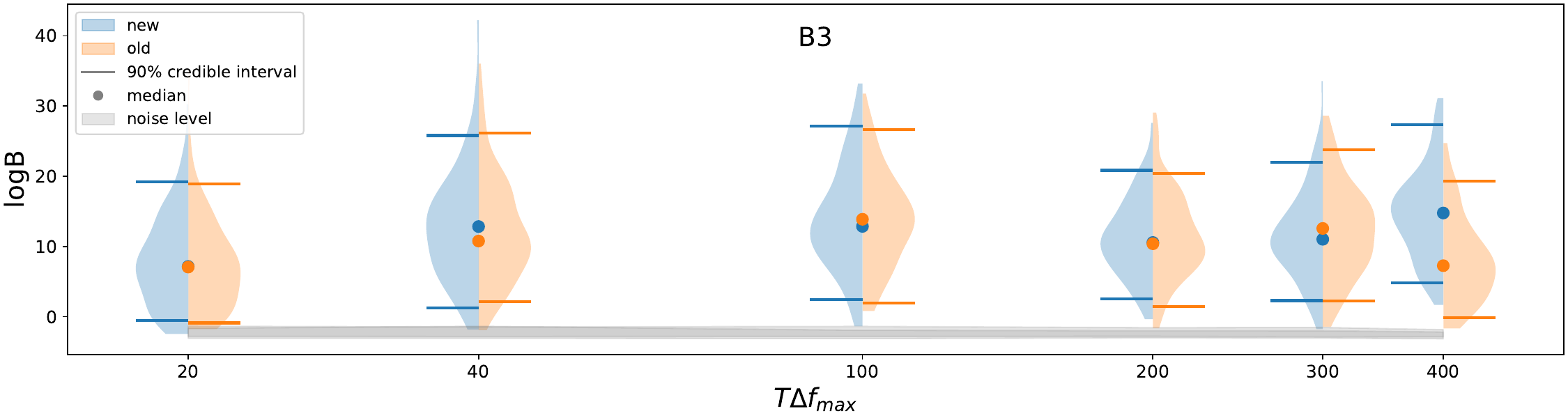} \,    \includegraphics[width=15cm]{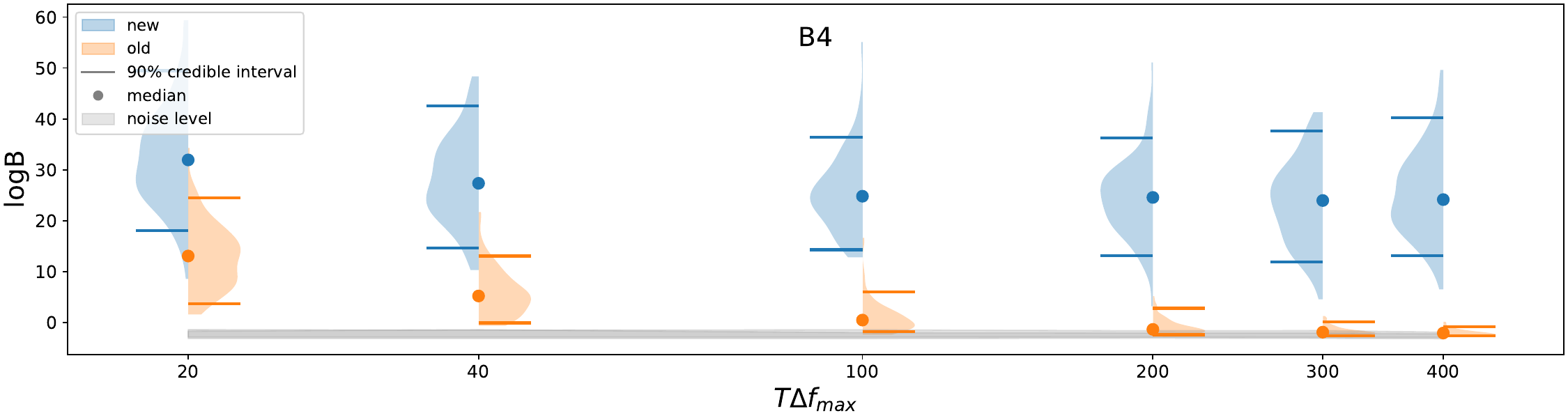}  
\caption{\label{fig:logBBmodelall} {
The log Bayes factor  distributions for the four benchmarks as a function of the time duration $T\Delta f_{\rm max}$. The signal amplitude is selected such that  $\snr_{\rm echo}\approx16$ at $T\Delta f_{\rm max}\approx 200$. The blue and orange colors represent the results obtained using the new and old likelihoods, respectively. The upper and lower bars represent the symmetric 90\% credible intervals, and the dots denote the median values. The gray band represents the 90\%
credible interval of the noise distributions.}
}
\end{figure}



\begin{figure}[!h]
	\centering  
 \includegraphics[width=16cm]{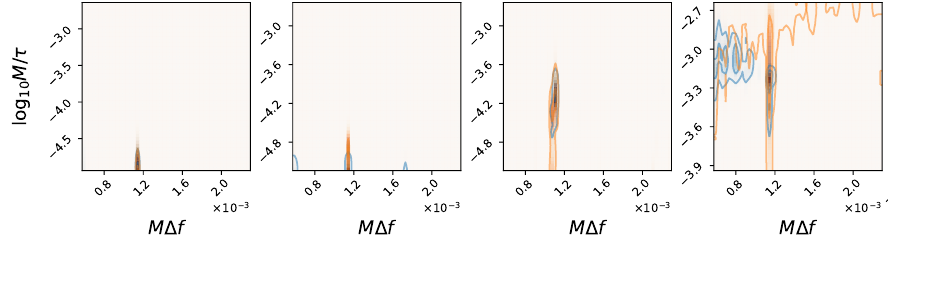}
	\\ 
 \includegraphics[width=16cm]{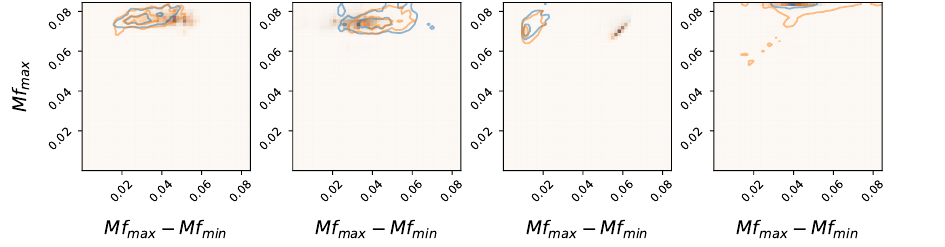}
\caption{\label{fig:Bmodelcorner} {
The overall posterior distributions of $M\Delta f$ and $\log_{10}M/\tau$ (top); $M f_{\rm max}-M f_{\rm min}$ and $Mf_{\rm max}$ (bottom) for one representative case of each benchmark, i.e. B1 with $T\Delta f_{\rm max}=200$ (first column), B2  with $T\Delta f_{\rm max}=400$ (second column), B3  with $T\Delta f_{\rm max}=400$ (third column), B4 
 with $T\Delta f_{\rm max}=20$ (fourth column), respectively. The full prior ranges are shown. The contours denote the $1\sigma$ and $2\sigma$ ranges.}
}
\end{figure}

\begin{figure}[!h]
	\centering
    \includegraphics[width=7.5cm]{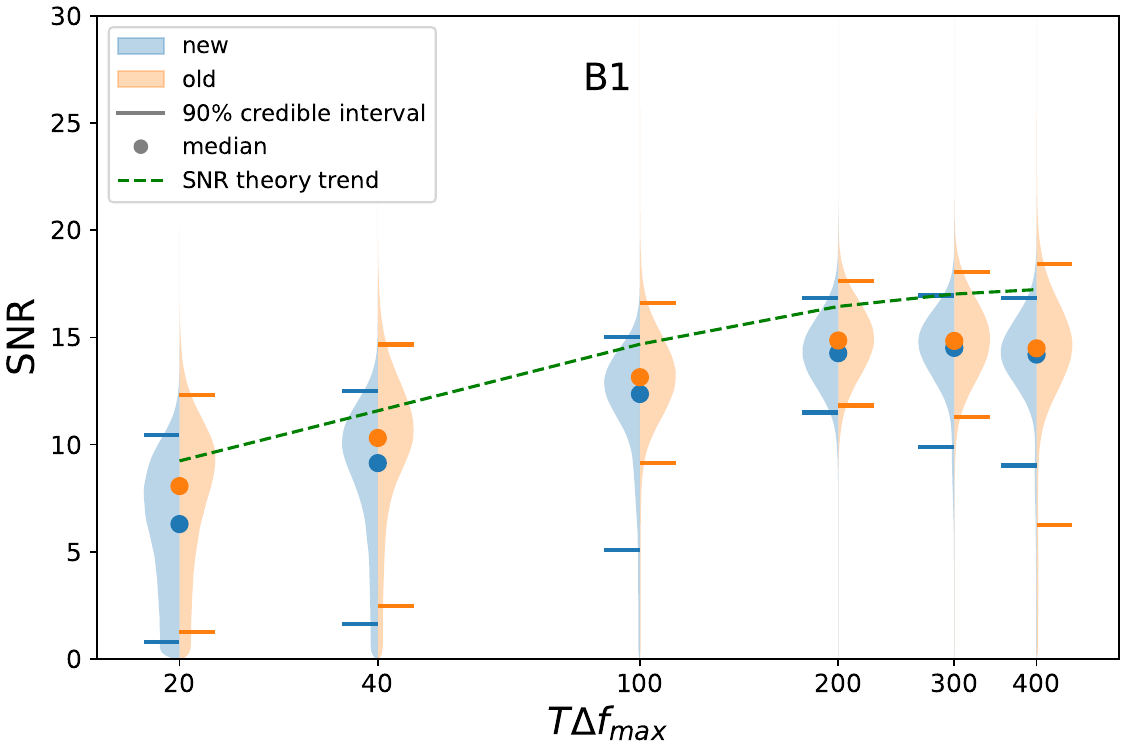} \,
    \includegraphics[width=7.5cm]{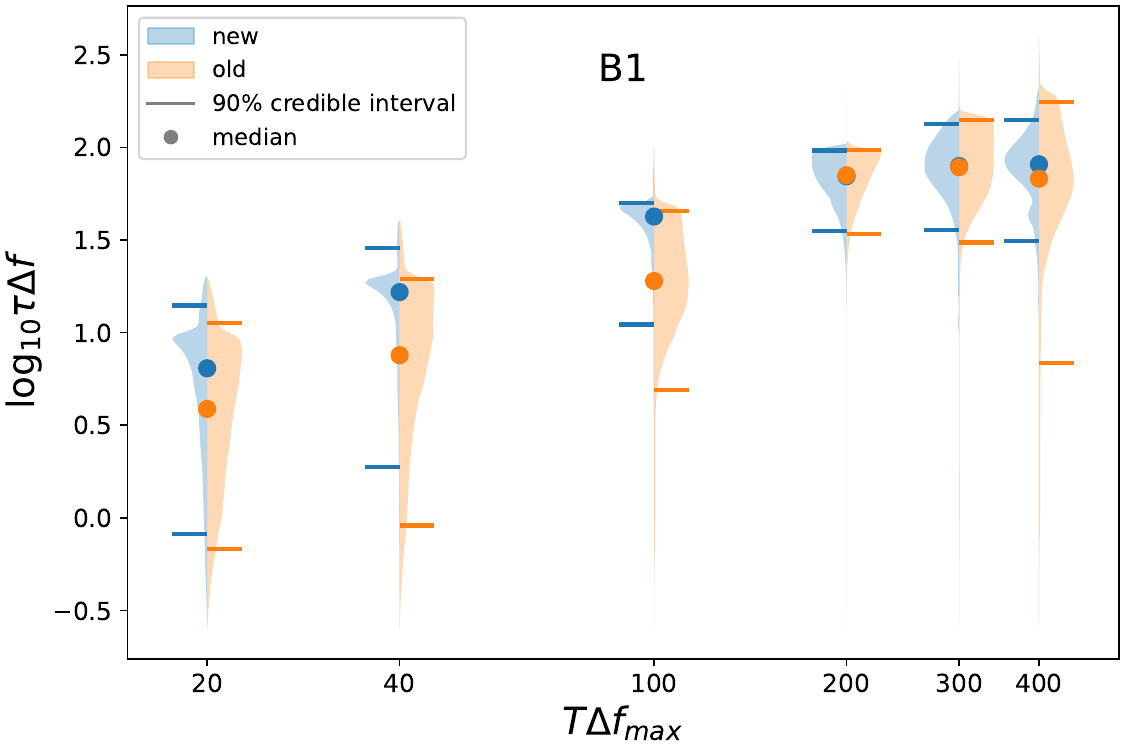} \\
	\includegraphics[width=7.5cm]{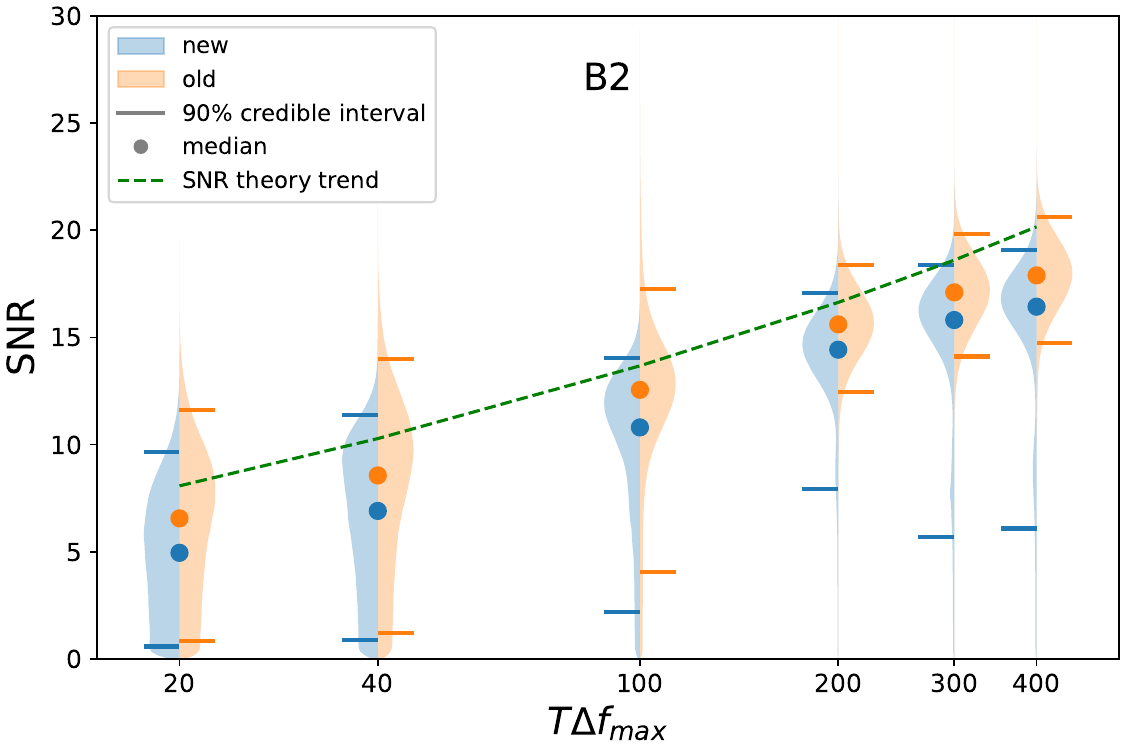} \,
    \includegraphics[width=7.5cm]{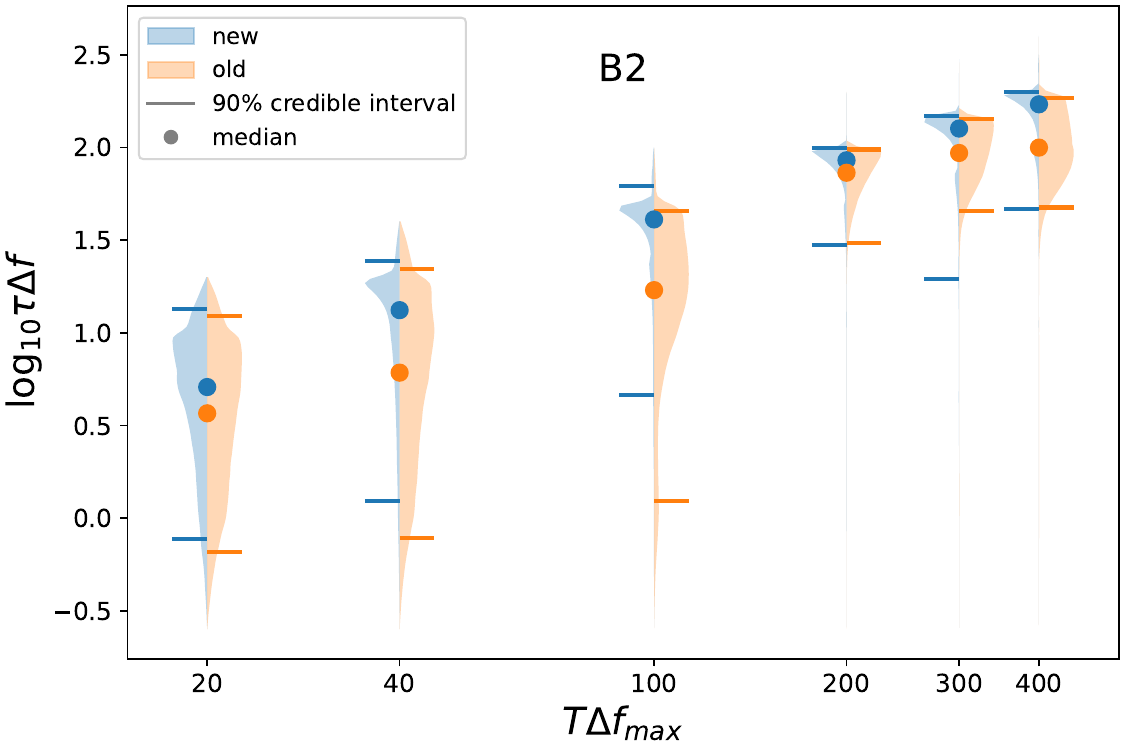} \\     
    \includegraphics[width=7.5cm]{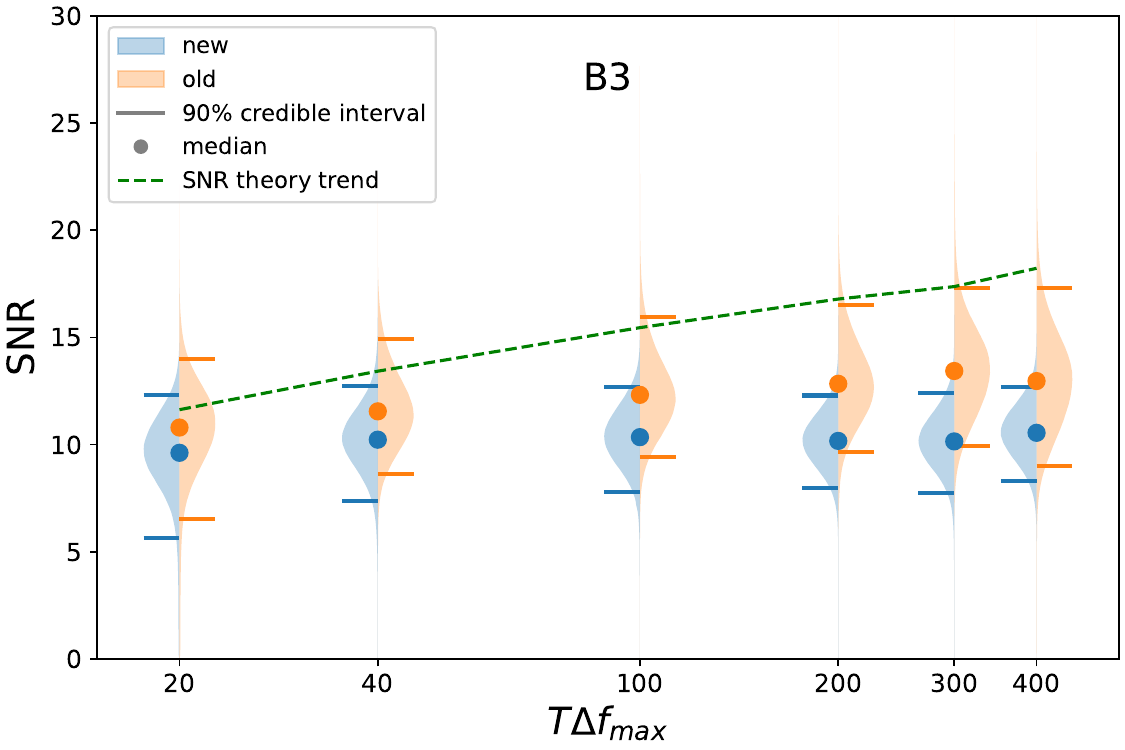} \,    
    \includegraphics[width=7.5cm]{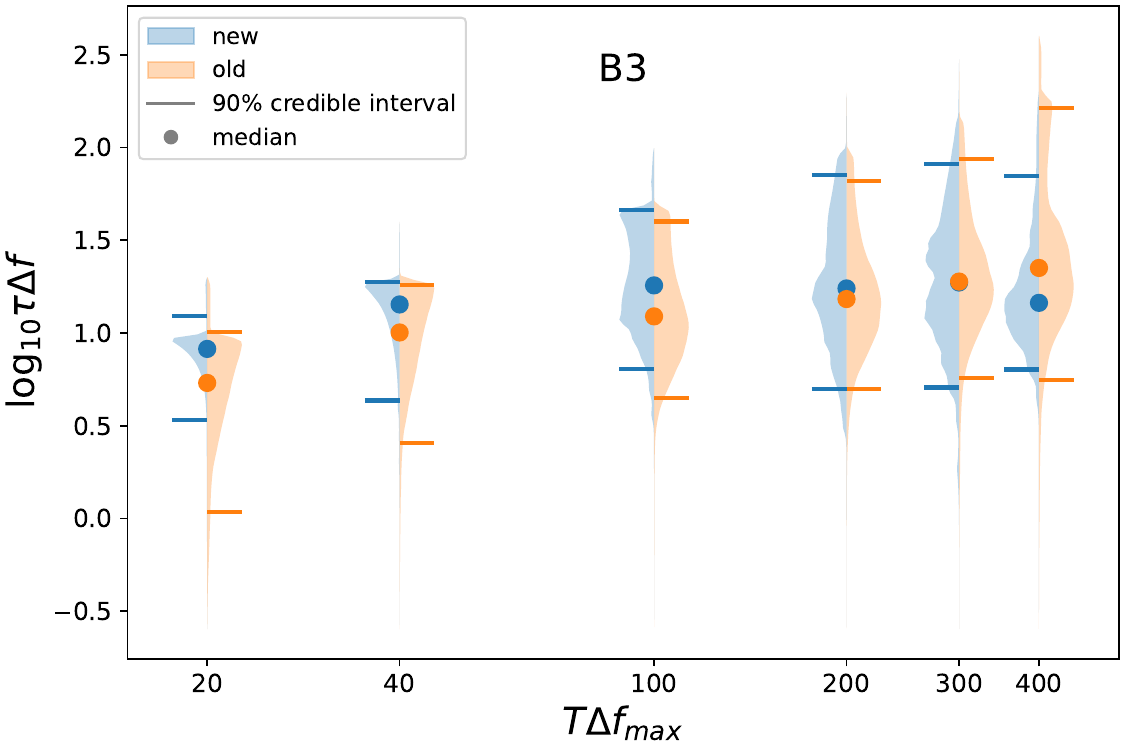} \\  	
	\includegraphics[width=7.5cm]{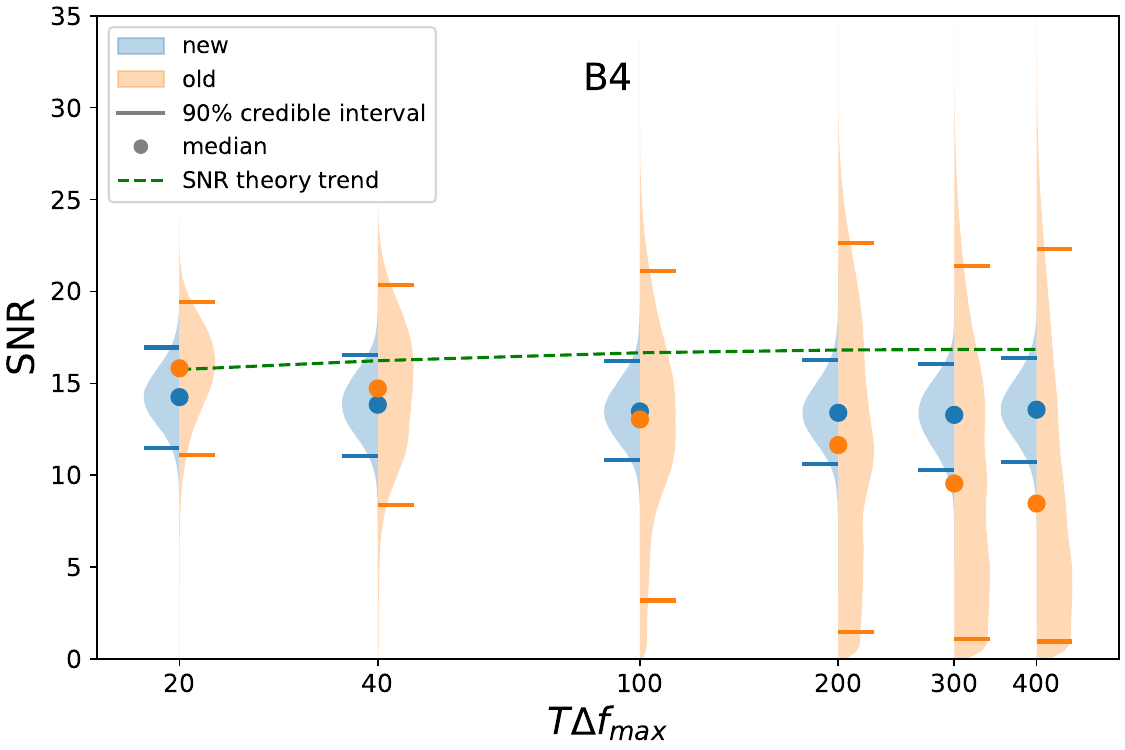} \,  
	\includegraphics[width=7.5cm]{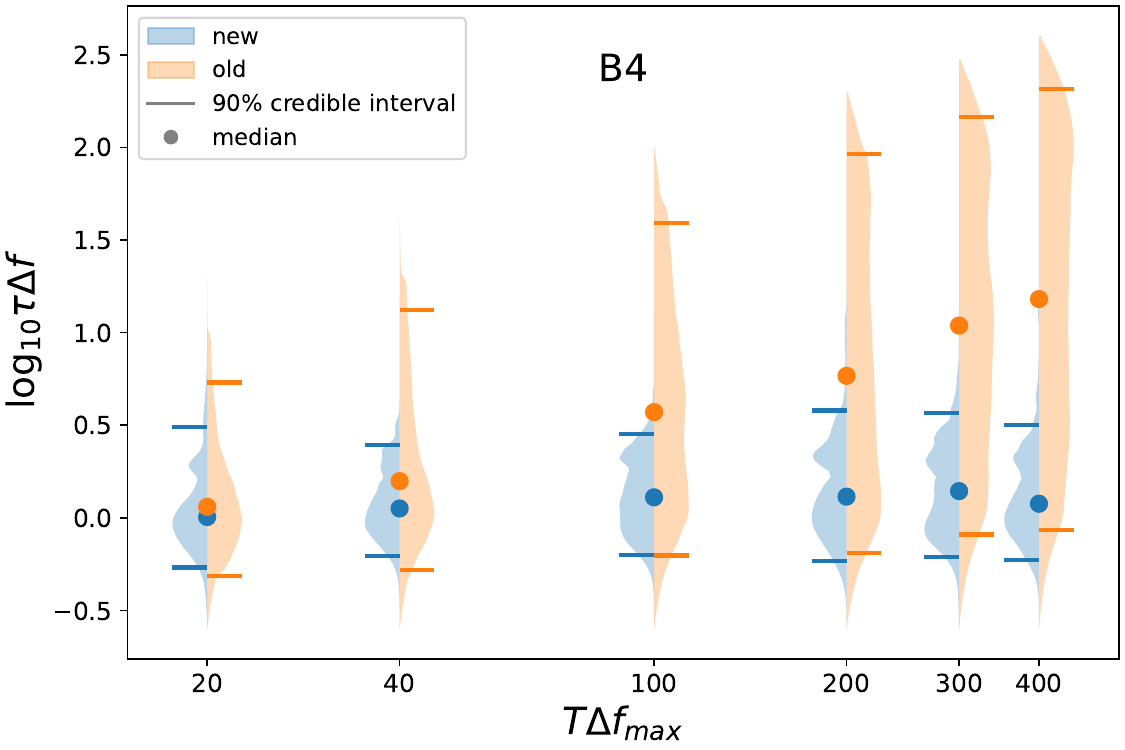}  
\caption{\label{fig:SNRBmodelall} {
The overall posterior distributions of the inferred SNR (left) and the combined reflectivity $1/|\ln\mathcal{R}_{\rm eff}|\approx \tau\Delta f$ (right) of the four benchmarks as a function of $T\Delta f_{\rm max}$. The green dashed line (left) denotes the theoretical prediction of $\snr_{\rm echo}$.  Other notations follow the same conventions as those in Fig.~\ref{fig:logBBmodelall}.}
}
\end{figure}

The performance for B2 is slightly worse, with a few failed cases where the log Bayes factor hovers around zero. This can be attributed to the unresolved QNMs at low frequency, where the current setting of $n_{\rm live}$ may not be sufficient. This observation is further supported by the overall posterior distribution of the width, which peaks at the lower end (i.e. $1/T$) of the prior range.
B3 exhibits the poorest performance due to the sharp decline in mode SNR away from $\omega_H$. Since this model is dominated by only a small number of QNMs, the average spacing cannot be determined as accurately as for B1 and B2. The search performance is also somewhat deteriorated by our requirement of no less than ten modes. 
The final example B4 is distinct from the other three due to the dominant presence of wide QNMs at high frequency, as illustrated in Fig.~\ref{fig:Bmodel}.\footnote{This is possible because we set the frequency band upper end by $\omega_{\rm RD}$ for $l=m=2$ mode, while B4 is dominated by the negative frequency component with a smaller fundamental mode frequency ($l=-m=2$).}
Although these wide modes were not the target of our search algorithm, they can still significantly overlap with the $\uew$ template. As a result, the evidence for the signal model is strong, as indicated by large log Bayes factors, but this comes at the cost of much longer search times and poorer parameter estimation, especially for the spacing.

Figures~\ref{fig:logBBmodelall}-\ref{fig:SNRBmodelall} summarize the full search results for the four benchmarks, revealing their distinct features from different perspectives (see Appendix.~\ref{sec:moreresults} for additional search results). 
The results of B1 again closely resemble those of the $\uew$ injections shown in Fig.~\ref{fig:violin1}. As the time duration increases, the signal model becomes more evident for both likelihoods due to an increase in $\snr_{\rm echo}$. When $\snr_{\rm echo}$  saturates the continuous limit for $T\Delta f_{\rm max}\gtrsim 200$, the new likelihood's performance remains stable as $T$ increases, while the old likelihood's performance deteriorates rapidly. The median value of the template SNR traces well the injected value of $\snr_{\rm echo}$ for all $T$. The inferred combined reflectivity $1/|\ln\mathcal{R}_{\rm eff}|$ peaks around the maximum value $T\Delta f$ for the small time duration case, as the frequency resolution is insufficient to resolve the dominant QNMs. When $T\Delta f_{\rm max}\gtrsim 200$, its posterior distribution is more symmetric and the median value reaches a saturation point, offering a reliable estimate of the average $1/|\ln\mathcal{R}_{\rm eff}|$ for the detected QNMs.  
For B2, both likelihoods show a faster increase in the log Bayes factor as $T$ increases, which is associated with the increasing $\snr_{\rm echo}$ shown in Fig.~\ref{fig:SpecBol}. The new likelihood is particularly effective when $T\Delta f_{\rm max}\gtrsim 300$, allowing for the detection of narrower modes at low frequency and the recovery of strong reflectivity associated with UCOs. The presence of a long tail of log Bayes factor distribution extending toward zero suggests that there may be insufficient sampling of the narrower QNMs at low frequency.
For B3, both likelihoods show stable performance with respect to $T$. However, due to the sharp decline of mode SNR away from $\omega_H$ and the requirement of ten modes within the band, the template SNRs tend to deviate more from the injected $\snr_{\rm echo}$. The old likelihood is more influenced by the highest modes, resulting in a larger inferred amplitude. The inferred $1/|\ln\mathcal{R}_{\rm eff}|$ is not much different and is more sensitive to the subdominant wider modes.
%
Thus far, the two likelihoods have captured almost the same QNMs for the first three benchmarks. However, this does not hold true for B4, where the differences between the QNMs identified by the two likelihoods become more pronounced. The new likelihood prefers wider modes at higher frequency, resulting in a stable performance for the list of $T$. Particularly, a much smaller value of spacing is allowed due to the large overlap for the wider modes. In contrast, the old likelihood tends to capture narrower modes at lower frequency, and its performance quickly deteriorates as $T$ increases. This is also evident from the larger inferred values of $1/|\ln\mathcal{R}_{\rm eff}|$ with the old likelihood. This highlights the complementary roles played by the two likelihoods in capturing different subsets of QNMs in the spectrum.   

As a final remark, while our discussions have centered around a specific choice of the echo SNR, i.e. $\snr_{\rm echo}\approx 16$ for $N_E\approx 100$, it is straightforward to generalize the analysis to benchmarks with a range of $\snr_{\rm echo}$ values. 
For models incorporating long-lived QNMs such as B1 and B2, a GW150914-like event with  $\snr_{\rm RD}\approx 8$
would predict a higher $\snr_{\rm echo}$ than the current choice from Fig.~\ref{fig:SpecBol}, thus indicating an increased detection probability with our method. 
On the other hand, $\snr_{\rm echo}$ with only the first few pulses would be significantly smaller, rendering the existing model-independent methods in Ref.~\cite{Tsang:2018uie,Tsang:2019zra,Miani:2023mgl} inefficient. This highlights the advantage of our method in capturing long-lived QNMs associated with echoes.

\section{Summary}
\label{sec:summary}

The identification of characteristic quasinormal modes in the postmerger signal is essential for discerning the nature of ultracompact objects, otherwise known as the ``quantum black hole seismology''.
When UCOs have a strong interior reflection and high compactness, they feature distinct quasiperiodic and long-lived QNMs that differentiate them from those expected in classical black holes.
The superposition of these modes then results in slowly damped echoes at a timescale much longer than that for the typical ringdown.
In this study, building upon our previous work that targeted these characteristic modes~\cite{Ren:2021xbe}, we incorporate phase information to further optimize the Bayesian search algorithm. This approach complements the existing model-independent searches for quasi-periodic bursts that rapidly decay in the time domain~\cite{Tsang:2018uie,Tsang:2019zra,Miani:2023mgl}.

We begin with a generic discussion of the echo waveform in Sec.~\ref{sec:generic}, and find that the phase for one QNM can be accurately approximated by the Lorentzian shape contribution up to a constant.
A new phase-marginalized likelihood is then derived in Eq.~(\ref{eq:newlnL}) by only marginalizing a constant phase per mode. Compared to the old likelihood in Eq.~(\ref{eq:oldlnL}) that discarded all phase information, the relative phases are now maintained without adding new parameters. Consequently, the new method can significantly improve the search sensitivity for fully resolved QNMs. A simple periodic and uniform model of echo waveform ($\uew$), defined in Eq.~(\ref{eq:uniformQNM}), is then adopted as the search template. This model, which involves only seven parameters, allows for an efficient detection of the variety of echo waveforms. By considering a four parameter search for a list of time duration, we verify the reliability of the Bayesian algorithm for the $\uew$ injections with the two likelihoods in Sec.~\ref{sec:BayesianUEW}. The influence of different noise realizations is also taken into account.
As shown in Fig.~\ref{fig:violin1}, the performance of the two likelihoods at low frequency resolution is similar due to the limited number of useful frequency bins. However, they exhibit different behaviors when the modes start to be resolved. Specifically, the new likelihood remains stable as the time duration increases, while the old likelihood deteriorates rapidly in the high resolution limit. Thus, incorporating the relative phases allows for the detection of QNMs with amplitudes well below the noise level. 
Efficient detection of long-lived QNMs also requires appropriate sampler settings.  Because of limited computational resources, the old likelihood can outperform the new one occasionally by reducing the occurrence of failed searches, as illustrated in Fig.~\ref{fig:uewresultnliv}.

Finally, in Sec.~\ref{sec:validate}, we validate our search algorithm using representative benchmarks of echo waveforms. 
The main uncertainties for echo waveforms arise from the effective reflection from the interior boundary and the frequency content of the initial pulse. To address these uncertainties, we construct four complementary benchmarks based on physical models, as depicted in Fig.~\ref{fig:SpecBol}. We then perform a more generic six-parameter Bayesian search with the two likelihoods for their injections, with the parameter settings  specified in Tab.~\ref{tab:parameter2}. The main search results are summarized in Figs.~\ref{fig:logBBmodelall}-\ref{fig:SNRBmodelall}, which display distinct features for various benchmarks with respect to the variations of time duration of the data segment. The use of both likelihoods can significantly increase the detection probability for echoes by capturing different subsets of QNMs, given that a real spectrum incorporates QNMs with different shapes and heights.  If the detection probability is sufficiently large, we can claim the detection of QNMs of UCOs and infer the credible interval of their average width, spacing, height and frequency range. Otherwise, we can impose an upper limit on the SNR or the emitted GW energy of the dominant QNMs.

These results demonstrate the robustness of our improved search algorithm in detecting the variety of long-lived QNMs associated with echoes. 
Furthermore, when considering more realistic UCO models beyond a truncated Kerr black hole, we anticipate that the search algorithm will continue to perform effectively, as long as the timescales fall within the capabilities of the current dataset. However, it is important to note that due to the intrinsic contributions arising from the UCO's interior, the priors for different parameters will need to be adjusted accordingly. Additionally, the inference of underlying physics for UCOs will become less straightforward.

Our next step is to apply the proposed search to strain data of the confirmed gravitational wave events from Advanced LIGO, Virgo, and KAGRA. Having validated the search algorithm with the old likelihood in real detector noise by notching out prominent instrumental lines~\cite{Ren:2021xbe}, we will then closely examine this procedure with the new likelihood. We will also investigate whether the additional information provided by the relative phases can help to further distinguish the signal from the instrumental lines. 
This endeavor will serve as a valuable complement to the ongoing model-independent search for echoes by the LIGO, Virgo, and KAGRA Collaborations, where only the BayesWave algorithm has been implemented~\cite{LIGOScientific:2021sio}.


\vspace{0.1cm}
\section*{Acknowledgements} 
\vspace{-0.1cm}
We would like to thank Naritaka Oshita for sharing his code for computing the reflection and transmission coefficients  of the light-ring potential barrier of black hole and Shuo Xin for communications on the calculation of echo waveform in Ref.~\cite{Xin:2021zir}. We thank Vitor Cardoso and Bob Holdom for useful comments on the manuscript. D. W., P. G. and J. R. are supported in part by the National Natural Science Foundation of China under Grant No. 12275276. N. A. is supported by the University of Waterloo, Natural Sciences and Engineering Research Council of Canada (NSERC) and the Perimeter Institute for Theoretical Physics.
Research at Perimeter Institute is supported in part by the Government of Canada through
the Department of Innovation, Science and Economic Development Canada and by the
Province of Ontario through the Ministry of Colleges and Universities.


\appendix

\section{Generic construction of the echo waveform}
\label{sec:formalism}

The spin weight $s=-2$ perturbations on a Kerr background spacetime are described by the Teukolsky equation. After separation of variables, the asymptotic solutions of the radial equation at the horizon and the spatial infinity are described by amplitudes $B_i$,
\begin{eqnarray}\label{eq:TEamp}
\mathfrak{R} \to
\left\{\begin{array}{cc}
B_\textrm{trans}\Delta^2 e^{-i \tilde\omega x}+B_\textrm{ref}e^{i \tilde\omega x},& x\to-\infty\\
B_\textrm{in}\frac{1}{r}e^{-i \omega x}+B_\textrm{out}r^3 e^{i \omega x},& x\to\infty\,,\end{array}
\right.
\end{eqnarray}
where we suppress the $lm\omega$ dependence for various variables for simplicity. 
$x$ is the tortoise coordinate with $dx/dr = (r^2+a^2)/\Delta$, $\Delta=r^2+a^2-2M r$ and $a=J/M\,(=\chi M)$. $\tilde\omega=\omega-\omega_H$ denotes the frequency close to inner boundary, where $\omega_H=m \Omega_H$ and $\Omega_H=a/(2Mr_+)$ is the horizon angular frequency. Since the potential of radial Teukolsky equation is not short ranged, the asymptotic behavior above take strange forms. 

To facilitate the numerical computation, it is convenient to transform to the Sasaki-Nakamura (SN) equation with the following transformation for the radial variable~\cite{Sasaki:2003xr}, 
\begin{eqnarray}\label{eq:TtoSN}
X= (r^2+a^2)^{1/2}r^2J_-J_-\left(\frac{1}{r^2}\mathfrak{R}\right)\;,
\end{eqnarray}
where $J_- = (d/dr) - i(K/\Delta)$ and $K=(r^2+a^2)\omega-ma$. With a short-ranged potential located around $x\sim 0$, the asymptotic form of the SN equation at spatial infinity and horizon simplifies as
\begin{align} \label{eq:SNasympto}
 \frac{d^{2} X}{d x^{2}}-V X=S'\,,
\end{align}
where $V\to -\omega^2$ and $-\tilde\omega^2$ when  $x\to \infty$ and $-\infty$, respectively. $S'$ denotes the source term in the SN formalism. The asymptotic solutions then take pure sinusoidal forms,
\begin{eqnarray}\label{eq:SNEamp}
X\to
\left\{\begin{array}{cc}
A_\textrm{trans}e^{-i \tilde\omega x}+A_\textrm{ref}e^{i \tilde\omega x},& x\to-\infty\\
A_\textrm{in}e^{-i \omega x}+A_\textrm{out}e^{i \omega x},& x\to\infty\,, \end{array}
\right.
\end{eqnarray}
with amplitudes $A_i$. Given the transformation in Eq.~(\ref{eq:TtoSN}), the two sets of amplitudes are linearly related by~\cite{Sasaki:2003xr} 
\begin{eqnarray}\label{eq:TSNtransf}
B_\textrm{in}=-\frac{1}{4 \omega ^2}A_\textrm{in},\quad
B_\textrm{out}=-\frac{4 \omega ^2}{c_0}A_\textrm{out},\quad
B_\textrm{trans}=\frac{1}{d}A_\textrm{trans},\quad
B_\textrm{ref}=\frac{1}{g}A_\textrm{ref}.
\end{eqnarray}
with $c_0=\lambda(\lambda+2)-12a\omega(a\omega-m)-i 12\omega M$, $d=-4(2M r_+)^{5/2}\left[(k_H^2-8\epsilon^2)+i 6 k_H \epsilon\right]$, $\epsilon=(r_+-M)/(4M r_+)$, $g=-b_0/(4k_H(2Mr_+)^{3/2}(k_H+i 2\epsilon))$ and $b_0=\lambda ^2+2 \lambda-96 k_H^2 M^2+72 k_H M r_+ \omega -12 r_+^2 \omega ^2-i [16 k_H M \left(\lambda+3-3M/r_+\right)-12 M \omega -8 \lambda  r_+ \omega]$. 

For a given source, we look for solutions that are outgoing at infinity, and satisfy the corresponding boundary conditions of 
BHs and UCOs near the horizon, namely, 
\begin{eqnarray}\label{eq:TBHECO}
\mathfrak{R}_{\rm BH} =
\left\{\begin{array}{cc}
Z_{\rm BH}^{-}\Delta^2 e^{-i \tilde\omega x},& x\to-\infty\\
Z_{\rm BH}^{+}r^3 e^{i \omega x},& x\to\infty\,,
\end{array}
\right.,\;
\mathfrak{R}_{\rm UCO} =
\left\{\begin{array}{cc}
Z_{\rm UCO}^{\rm trans}\Delta^2 e^{-i \tilde\omega x}+Z_{\rm UCO}^{\rm ref}e^{i \tilde\omega x},& x\to-\infty\\
Z_{\rm UCO}^{+}r^3 e^{i \omega x},& x\to\infty
\end{array}
\right.
\end{eqnarray}
in the Teukolsky formalism, and 
\begin{eqnarray}\label{eq:SNBHECO}
X_{\rm BH} =
\left\{\begin{array}{cc}
\xi_{\rm BH}^{-}e^{-i \tilde\omega x},& x\to-\infty\\
\xi_{\rm BH}^{+}e^{i \omega x},& x\to\infty\,,
\end{array}
\right.,\;
X_{\rm UCO} =
\left\{\begin{array}{cc}
\xi_{\rm UCO}^{\rm trans} e^{-i \tilde\omega x}+\xi_{\rm UCO}^{\rm ref}e^{i \tilde\omega x},& x\to-\infty\\
\xi_{\rm UCO}^{+}e^{i \omega x},& x\to\infty
\end{array}
\right.
\end{eqnarray}
in the SN formalism. The gravitational wave strain at infinity is directly related to the Newman-Penrose scalar curvature $\psi_4$. If ignoring the mode mixing in the spin-weighted spheroidal harmonics, the GW strain at distance $D$ can be mapped to the UCO response at infinity for Teukolsky variable as
\begin{eqnarray}\label{eq:hZtransform}
h_{\rm UCO \,(BH)} =-\frac{2}{D} \frac{1}{\omega^2} Z_{\rm UCO \, (BH)}^+\,.
\end{eqnarray}

With the Green function method, the response of UCO at infinity can be written as 
\begin{align}\label{eq:ximaster}
\xi^{+}_{\rm UCO}=\xi_{\mathrm{BH}}^{+}+\mathcal{K}\,\xi_{\mathrm{BH}}^{-} \,.
\end{align}
The BH responses to the source $S'$ in the SN formalism are given by~\cite{Conklin:2017lwb,Conklin:2019fcs}
\begin{eqnarray}\label{eq:xiBHpm}
\xi_{\mathrm{BH}}^{\pm}(\omega)= \int_{-\infty}^{+\infty} d x \,\frac{1}{W_{\mathrm{BH}}(\omega,x)}  S'(\omega,x) X^{\mp}(\omega,x)\,, 
\end{eqnarray}
where the Wronskian $W_{\mathrm{BH}} $ is defined as 
$ W_{\mathrm{BH}}=\frac{d X^{+}}{d x} X^{-}-X^{+} \frac{d X^{-}}{d x}  $.
$X^{\pm}$ are two independent solutions of the homogeneous SN equation
\begin{eqnarray}\label{eq:SNpm}
X^+=
\left\{\begin{array}{cc}
A^+_\textrm{trans}e^{-i \tilde\omega x}+A^+_\textrm{ref}e^{i \tilde\omega x},& x\to-\infty\\
e^{i \omega x},& x\to\infty \end{array}
\right.,\;
X^-=
\left\{\begin{array}{cc}
e^{-i \tilde\omega x},& x\to-\infty\\
A^-_\textrm{in}e^{-i \omega x}+A^-_\textrm{out}e^{i \omega x},& x\to\infty \end{array}
\right.,
\end{eqnarray}
which satisfy the out-going and in-going boundary conditions at infinity and horizon, respectively.
Here, the two solutions are normalized with $A^+_{\rm out}=1$ and $A^-_{\rm trans}=1$.

The transfer function $\mathcal{K} $ is given by 
\begin{align}\label{eq:transfer}
 \mathcal{K}=\frac{  T_{\rm BH }  R_{\rm wall}}{1-R_{\rm BH } R_{\rm wall} } \,,
\end{align}
where $R_{\rm wall}=\xi_{\rm UCO}^{\rm ref}/\xi_{\rm UCO}^{\rm trans}$ is the reflection coefficient at the inner boundary given in Eq.~(\ref{eq:SNBHECO}). $R_{\rm BH}=A^+_\textrm{trans}/A^+_\textrm{ref}$ and $T_{\rm BH}=1/A^+_\textrm{ref}$ are the reflection and transmission coefficients of BH for waves coming from the left from Eq.~(\ref{eq:SNpm}). 
The response $ \xi^{+}_{\rm UCO}$ reduces to the BH counterpart $ \xi_{\mathrm{BH}}^+$ when interior reflection is absent, i.e. $ \mathcal{K}(\omega)=0$.
For a merger product, $\xi_{\mathrm{BH}}^+$ and $\mathcal{K}\,\xi_{\mathrm{BH}}^{-}$ thus give rise to the ringdown signal and echoes, respectively.

Now, we define the counterpart of $X^{\pm}$ for the homogeneous Teukolsky equation with the transformation Eq.~(\ref{eq:TSNtransf}),
\begin{eqnarray}\label{eq:TEUpm}
\mathfrak{R}^+=
\left\{\begin{array}{cc}
B^+_\textrm{trans}\Delta^2 e^{-i \tilde\omega x}+B^+_\textrm{ref}e^{i \tilde\omega x},& x\to-\infty\\
B^+_\textrm{out}r^3 e^{i \omega x},& x\to\infty \end{array}
\right.,\;
\mathfrak{R}^-=
\left\{\begin{array}{cc}
B^-_\textrm{trans}\Delta^2 e^{-i \tilde\omega x},& x\to-\infty\\
B^-_\textrm{in}\frac{1}{r}e^{-i \omega x}+B^-_\textrm{out}r^3 e^{i \omega x},& x\to\infty\end{array}
\right..
\end{eqnarray}
In terms of the Teukolsky variables, the master equation Eq.~(\ref{eq:ximaster}) then becomes:  $Z^{+}_{\rm UCO}=Z^{+}_{\mathrm{BH}}+Z^+_{\rm echo}$, with 
\begin{eqnarray}\label{eq:Zmaster}
Z^{+}_{\rm echo}&=&\mathcal{K}\, B^+_{\rm out} \frac{A^+_{\rm trans}}{B^+_{\rm trans}} Z^{-}_{\mathrm{BH}}
=\frac{  T_{\rm BH }  R_{\rm wall}}{1-R_{\rm BH } R_{\rm wall} } \frac{R_{\rm BH}}{T_{\rm BH}} \frac{B^+_{\rm out}}{B^+_{\rm trans}}Z^{-}_{\mathrm{BH}}\nonumber\\
&\equiv& \frac{  R_{\rm BH }  R_{\rm wall}}{1-R_{\rm BH } R_{\rm wall} } Z_{\rm eff}^+,\;
\textrm{with }Z_{\rm eff}^+\equiv  (B^+_{\rm out}/B^+_{\rm trans})Z^{-}_{\mathrm{BH}}\,,
\end{eqnarray}
where $Z^+/\xi^+=B^+_{\rm out}$,  $Z^-/\xi^-=B_{\rm trans}^+/A_{\rm trans}^+$ and  $A_{\rm trans}^+=R_{\rm BH}/T_{\rm BH}$ are used. $Z_{\rm eff}^+$ can be viewed as the counterpart of $Z^{-}_{\mathrm{BH}}$ at infinity.

With Eq.~(\ref{eq:hZtransform}), the observed GW strain can be put as: $h_{\rm UCO}=h_{\rm RD}+h_{\rm echo}$, where 
\begin{eqnarray}\label{eq:hmaster}
h_{\rm echo}=  \frac{  R_{\rm BH }  R_{\rm wall}}{1-R_{\rm BH } R_{\rm wall} } h_{\rm eff}^+,\; 
\textrm{with } h_{\rm eff}^+=-\frac{2}{D} \frac{1}{\omega^2} Z_{\rm eff}^+\,.
\end{eqnarray}
Thus, the theoretical modeling of the echo waveform reduces to two parts: the reflection of cavity $R_{\rm BH}R_{\rm wall}$ and the source related term $h^+_{\rm eff}$. 

For certain cases, the BH response $\xi^-_{\rm BH}$ at horizon can be further simplified. 
If the source $S'$ has the support only in the interior, i.e. $x\ll 0$, from Eqs. (\ref{eq:xiBHpm}) and (\ref{eq:SNpm}), we have 
\begin{align}\label{eq:ZBHm_in}
\xi_{\mathrm{BH}}^{-} 
& \approx  \int_{-\infty}^{+\infty} d x \frac{S'}{ W_{\mathrm{BH}}}   \left( \frac{R_{\mathrm{BH}}}{T_{\mathrm{BH}} }   e^{-i \tilde{\omega} x}+ \frac{1}{T_{\mathrm{BH}} }   e^{i \tilde{\omega} x}\right) \nonumber\\
& \approx \frac{R_{\mathrm{BH}}}{T_{\mathrm{BH}}} \xi^+_{\mathrm{BH}}+\frac{1}{T_{\mathrm{BH}}}\hat{\xi}_{\mathrm{BH}}^{+} \,,\quad \textrm{(inside)}
\end{align}
where in the last step we have used $\xi_{\mathrm{BH}}^{+} \approx \int_{-\infty}^{+\infty} d x \frac{S'}{ W_{\mathrm{BH}}}  e^{-i \tilde{\omega} x}$ and defined 
\begin{align}\label{eq:xihatp}
   \hat{\xi}_{\mathrm{BH}}^{+} \equiv \int_{-\infty}^{+\infty} d x \frac{S'}{ W_{\mathrm{BH}}}  e^{i \tilde{\omega} x}
   \approx \int_{-\infty}^{+\infty} d x  \frac{ \hat{S}^+}{ W_{\mathrm{BH}}}   X^-\,, 
\end{align}
which is the BH response at infinity to an effective source $ \hat{S}^+=S' e^{2 i \tilde{\omega} x} $ within the cavity. 
On the other hand, if the source is far away from the object and the support is mainly in the exterior, i.e. $x\gg 0$, we have
$\xi_{\mathrm{BH}}^{+}= A^-_{\rm out}\xi_{\mathrm{BH}}^{-} +A^-_{\rm in}\hat{\xi}_{\mathrm{BH}}^{-}=\frac{R'_{\rm BH}}{T'_{\rm BH}}\xi_{\mathrm{BH}}^{-}+\frac{1}{T'_{\rm BH}}\hat{\xi}_{\mathrm{BH}}^{-}$ similar from Eqs. (\ref{eq:xiBHpm}) and (\ref{eq:SNpm}). 
Here, $T'_{\rm BH}=1/A^-_{\rm in}$ and $R'_{\rm BH}=A^-_{\rm out}/A^-_{\rm in}$ are the BH reflection and transmission coefficients for waves coming from the right. $\hat{\xi}_{\mathrm{BH}}^{-}$ is defined as 
\begin{align}
   \hat{\xi}_{\mathrm{BH}}^{-} \equiv  \int_{-\infty}^{+\infty} d x \frac{S'}{ W_{\mathrm{BH}}}  e^{ - i \omega x}
   \approx  \int_{-\infty}^{+\infty} d x    \frac{\hat{S}^-}{ W_{\mathrm{BH}}}
    \tilde\Psi_+\ .
\end{align}
which is the BH response at horizon to an effective source $ \hat{S}^{-} = S'e^{-2 i \omega x} $ outside the cavity. Thus, we have 
\begin{eqnarray}\label{eq:ZBHm_out}
\xi_{\mathrm{BH}}^{-} 
\approx \frac{T'_{\mathrm{BH}}}{R'_{\mathrm{BH}}} \xi_{\mathrm{BH}}^{+}-\frac{1}{R'_{\mathrm{BH}}}\hat{\xi}_{\mathrm{BH}}^{-} \,,\quad\textrm{(outside)}\,.
\end{eqnarray}
Using the relation $T'_{\rm BH}T_{\rm BH}/(R'_{\rm BH}R_{\rm BH})=-\mathcal{T}_{\rm BH}^2/\mathcal{R}_{\rm BH}^2$ \footnote{The lhs is invariant under different perturbation variables. Using the Chandrasekhar-Detweiler (CD) expression of the lhs one can easily obtain the rhs.}, where $\mathcal{R}_{\rm BH}^2$ and $\mathcal{T}_{\rm BH}^2$ denote the energy flux reflection and transition for BH, and substituting Eqs.~(\ref{eq:ZBHm_in})  and (\ref{eq:ZBHm_out}) into Eq.~(\ref{eq:Zmaster}), we obtain the source related term for the two cases,
\begin{eqnarray}\label{eq:Zeff2case}
Z_{\rm eff}^+ \approx 
\left\{
\begin{array}{ll}
Z_{\mathrm{BH}}^{+}+ \frac{1}{R_{\rm BH}}\hat{Z}_{\mathrm{BH}}^{+}\,, & \textrm{inside}\\
-\frac{\mathcal{T}_{\rm BH}^2}{\mathcal{R}_{\rm BH}^2}\left( Z_{\mathrm{BH}}^{+}- \frac{1}{T'_{\rm BH}}\hat{Z}_{\mathrm{BH}}^{-}\right), & \textrm{outside}
\end{array}\right.\,.
\end{eqnarray}
where $\hat{Z}_{\mathrm{BH}}^{\pm}$ are proportional to $\hat{\xi}_{\mathrm{BH}}^{\pm}$.

To further simply Eq.~(\ref{eq:Zeff2case}), let's consider the special cases in which echoes are produced by an in- or outgoing pulse inside or outside the cavity. The source term for the SN equation is then obtained from the initial condition of the pulses through a Laplace transform of the homogeneous equation in the time domain. 
More specifically, the equation at $|x|\gg M$ is approximately
\begin{align}\label{eq:waveEtime}
 \partial_{x}^{2} \psi-\partial_{t}^{2} \psi - 2 i \omega_{0} \partial_{t} \psi+\omega_{0}^{2} \psi=0  \ ,
\end{align}
where $\omega_0=0$ for $x\to \infty$ and $\omega_0=\omega_H$ for $x\to -\infty$, respectively. The Laplace transform  from a starting time $t_0$ is $X(\omega, x)=\int_{t_0}^{\infty} \psi(t, x) e^{i\omega t} d t $ at $s=-i\omega$. 
Applying the Laplace transform to Eq.~(\ref{eq:waveEtime}), the source term in Eq.~(\ref{eq:SNasympto}) can be obtained as
\begin{align}
 S'(\omega, x)=i\left(\omega-2 \omega_{0}\right) \psi (t_0, x) - \psi'(t, x) |_{t=t_0} \ .
\end{align}
Considering the initial pulse sufficiently away from the light-ring potential barrier, we can write $ \psi(t, x)=\int_{-\infty}^{\infty} f(\omega) e^{-i \omega (t-t_{0} ) \pm  i (\omega-\omega_{0} ) (x-x_{s} )} d \omega$ at $t\gtrsim t_0$. Here,  $\pm$ corresponds to outgoing and ingoing pulse, $f(\omega)$ denotes its frequency content, and $x_s$ is the initial position with $|x_s| \gg M $. 
The source term is then given by 
\begin{align}
{S'}_{\pm}(\omega, x)=i \int_{-\infty}^{\infty} (\tilde{\omega}+\tilde{\omega}^{\prime})  f(\omega^{\prime}) e^{\pm  i \tilde{\omega}^{\prime} (x-x_{s})} d \tilde{\omega}^{\prime}  \ ,
\end{align}
where we have defined $\tilde{\omega} \equiv \omega - \omega_0$ and $\tilde{\omega}^{\prime} \equiv \omega^{\prime} - \omega_0$.
If the pulse starts from the interior, from Eq.~(\ref{eq:xihatp}), we have 
\begin{align}
\hat{\xi}_{\mathrm{BH}}^{+} 
 & \approx \frac{1}{W_{\mathrm{BH}} (\omega, -\infty) }\int_{-\infty}^{+\infty} d x {S'}_{\pm} e^{i \tilde{\omega} x} 
 = \begin{cases}
    0 & \text{outgoing} \\
    \frac{4 \pi  \tilde{\omega}  f(\omega) i }{W_{\mathrm{BH}} (\omega, -\infty) } e^{+ i \tilde{\omega} x_{s}} & \text{ingoing}
\end{cases} .   \label{eq:Z+_hat_inside}  \ 
\end{align}
where we take $W_{\rm BH} (\omega ,x ) \approx  W_{\rm BH}(\omega, -\infty) $ for $-x\gg M$. 
Similarly, if the pulse starts from the exterior, we have 
\begin{align}
\hat{\xi}_{\mathrm{BH}}^{-} 
 & \approx \frac{1}{W_{\mathrm{BH}} (\omega, +\infty) }\int_{-\infty}^{+\infty} d x {S'}_{\pm} e^{-i \omega x} 
 = \begin{cases}
    \frac{4 \pi  \omega f(\omega) i }{W_{\mathrm{BH}} (\omega, +\infty) } e^{+ i \omega x_{s}} & \text{outgoing}\\
    0 & \text{ingoing}    
\end{cases} .   \label{eq:Z-_hat_inside}  \ 
\end{align}
with $W_{\rm BH} (\omega ,x ) \approx  W_{\rm BH}(\omega, +\infty) $ for $x\gg M$. 
Notably, as long as the pulse is sufficiently localized in space, the additional contribution $\hat\xi_{\rm BH}^+$ for the outgoing pulse from inside and $\hat\xi_{\rm BH}^-$ for the ingoing pulse from outside vanish, regardless of the frequency content of the pulses. 
Thus, $\xi_{\rm BH}^-$ in Eqs.~(\ref{eq:ZBHm_in}) and (\ref{eq:ZBHm_out}) are fully specified by $\xi_{\rm BH}^+$.  
This means that the Teukolsky variable $Z_{\rm eff}^+$ in Eq.~(\ref{eq:Zeff2case}) can be uniquely determined by $Z_{\mathrm{BH}}^{+}$, and Eq.~(\ref{eq:hmaster}) becomes
\begin{eqnarray}\label{eq:heff2case}
h_{\rm echo} \approx \frac{  R_{\rm BH }  R_{\rm wall}}{1-R_{\rm BH } R_{\rm wall} }
\left\{
\begin{array}{ll}
h_{\rm RD}\,, & \textrm{inside}\\
-\frac{\mathcal{T}_{\rm BH}^2}{\mathcal{R}_{\rm BH}^2}h_{\rm RD}, & \textrm{outside}
\end{array}\right.\,.
\end{eqnarray}
These match exactly the inside and outside prescriptions in the geometric optics picture~\cite{Wang:2019rcf}.

At sufficiently late time of the ringdown stage, the waveform can be modeled as a linear superposition of two polarization modes of the fundamental QNM ($l=m=2$), with~\cite{Maggio:2019zyv} 
\begin{eqnarray}\label{eq:ZBH}
h_{\rm RD}(\omega) 
 &=& \int_{-\infty}^{+\infty} \frac{d t}{\sqrt{2 \pi}} e^{i \omega t}   \Theta\left(t\right)\left(\mathcal{A}_{+} \cos \left(\omega_{\rm RD} t+\phi_{+}\right)+i \mathcal{A}_{\times} \sin \left(\omega_{\rm RD} t+\phi_{\times}\right)\right) e^{-t / \tau}\nonumber\\
 &=& \frac{1}{2 \sqrt{2 \pi}} \left(\frac{\alpha_{1+} \mathcal{A}_{+}-\alpha_{1 \times} \mathcal{A}_{\times}}{ \omega- \omega_{\rm QNM}  }\right. \left.+\frac{\alpha_{2+} \mathcal{A}_{+}+\alpha_{2 \times} \mathcal{A}_{\times}}{\omega+\omega^*_{\rm QNM}    }\right), 
\end{eqnarray}
where  $ \omega_{\rm QNM}=\omega_{\rm RD} - i/\tau_{\rm RD}, \alpha_{1+, \times} = i e^{-i\phi_{+, \times}}$,
 $\alpha_{2+, \times} = -\alpha_{1+, \times}^{*}$, and the start time is at $t=0$.\footnote{The abrupt change at the start time $t_0$ in Eq.~(\ref{eq:ZBH}) may bring in artificial changes for the high frequency modes above $ \omega_{\rm RD}$, but these modes will quickly transmit to the outside and are irrelevant to our echo search that targets low-frequency modes with long lifetimes}
For simplicity, we consider only the plus mode in the main text.

\section{Superposition of QNMs in the time domain}
\label{eq:echoQNMtime}

In the context of ``quantum black hole seismology'', the postmerger echoes can be described as a linear superposition of the characteristic QNMs of UCOs, with the time domain waveform    
\begin{eqnarray}\label{eq:htQNM}
	\tilde{h}(t) 
 =\sum_{n} \tilde{h}_n(t_n) \ee^{-\ii \omega_n (t-t_n)} \, \ee ^{-(t-t_n)/\tau_n} \,\Theta(t-t_n)\,,
\end{eqnarray}
where we suppress the subscript of $h_{\rm echo}$ in this section for simplicity. Here, the overall amplitude and phase are encoded in $ \tilde{h}_n(t_n)$, and $\omega_n$, $\tau_n$, $t_n$ denote the angular frequency, damping time, and start time for the $n$-th mode. 
In the frequency domain, the waveform is given by
\begin{eqnarray}\label{eq:hfQNM}
	h(f) = \int_{-\infty}^{\infty} \tilde h(t) \ee^{\ii 2\pi f t} \dd t 
	= \sum_{n} \tilde{h}_n(t_n)\, \ee^{\ii 2\pi f t_n}  \frac{\ii}{2\pi(f-f_n)+\ii /\tau_n} \label{CFT_origin}\,.
\end{eqnarray}
It is a sum over Lorentzians for QNMs with location $f_n=\omega_n/(2\pi)$ and width $1/\tau_n$~\cite{Conklin:2021cbc}.
The start time brings in an additional phase $2\pi f t_n$. 

\begin{figure}[!h]
	\centering
	\includegraphics[width=15cm]{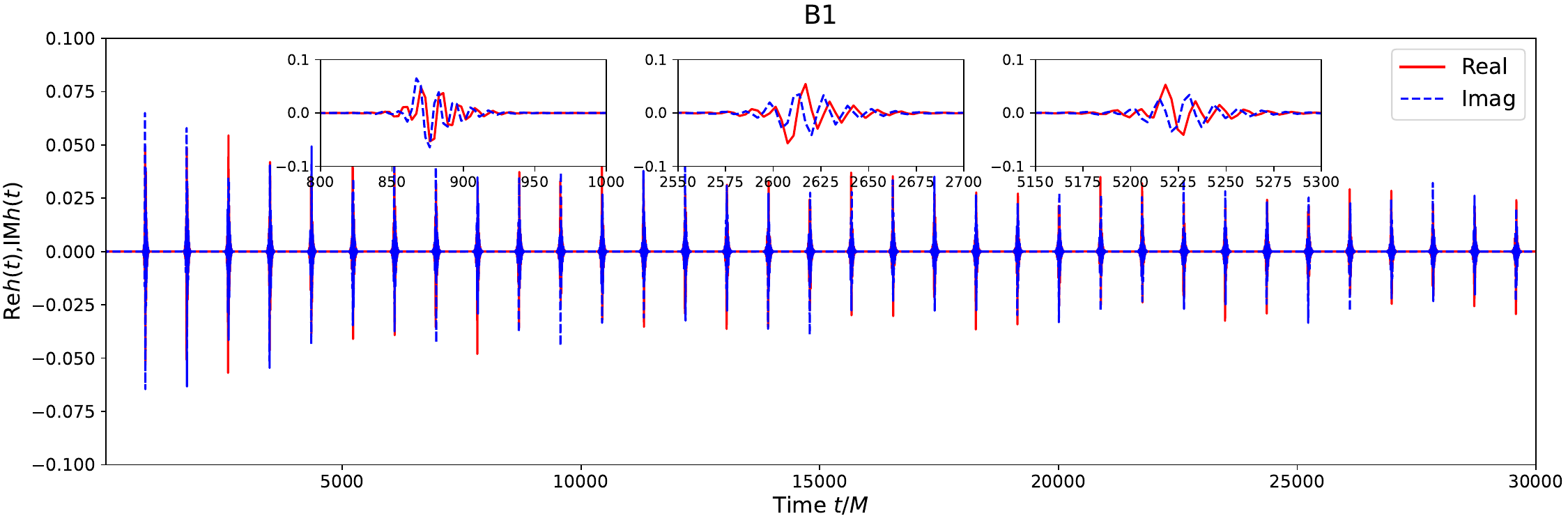}\\
 \includegraphics[width=15cm]{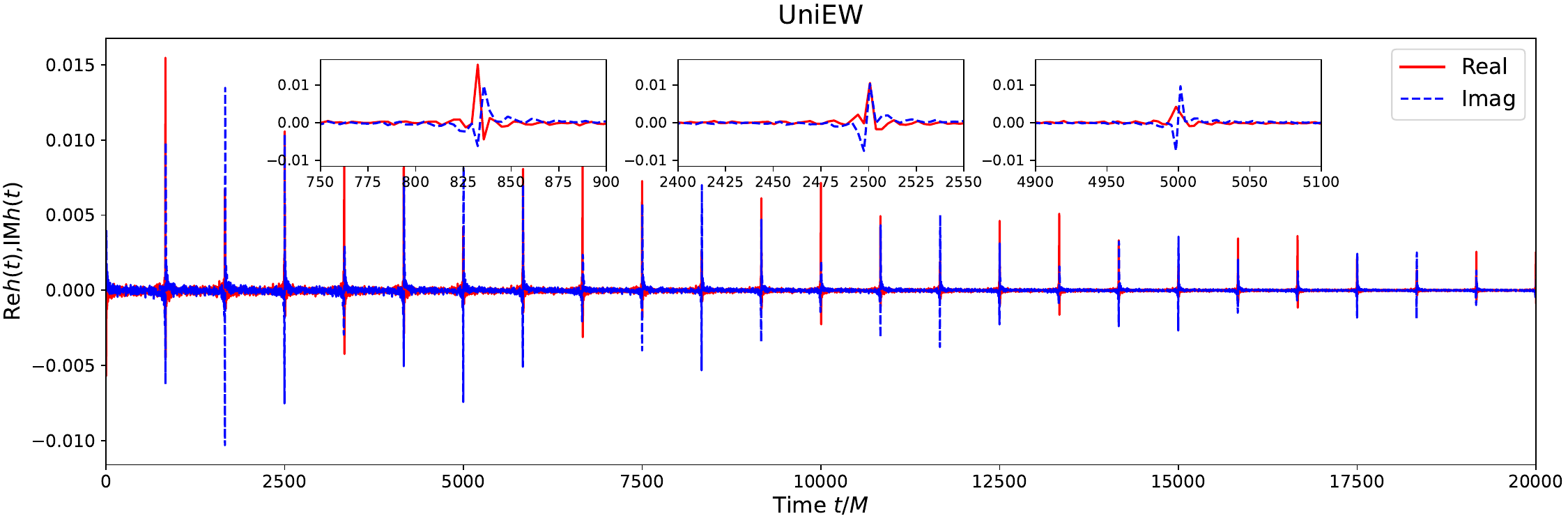}\\
\caption{\label{fig:EWFtime} Examples of time domain waveforms with 200 pulses. Top: the benchmark B1 considered in Sec.~\ref{sec:validate}. Bottom: our search template model ($\uew$) with $t_n/M=0.1$ and $\delta_n$ randomly sampled in $[0,0.2\pi]$. Red and blue lines correspond to the real and imaginary parts of $\tilde{h}(t)$. Insets: the profiles of the 3rd, $\sim20$th, and $\sim100$th pulse from left to right, respectively.}
\end{figure}

In practice, we usually analyze a finite segment of strain data. Suppose the start and finish times of the segment are $t_0$ and $t_T$, with the duration $t_T-t_0$ sufficiently longer than the time delay $t_d$, the frequency domain waveform becomes
\begin{eqnarray}\label{eq:hfQNMT}
	h^{(T)}(f) &=& \int_{t_0}^{t_T} \tilde h(t) \ee^{\ii 2\pi f t} \dd t  \nonumber \\
	& = &\sum_{n} A_n \ee^{i \delta_n} \ee^{\ii 2\pi f t'_n} \frac{1}{2\pi(f-f_n)+ \ii /\tau_n} \left[1-\ee^{-T_n/\tau_n}\ee^{\ii 2\pi (f-f_n)T_n}\right]\,,
\end{eqnarray}
where $A_n=\abs{\tilde{h}(t'_n)}$, $\delta_n=\arg(\tilde{h}(t'_n))+\pi/2$, $t'_n\equiv \max(t_0, t_n)$ and $T_n\equiv t_T-t'_n\gg t_d$ denote the possibly different start time and time duration for the $n$-th mode. The correction term  $1-\ee^{-T_n/\tau_n}\ee^{\ii 2\pi (f-f_n)T_n}$  compared to  Eq.~(\ref{eq:hfQNM}) denotes the finite duration effects. 
For demonstration purposes, we present in Fig.~\ref{fig:EWFtime} the time domain waveforms for a benchmark example and the search template being discussed in the main text.

When the UCO has high compactness and features strong combined reflectivity, the QNMs are well separated in the frequency space with $1/\tau_n\ll \Delta f$.
The peak region around each mode is then expected to be dominated by this one mode, where the contributions from other modes are safely negligible. 
To justify this approximation, we study the influence of QNMs interference around the resonance peak by  considering a simple model of periodic and uniform QNMs, with  $A_n =A$, $\tau_n = \tau$, $f_n = n (\Delta f +\rshift)$ and different choices of $\{t_n,\,\delta_n\}$ for the waveform in Eq.~(\ref{eq:hfQNMT}). A sufficiently long time duration is used to ensure that the peak region is well resolved.



\begin{figure}[!h]
	\centering
	\includegraphics[width=15cm]{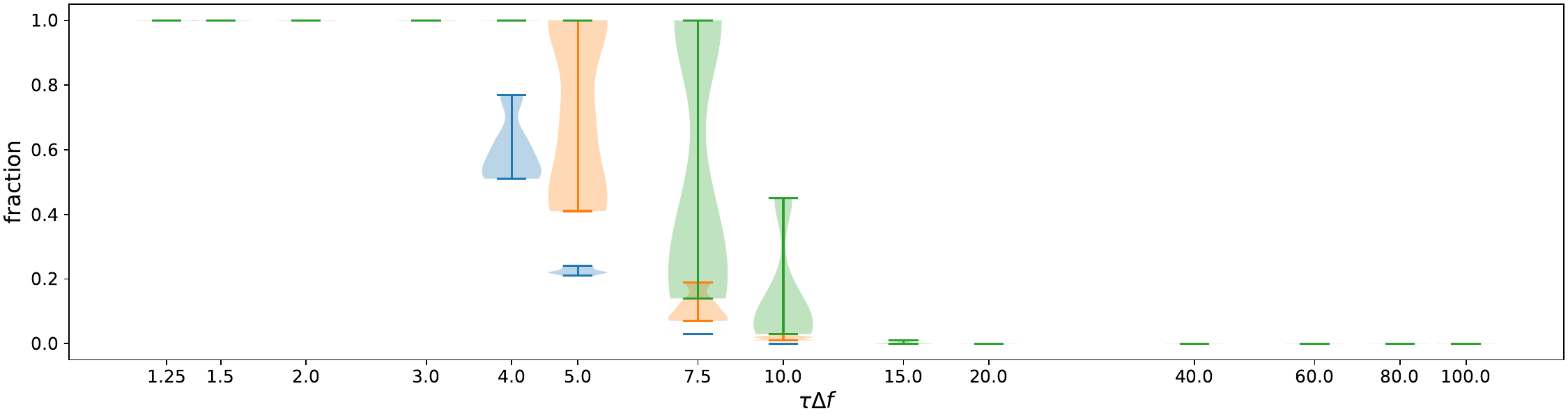}
	\includegraphics[width=15cm]{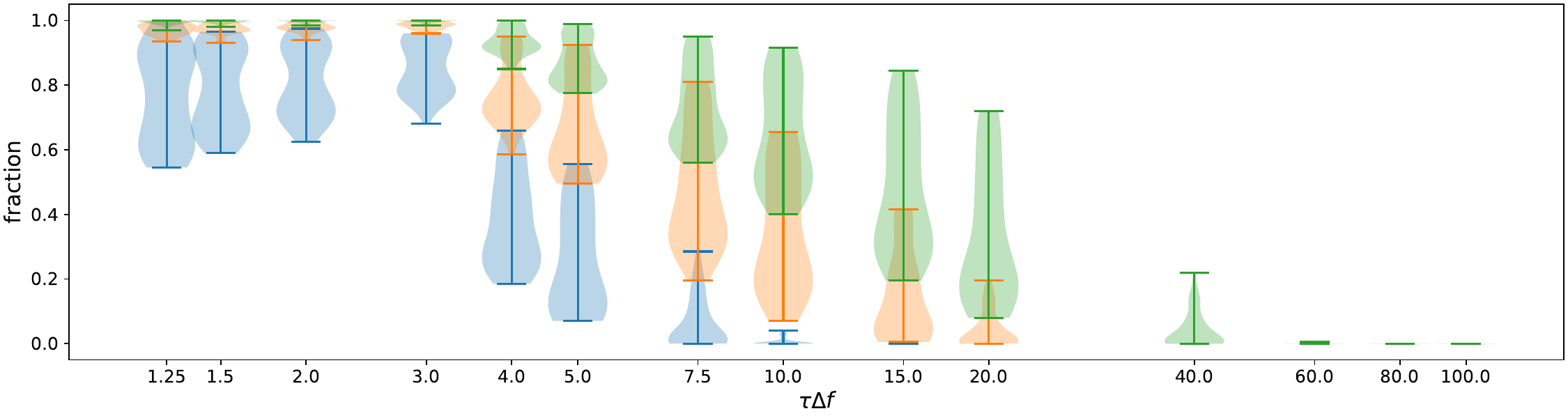}\\
	\includegraphics[width=15cm]{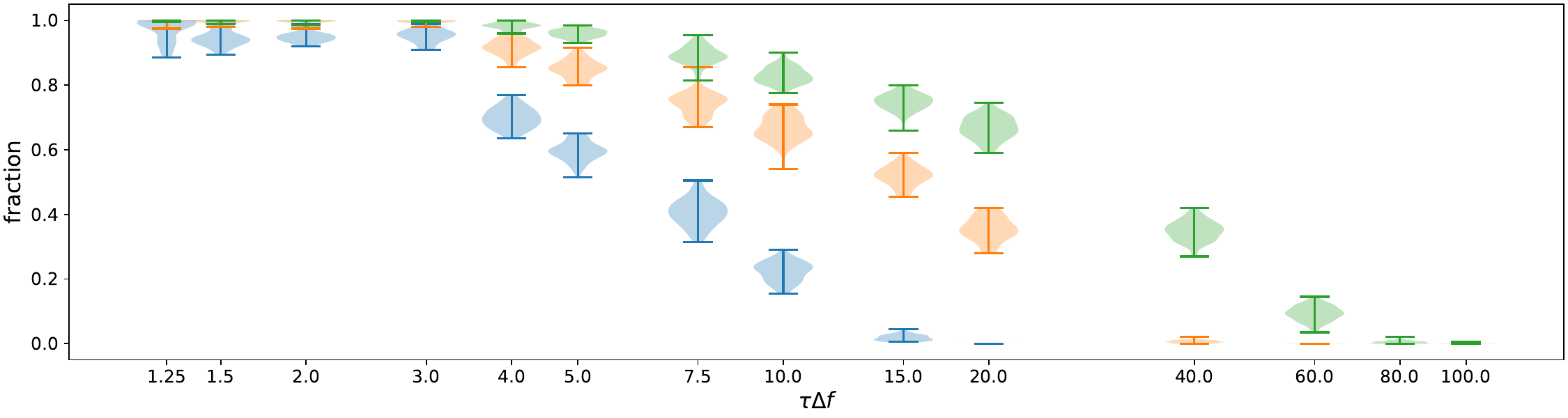}	
	\caption{\label{fig:violin} {The fraction of QNMs with $\Delta \phi_n/\pi$ larger than 0.5 (blue), 0.2 (orange), 0.1 (green) as a function of $\tau\, \Delta f$ under different assumptions of $t_n$ and $\delta_n$. Top: $t_n=0$, $\delta_n=0$ fixed. Middle: $t_n=0$ fixed, $\delta_n$ randomly sampled in $[0,\pi]$. Bottom: $t_n$ randomly sampled in $[0,t_d]$, $\delta_n$ randomly sampled in $[0,\pi]$. }}
\end{figure}

For our improved QNM search, the main task is to examine the influence of modes interference on the constant phase assumption.
For this purpose, we use $\Delta \phi_n \equiv \max(\arg \qty(h(f_{n,i}) ) )-\min(\arg \qty(h(f_{n,i}) ) )$ to measure the phase variation for each mode $n$, where $f_{n,i}\in [f_n-\mathcal{O}(1)/\tau, f_n+\mathcal{O}(1)/\tau]$ is the frequency bin around the resonance. We then evaluate $\Delta \phi_n$ for a large number of QNMs,  and derive the fraction of QNMs with $\Delta \phi_n/\pi$ larger than some threshold. Here, $\pi$ denotes the phase variation for the pole contribution in Eq.~(\ref{eq:arghf}).
We find that this fraction is most sensitive to the dimensionless quantity $\tau \Delta f$, i.e. the spacing to width ratio, and varies little with either the duration $T$ or total number $N$.
Fig.~\ref{fig:violin} shows the $\tau\, \Delta f$ dependence of the fraction for various choices of the start time $t_n$ and overall phase $\delta_n$. 
The phase varies a lot when the width is comparable to the spacing, i.e. $\tau\, \Delta f\sim 1$. As the spacing to width ratio increases, different QNMs interfere less and the fraction drops significantly at some typical value of $\tau\, \Delta f\gg 1$.
This typical value relies on the choice of $t_n$ and $\delta_n$, and a more random sampling of these two parameters pushes the typical value larger. 
The interference of QNMs then adds more uncertainties to a simple modeling of the phase around the resonance. 

\begin{figure}[!h]
	\centering
	\includegraphics[width=7.1cm]{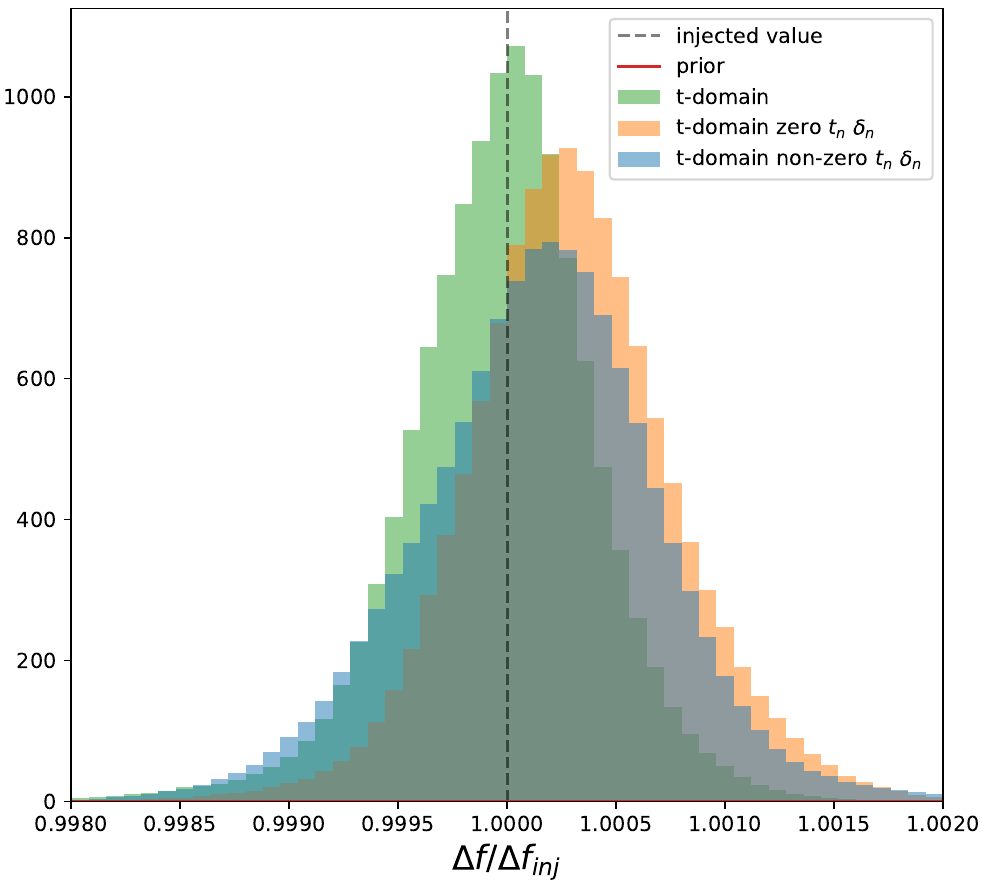}\quad
	\includegraphics[width=6.8cm]{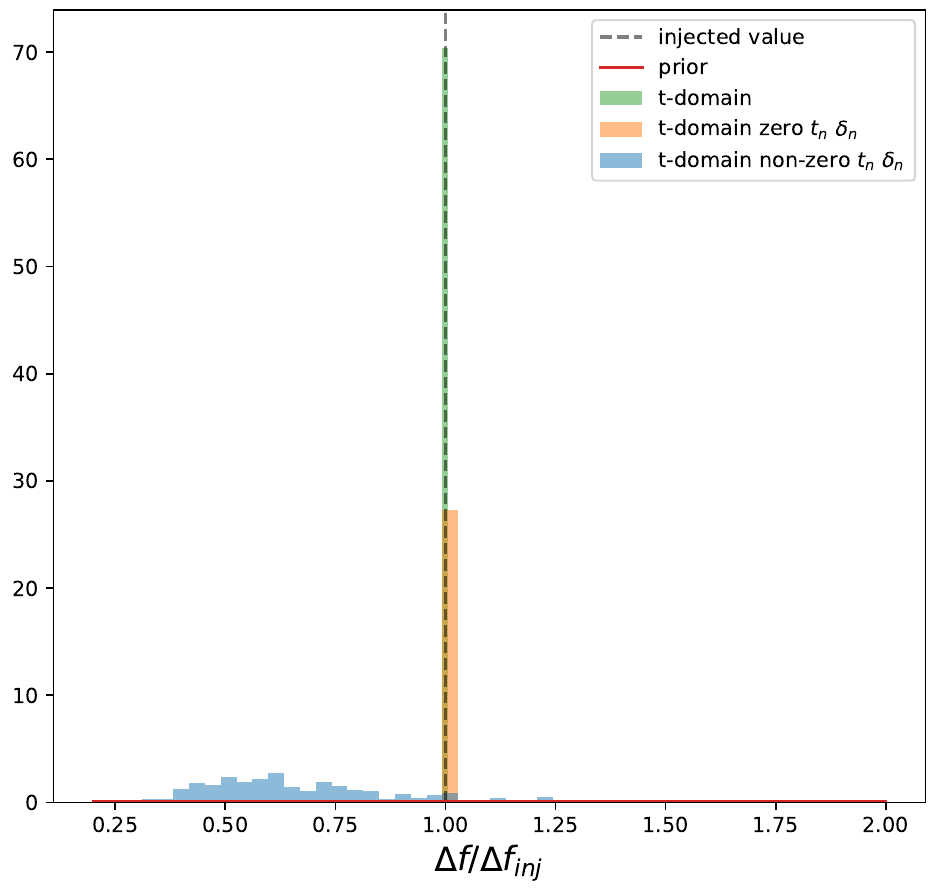}	
	\caption{\label{fig:phaseSearch} {The overall posterior of the spacing for injections of the periodic and uniform model of QNMs with the new likelihood. For the injected signal, we set $T \Delta f_{\rm inj}=250$, $N=49$ and $\tau_{\rm inj} \Delta f_{\rm inj}=5$ (left), $\tau_{\rm inj} \Delta f_{\rm inj}=1$ (right). The green is for the $\uew$ injections in the frequency space. The orange is for injections of periodic and uniform QNMs in the time domain with $t_n$ and $\delta_n$ set to zero. The blue is for the time domain injections with a pair of nonzero $\{t_n, \delta_n\}$ saturating the largest value of  $\Delta\phi/\pi$ in the bottom panel of Fig.~\ref{fig:violin}.}}
\end{figure}

To determine the influence of phase variation on the Bayesian search, we consider injections of periodic and uniform QNMs in the time domain for various choices of $t_n$ and $\delta_n$, and compare the results with those obtained for injections of $\uew$ in the frequency space. 
Fig.~\ref{fig:phaseSearch} presents the overall posterior of the spacing for various cases using the new likelihood. 
We find that when $\tau_{\rm inj}\Delta f_{\rm inj}=5$, the search results are relatively stable, although one third of the QNMs have $\Delta \phi_n/\pi\gtrsim 0.5$. Only for $\tau_{\rm inj}\Delta f_{\rm inj}=1$ do we observe a strong deviation, with $\Delta \phi/\sim\pi$ for most cases.  This demonstrates the limited impact of phase variations on search performance, particularly when $\tau_n \Delta f$ is considerably larger than 1. 

Thus, the assumption of one-mode dominance is a good approximation for the majority of cases. 
The $n$-th mode contribution in Eq.~(\ref{eq:hfQNMT}) is then directly related to the theoretical prediction of $h_{\rm echo}^{(T)}$ in Eq.~(\ref{eq:hechosumT}) at $f\sim f_n$. In particular, the finite term corrections in Eq.~(\ref{eq:hechosumT}) match exactly the finite range effects for the Fourier transform in Eq.~(\ref{eq:hfQNMT}). The free parameters in Eq.~(\ref{eq:hfQNMT}) are then determined as follows
\begin{eqnarray}\label{eq:coefficient}
A_n\approx|h_{\rm eff}(f_n)|\mathcal{R}_{\rm eff}(f_n)/t_d,\quad
\delta_n\approx \delta_0-2\pi f_n t_d+\pi,\quad 
t'_n\approx t_d+t_0,\quad
T_n\approx T\,, 
\end{eqnarray}
where $\arg(h_{\rm eff}(f))\approx \delta_0+2\pi f t_0$. Under this approximation, all modes are excited at a time $t_d$ after the initial time $t_0$ defined by $h_{\rm eff}$.

\section{Echo SNR and new likelihood dependence}

The optimal signal-to-noise ratio (SNR) of echoes within a frequency band $[f_{\rm min}, f_{\rm max}]$ is given by $\snr_{\rm echo}^2=\int_{f_{\rm min}}^{f_{\rm max}} \frac{|h_{\rm echo}(f)|^2}{P(f)}df$, where $P(f)$ is the one-sided power spectral density. Under the approximation of one-mode dominance around each resonance peak, the total $\snr_{\rm echo}^2$ is roughly a sum of $\snr_n^2$ over modes. In the high frequency resolution limit, it is approximately 
\begin{eqnarray}\label{eq:SNRopt}
\snr_{\rm echo}^2 
\approx \sum_{n=N_{\rm min}}^{N_{\rm max}}  \int_{f_n-\Delta f/2}^{f_n+\Delta f/2} \frac{|h_n(f)|^2}{P(f)}df
\approx \frac{1}{2P_{\rm RD}}\sum_{n=N_{\rm min}}^{N_{\rm max}}  r_n A_n^2 \tau_n\,,
\end{eqnarray}
where $h_n(f)$ denotes the waveform for the $n$-th mode. In the last step, we assume $P(f)$ does not vary much for one mode and put $P(f_n)=P_{\rm RD}/r_n$, where $P_{\rm RD}$ is the PSD at ringdown frequency and $r_n$ is the weight for each mode. 
Taking into account the effect of finite time duration effect in  Eq.~(\ref{eq:hfQNMT}), the mode sum in Eq.~(\ref{eq:SNRopt}) will be suppressed by $1-e^{-2T/\tau_n}$.

Given the theoretical prediction of QNM in Eqs.~(\ref{eq:QNMn}) and (\ref{eq:coefficient}), we find
\begin{eqnarray}\label{eq:SNRopt1}
\snr_{\rm echo}^2 \approx \frac{1}{2P_{\rm RD}}\sum_{n=N_{\rm min}}^{N_{\rm max}}  r_n |h_1(f_n)|^2\Delta f \frac{1-e^{-2T/\tau_n}}{|\ln\mathcal{R}_{\rm eff}(f_n)|}
\approx \sum_{n=N_{\rm min}}^{N_{\rm max}}\snr_{1,n}^2 \frac{\min\{\tau_n, T\}}{t_d}\,,
\end{eqnarray}
where $\snr^2_{1,n}\equiv (|h_1(f_n)|^2/P(f_n))\Delta f$ in the last step. In the case of very compact UCOs with a small $\Delta f$, 
the sum $\sum_{n=N_{\rm min}}^{N_{\rm max}}\snr^2_{1,n}$ gives approximately $\snr^2$ for the first pulse within the band (i.e. $\snr_1^2$). The enhancement $\snr_{\rm echo}^2/\snr_1^2$ resulting from the long-term accumulation of the echo signal is then governed by $\sim\min\{\tau_n, T\}/t_d$. 

The SNR is also closely related to the energy emitted associated with echoes. 
The leading order GW flux for the QNMs in Eq.~(\ref{eq:htQNM}) is given as~\cite{Maggiore:2007ulw, Westerweck:2021nue}
\begin{eqnarray}
\dot{E}_\textrm{GW} (t)
\approx \frac{1}{8 G}D_L^2 \langle |\dot{\tilde h} (t)|^2 \rangle
\approx \frac{1}{8 G}D_L^2\left|\sum_{n=N_{\rm min}}^{N_{\rm max}} \omega_n^2 \tilde{h}_n(t)\right|^2\,,
\end{eqnarray} 
where $D_L$ is the luminosity distance and $\omega_n\tau_n\gg1$ is assumed in the last expression. The emitted energy is then 
\begin{eqnarray}\label{eq:DeltaE}
\Delta E_{\rm echo} &=& \int_0^\infty \dot{E}_\textrm{GW} \,dt\approx \frac{D_L^2}{8 G} \int _0^\infty\left|\sum_{n=N_{\rm min}}^{N_{\rm max}} \omega_n^2 \tilde{h}_n(t)\right|^2 dt=  \frac{D_L^2}{8 G} \int _0^\infty\left|\sum_{n=N_{\rm min}}^{N_{\rm max}} \omega_n^2 h_n(f)\right|^2 df\nonumber\\
&\approx&  \frac{D_L^2}{8 G}\sum_{n=N_{\rm min}}^{N_{\rm max}}   \int _0^\infty \omega_n^2\left|h_n(f)\right|^2 df
= \frac{D_L^2}{16 G}\sum_{n=N_{\rm min}}^{N_{\rm max}} (2\pi f_n)^2 A_n^2 \tau_n\,,
\end{eqnarray}
which assumes negligible overlap of QNMs in the frequency domain. 

It is useful to compare the  GW energy emitted through echoes with that emitted during the ringdown stage. Due to the dominance of the fundamental mode, the later is given approximately as 
\begin{eqnarray}\label{eq:DeltaESNRRD}
\Delta E_\textrm{RD}\approx \frac{D_L^2}{16 G}A_\textrm{RD}^2 \tau_\textrm{RD} (2\pi f_\textrm{RD})^2
\approx \frac{D_L^2P_{\rm RD}}{16 G} (2\pi f_\textrm{RD})^2\snr^2_\textrm{RD}\,,
\end{eqnarray}
where $f_\textrm{RD}$ is the fundamental mode frequency in Eq.~(\ref{eq:fRD}). For echoes, we can rewrite Eq.~(\ref{eq:DeltaE}) in a similar form 
\begin{eqnarray}\label{eq:DeltaESNRecho}
\Delta E_\textrm{echo}\approx \frac{D_L^2 P_{\rm RD}}{16 G} (2\pi f_\textrm{echo})^2 \snr^2_\textrm{echo}\,,
\end{eqnarray}
by defining $f_\textrm{echo}^2\equiv \left(\sum_{n=N_{\rm min}}^{N_{\rm max}} f_n^2 A_n^2 \tau_n\right)/\left(\sum_{n=N_{\rm min}}^{N_{\rm max}} r_n A_n^2 \tau_n\right)$.
Taking the ratio of these two expressions, we obtain  
\begin{eqnarray}\label{eq:DeltaEratio}
\frac{\Delta E_\textrm{echo}}{\Delta E_\textrm{RD}}=\frac{ f^2_\textrm{echo}}{f^2_\textrm{RD}} \frac{\snr^2_\textrm{echo} }{\snr^2_\textrm{RD}}\,,
\end{eqnarray}
where $D_L$ and $P_{\rm RD}$ dependences are factorized out. If echoes carry away similar amount of energy as the detected ringdown, $\snr_\textrm{echo}$ could be larger than $\snr_\textrm{RD}$ since the frequencies of the trapped QNMs $f_n$ is smaller than $f_\textrm{RD}$ in general, i.e. $f_\textrm{echo}\lesssim f_\textrm{RD}$. 
Taking into account the finite time duration effect, the mode sum in Eq.~(\ref{eq:DeltaE}) will be suppressed by $1-e^{-2T/\tau_n}$. The energy ratio then takes the same expression as in Eq.~(\ref{eq:DeltaEratio}), but with a more general definition of the echo frequency 
\begin{eqnarray}
f_\textrm{echo}^2\equiv \frac{\sum_{n=N_{\rm min}}^{N_{\rm max}} f_n^2 A_n^2 \tau_n (1-e^{-2T/\tau_n})}{\sum_{n=N_{\rm min}}^{N_{\rm max}} A_n^2 \tau_n (1-e^{-2T/\tau_n})}\,.
\end{eqnarray}
Poorly resolved QNMs with $\tau_n\gg T$ make negligible contribution to both SNR and $\Delta E_{\rm echo}$.

\begin{figure}[!h]
	\centering
	\includegraphics[width=8cm]{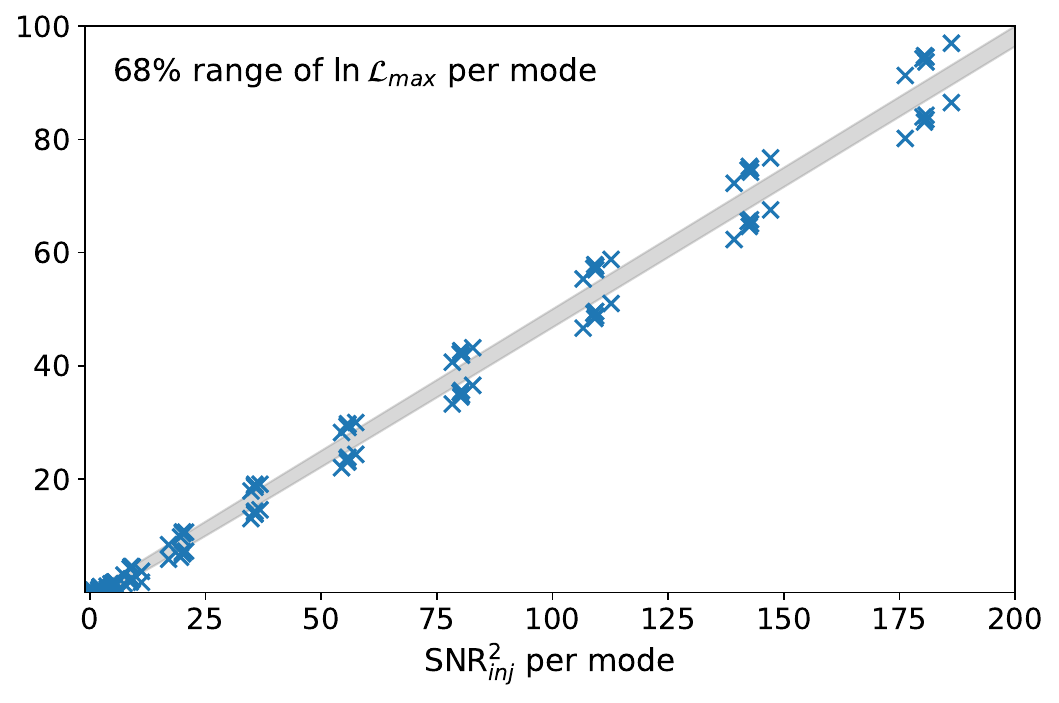}
	\includegraphics[width=8cm]{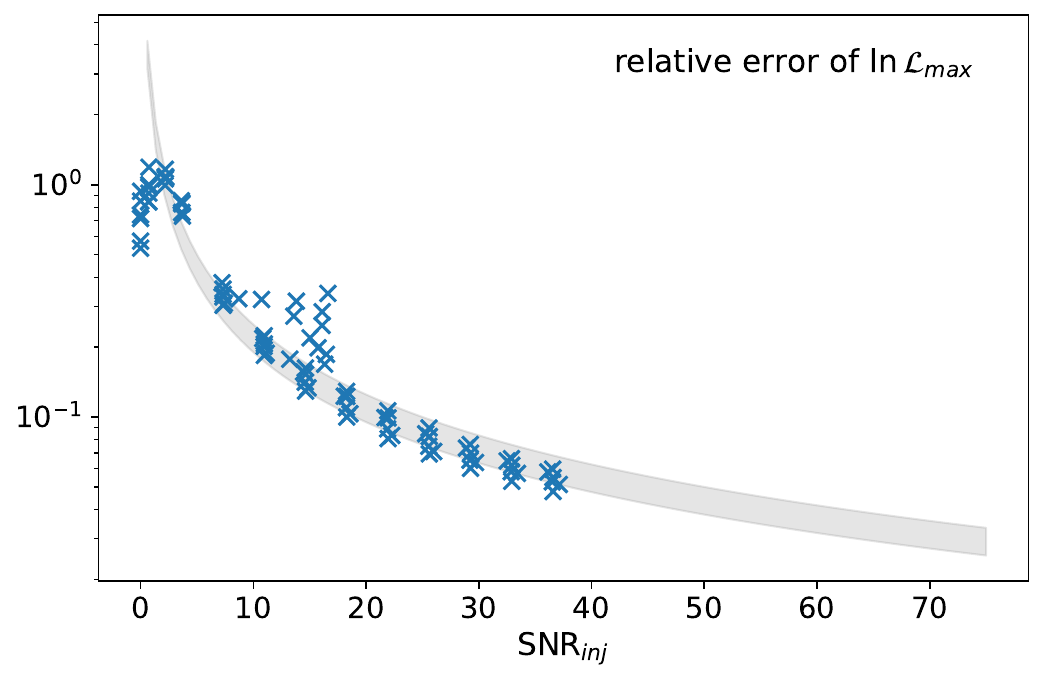}\\
	\includegraphics[width=8cm]{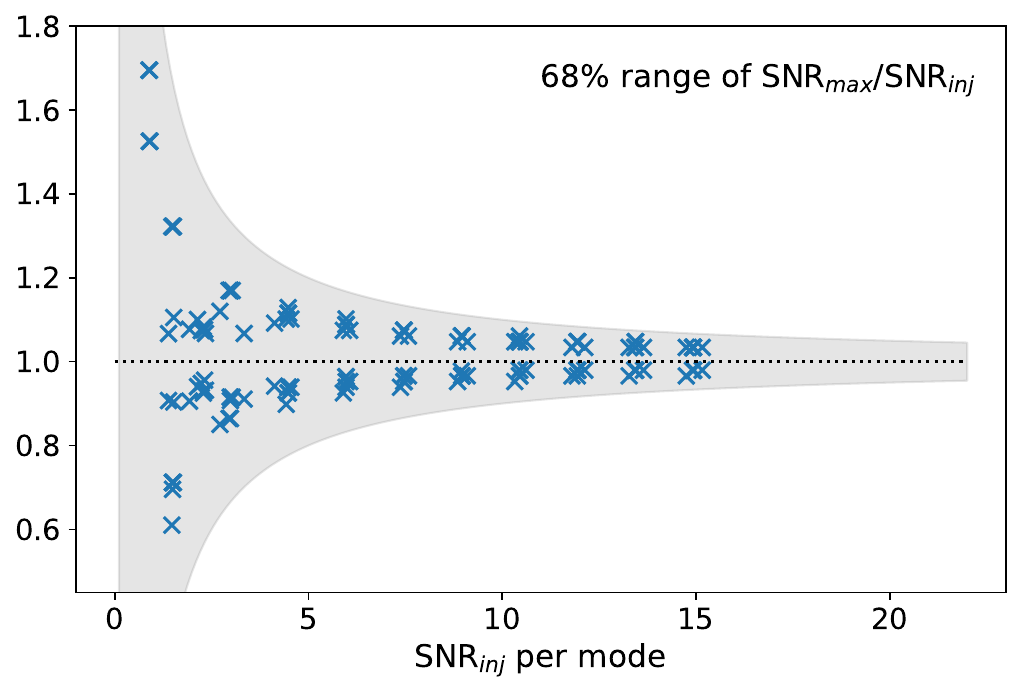}
	\includegraphics[width=8cm]{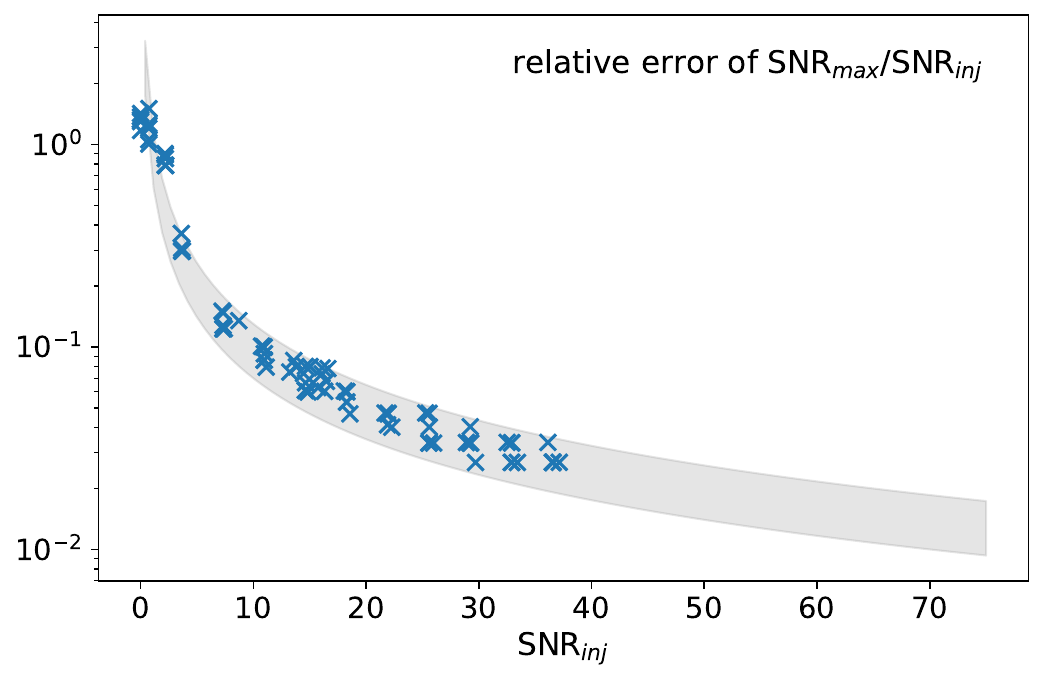}
	\caption{\label{fig:SNRLogLikelihood} {Search results as a function of the injected signal $\snr_{\rm inj}$ (per mode). The cross points include the simple one-parameter search in Sec.~\ref{sec:simpleUEW} and BILBY searches in Sec.~\ref{sec:BayesianUEW}. The gray band denotes the rough range of scattered points. Top left: the 68\% range of maximum log-likelihood per mode (gray band defined by $\frac{1}{2}x$ and $\ln I_0(x)-\frac{1}{2}x$). Top right: maximum log-likelihood error normalized by its mean (gray band defined by $(1.9,\,2.5)/x$). Bottom left: the 68\% range of  $\snr_{\rm max}$ to $\snr_{\rm inj}$ ratio (gray band defined by $1\pm 0.8/x$), where $\snr_{\rm max}$ denotes SNR at the maximum log-likelihood. Bottom right: $\snr_{\rm max}$ error normalized by $\snr_{\rm inj}$ (gray band defined by $(0.7,\,1.3)/x$).}}
\end{figure}

Finally, we discuss the injected signal-to-noise ratio  dependence of the new likelihood. As detailed in Sec.~\ref{sec:likelihood}, the new likelihood in Eq.~(\ref{eq:newlnL}) is mostly sensitive to $\snr_{\rm inj}$ per mode. To demonstrate the robustness of this dependence, we consider the search of $\uew$ injections with Gaussian noise. Fig.~\ref{fig:SNRLogLikelihood} shows the $\snr_{\rm inj}$ dependence of the maximum log-likelihood and the corresponding SNR for the generic searches, where we vary all parameters in Eq.~(\ref{eq:parameters}) that may influence the $\snr_{\rm inj}$ and also the number of search parameters.

With the uniform contribution for all modes, the maximum log-likelihood $\ln \mathcal{L}_\textrm{new}$ per mode and the searched SNR over injected SNR ratio exhibit a simple dependence on the SNR per mode. In the large signal limit, the searched $\snr$ (amplitude) approaches the injected values, i.e. $\snr_\textrm{max}\approx \snr_{\rm inj}$, and the maximum log-likelihood per mode $\frac{1}{N}\ln \mathcal{L}_\textrm{new}$ is approximately half of the $\snr_{\rm inj}^2$ per mode.  
The relative errors for $\ln \mathcal{L}_\textrm{new}$ and $\snr_\textrm{max}$ are more sensitive to the total injected $\snr_{\rm inj}$. As shown by the right columns, they scale roughly as $2/\snr_{\rm inj}$ and $1/\snr_{\rm inj}$, regardless of the parameter settings.

\section{Additional Bayesian search results}
\label{sec:moreresults}

\begin{figure}[!h]
	\centering
	\includegraphics[width=15cm]{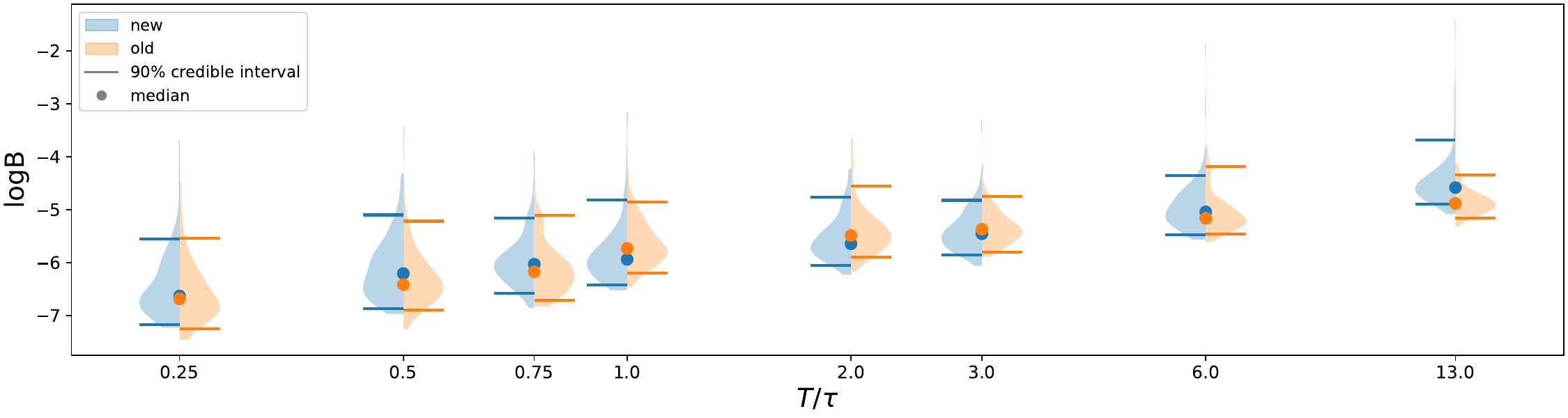}\\
 	\includegraphics[width=15cm]{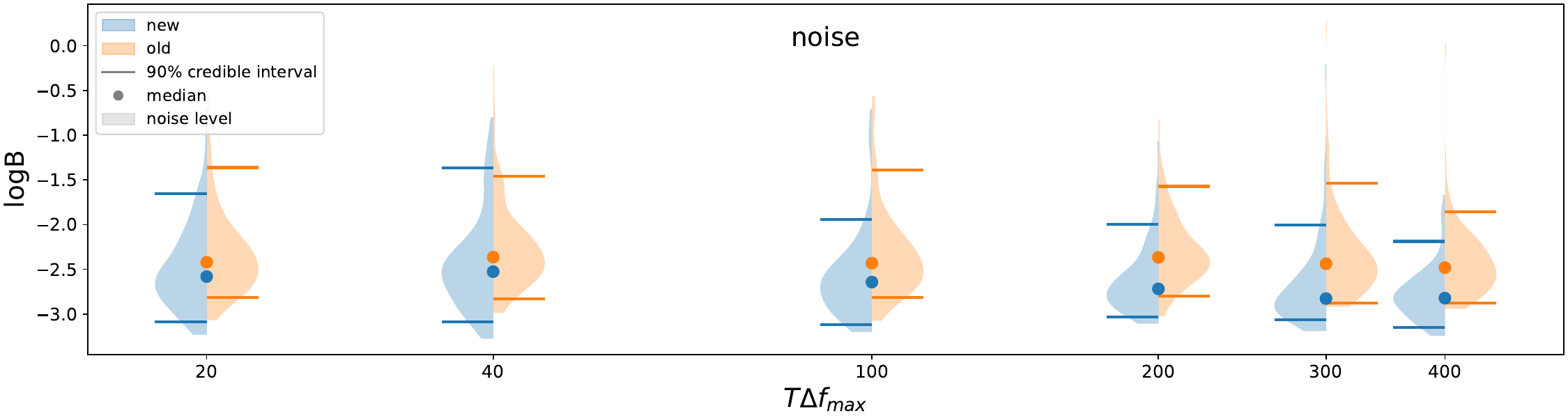}	\caption{\label{fig:violin_noise} {The log Bayes factor distributions for Gaussian noises with the two likelihoods. Top: four-parameter search in Sec.~\ref{sec:BayesianUEW} with parameter settings in Tab.~\ref{tab:parameter1}. Bottom: six-parameter search in Sec.~\ref{sec:searchbenchmark} with parameter settings in Tab.~\ref{tab:parameter2}.}}
\end{figure}

We begin by discussing the Bayesian search results for Gaussian noises. Fig.~\ref{fig:violin_noise} shows the log Bayes factor distributions for $\mathcal{N}=100$ noise realizations with different settings. For both likelihoods, the log Bayes factor is negative, indicating that the data prefer the model without QNMs associated with echoes. The results obtained using the new likelihood are generally more negative than those obtained using the old one, and the difference becomes more pronounced as the time duration or the number of frequency bins increases. Furthermore, the distribution shifts toward zero as the time duration increases. Despite these variations, the background search results stay relatively stable for different parameter settings.

\begin{table}[h]
\begin{center}
\resizebox{\textwidth}{60mm}{
\begin{tabular}{c|c||ccccccc}
\hline\hline
&
\\[-3mm]
 & $T \Delta f_{\rm max}$ & logB & $M \Delta f$ & $q_0$ & $A/\langle \tilde{P}\rangle^{1/2}$ & $\log_{10} M/\tau$ & $M f_{\rm min}$ & $M f_{\rm max}$
\\
&
\\[-3.5mm]
\hline
&
\\[-4.0mm]
\multirow{6}{*}{B1} 
& 20 & $-1.6^{+6.4}_{-1.1}$ & $0.0012^{+0.0011}_{-0.0005}$ & $0.6^{+0.4}_{-0.5}$ & $1.3^{+2.0}_{-1.2}$ & $-3.7^{+0.9}_{-0.2}$ & $0.04^{+0.02}_{-0.04}$ & $0.077^{+0.007}_{-0.042}$ \\
& 40 & $1^{+14}_{-3}$ & $0.0011^{+0.0011}_{-0.0004}$ & $0.8^{+0.2}_{-0.7}$ & $2.5^{+1.5}_{-2.2}$ & $-4.14^{+0.95}_{-0.09}$ & $0.05^{+0.02}_{-0.04}$ & $0.080^{+0.005}_{-0.037}$ \\
& 100 & $20.8^{+15.6}_{-23.1}$ & $0.0011^{+0.0006}_{-0.0004}$ & $0.83^{+0.05}_{-0.64}$ & $3.3^{+1.2}_{-2.6}$ & $-4.56^{+0.58}_{-0.07}$ & $0.05^{+0.01}_{-0.03}$ & $0.078^{+0.005}_{-0.008}$ \\
& 200 & $40^{+20}_{-20}$ & $11492^{+3}_{-5746} 10^{-7}$ & $0.83^{+0.01}_{-0.16}$ & $3.3^{+1.3}_{-1.2}$ & $-4.8^{+0.2}_{-0.1}$ & $0.046^{+0.010}_{-0.013}$ & $0.076^{+0.005}_{-0.003}$ \\
& 300 & $40^{+20}_{-40}$ & $11493^{+3}_{-5746} 10^{-7}$ & $0.83^{+0.01}_{-0.20}$ & $3.0^{+1.2}_{-1.4}$ & $-4.9^{+0.3}_{-0.2}$ & $0.04^{+0.01}_{-0.02}$ & $0.076^{+0.005}_{-0.005}$ \\
& 400 & $40^{+20}_{-30}$ & $11493^{+4}_{-5746} 10^{-7}$ & $0.831^{+0.010}_{-0.188}$ & $2.5^{+1.0}_{-1.3}$ & $-4.9^{+0.4}_{-0.2}$ & $0.04^{+0.01}_{-0.02}$ & $0.075^{+0.006}_{-0.005}$ \\
&
\\[-4.5mm]
\hline 
&
\\[-4.mm]
\multirow{6}{*}{B2} 
& 20 & $-2.1^{+2.8}_{-0.8}$ & $0.0012^{+0.0010}_{-0.0006}$ & $0.5^{+0.4}_{-0.5}$ & $0.9^{+1.7}_{-0.8}$ & $-3.6^{+0.8}_{-0.3}$ & $0.03^{+0.03}_{-0.03}$ & $0.07^{+0.01}_{-0.04}$ \\
& 40 & $-1.7^{+8.2}_{-1.0}$ & $0.0012^{+0.0010}_{-0.0005}$ & $0.7^{+0.2}_{-0.6}$ & $1.4^{+2.0}_{-1.3}$ & $-4.0^{+1.0}_{-0.2}$ & $0.04^{+0.03}_{-0.03}$ & $0.077^{+0.007}_{-0.042}$ \\
& 100 & $4^{+21}_{-7}$ & $0.0011^{+0.0010}_{-0.0006}$ & $0.82^{+0.06}_{-0.70}$ & $2.7^{+1.2}_{-2.4}$ & $-4.55^{+0.94}_{-0.09}$ & $0.04^{+0.02}_{-0.04}$ & $0.078^{+0.006}_{-0.032}$ \\
& 200 & $30^{+30}_{-40}$ & $0.0011^{+0.0001}_{-0.0006}$ & $0.831^{+0.009}_{-0.392}$ & $3.4^{+1.0}_{-2.1}$ & $-4.87^{+0.40}_{-0.06}$ & $0.04^{+0.01}_{-0.02}$ & $0.076^{+0.005}_{-0.004}$ \\
& 300 & $50^{+20}_{-50}$ & $0.0011^{+0.0006}_{-0.0006}$ & $0.830^{+0.005}_{-0.596}$ & $3.6^{+1.0}_{-3.0}$ & $-5.05^{+0.88}_{-0.06}$ & $0.03^{+0.02}_{-0.02}$ & $0.076^{+0.005}_{-0.012}$ \\
& 400 & $60^{+30}_{-60}$ & $0.0011^{+0.0007}_{-0.0006}$ & $0.831^{+0.005}_{-0.535}$ & $3.7^{+1.0}_{-2.9}$ & $-5.18^{+0.69}_{-0.05}$ & $0.03^{+0.02}_{-0.02}$ & $0.074^{+0.007}_{-0.005}$ \\
&
\\[-4.5mm]
\hline 
&
\\[-4.mm]
\multirow{6}{*}{B3} 
& 20 & $7^{+12}_{-8}$ & $110.9^{+109.6}_{-1.4} 10^{-5}$ & $0.5^{+0.5}_{-0.4}$ & $3.0^{+1.3}_{-2.1}$ & $-3.86^{+0.39}_{-0.07}$ & $0.057^{+0.009}_{-0.033}$ & $0.079^{+0.005}_{-0.009}$ \\
& 40 & $12.8^{+13.0}_{-11.6}$ & $110.7^{+1.4}_{-1.5} 10^{-5}$ & $0.5^{+0.5}_{-0.4}$ & $3.3^{+1.7}_{-1.9}$ & $-4.1^{+0.5}_{-0.1}$ & $0.059^{+0.006}_{-0.015}$ & $0.076^{+0.007}_{-0.007}$ \\
& 100 & $12.9^{+14.3}_{-10.4}$ & $110.5^{+2.4}_{-1.6} 10^{-5}$ & $0.6^{+0.4}_{-0.5}$ & $2.6^{+2.2}_{-1.4}$ & $-4.2^{+0.4}_{-0.4}$ & $0.059^{+0.005}_{-0.013}$ & $0.075^{+0.007}_{-0.007}$ \\
& 200 & $11^{+10}_{-8}$ & $110.1^{+1.9}_{-1.4} 10^{-5}$ & $0.5^{+0.4}_{-0.5}$ & $1.8^{+2.0}_{-1.1}$ & $-4.2^{+0.6}_{-0.6}$ & $0.059^{+0.006}_{-0.015}$ & $0.074^{+0.009}_{-0.006}$ \\
& 300 & $11^{+11}_{-9}$ & $110.0^{+1.9}_{-1.3} 10^{-5}$ & $0.5^{+0.4}_{-0.4}$ & $1.6^{+1.7}_{-0.9}$ & $-4.2^{+0.6}_{-0.6}$ & $0.059^{+0.006}_{-0.008}$ & $0.073^{+0.010}_{-0.005}$ \\
& 400 & $15^{+13}_{-10}$ & $109.8^{+1.5}_{-1.2} 10^{-5}$ & $0.5^{+0.4}_{-0.5}$ & $1.3^{+1.6}_{-0.6}$ & $-4.1^{+0.4}_{-0.7}$ & $0.059^{+0.005}_{-0.006}$ & $0.073^{+0.008}_{-0.004}$ \\
&
\\[-4.5mm]
\hline 
&
\\[-4.mm]
\multirow{6}{*}{B4} 
& 20 & $32.0^{+17.5}_{-13.9}$ & $0.0008^{+0.0004}_{-0.0002}$ & $0.5^{+0.5}_{-0.4}$ & $1.3^{+0.6}_{-0.3}$ & $-3.1^{+0.2}_{-0.3}$ & $0.044^{+0.009}_{-0.012}$ & $0.0839^{+0.0005}_{-0.0027}$ \\
& 40 & $27.4^{+15.2}_{-12.7}$ & $0.0008^{+0.0004}_{-0.0002}$ & $0.5^{+0.4}_{-0.5}$ & $0.9^{+0.4}_{-0.2}$ & $-3.2^{+0.2}_{-0.2}$ & $0.043^{+0.009}_{-0.012}$ & $0.0837^{+0.0007}_{-0.0105}$ \\
& 100 & $24.8^{+11.6}_{-10.5}$ & $0.0008^{+0.0004}_{-0.0002}$ & $0.4^{+0.5}_{-0.4}$ & $0.6^{+0.2}_{-0.2}$ & $-3.2^{+0.3}_{-0.2}$ & $0.043^{+0.010}_{-0.012}$ & $0.083^{+0.001}_{-0.010}$ \\
& 200 & $24.6^{+11.7}_{-11.4}$ & $0.0008^{+0.0004}_{-0.0002}$ & $0.4^{+0.5}_{-0.4}$ & $0.4^{+0.2}_{-0.1}$ & $-3.2^{+0.3}_{-0.3}$ & $0.042^{+0.010}_{-0.012}$ & $0.083^{+0.002}_{-0.016}$ \\
& 300 & $24.0^{+13.6}_{-12.0}$ & $0.0009^{+0.0003}_{-0.0003}$ & $0.4^{+0.5}_{-0.4}$ & $0.34^{+0.20}_{-0.09}$ & $-3.2^{+0.3}_{-0.3}$ & $0.04^{+0.01}_{-0.02}$ & $0.083^{+0.002}_{-0.015}$ \\
& 400 & $24.2^{+16.1}_{-11.0}$ & $0.0008^{+0.0004}_{-0.0002}$ & $0.4^{+0.5}_{-0.4}$ & $0.29^{+0.15}_{-0.08}$ & $-3.2^{+0.2}_{-0.3}$ & $0.043^{+0.008}_{-0.012}$ & $0.083^{+0.001}_{-0.010}$ \\
\hline\hline

\end{tabular}}
\caption{Search results for the four benchmarks with the new likelihood. The first column denotes the median value and the symmetric 90\% credible interval of the log Bayes factor distribution. The remaining columns indicate the median values and the 90\% credible regions \cite{LIGOScientific:2013yzb} of the six search parameters from the overall posterior distributions.
}
\label{tab:BMresults_new}
\end{center}
\end{table}

\begin{table}[h]
\begin{center}
\resizebox{\textwidth}{60mm}{
\begin{tabular}{c|c||ccccccc}
\hline\hline
&
\\[-3mm]
 & $T \Delta f_{\rm max}$ & logB & $M \Delta f$ & $q_0$ & $A/\langle \tilde{P}\rangle^{1/2}$ & $\log_{10} M/\tau$ & $M f_{\rm min}$ & $M f_{\rm max}$
\\
&
\\[-3.5mm]
\hline
&
\\[-4.0mm]
\multirow{6}{*}{B1} 
& 20 & $-1.1^{+6.1}_{-1.1}$ & $0.0012^{+0.0010}_{-0.0005}$ & $0.6^{+0.4}_{-0.5}$ & $1.3^{+1.8}_{-1.1}$ & $-3.5^{+0.7}_{-0.4}$ & $0.04^{+0.02}_{-0.04}$ & $0.078^{+0.006}_{-0.040}$ \\
& 40 & $0^{+10}_{-3}$ & $0.0012^{+0.0010}_{-0.0004}$ & $0.7^{+0.2}_{-0.7}$ & $1.7^{+1.6}_{-1.4}$ & $-3.8^{+0.9}_{-0.4}$ & $0.05^{+0.02}_{-0.04}$ & $0.079^{+0.005}_{-0.030}$ \\
& 100 & $10.0^{+13.5}_{-11.1}$ & $1149^{+15}_{-3} 10^{-6}$ & $0.83^{+0.08}_{-0.47}$ & $2.4^{+1.5}_{-1.5}$ & $-4.2^{+0.6}_{-0.4}$ & $0.05^{+0.01}_{-0.03}$ & $0.078^{+0.005}_{-0.007}$ \\
& 200 & $31.3^{+17.6}_{-20.2}$ & $11493^{+5}_{-5} 10^{-7}$ & $0.83^{+0.02}_{-0.03}$ & $3.5^{+1.3}_{-1.2}$ & $-4.8^{+0.3}_{-0.1}$ & $0.04^{+0.01}_{-0.02}$ & $0.076^{+0.005}_{-0.004}$ \\
& 300 & $26.1^{+13.6}_{-22.5}$ & $11492^{+5}_{-4} 10^{-7}$ & $0.83^{+0.02}_{-0.03}$ & $3.1^{+1.6}_{-1.2}$ & $-4.8^{+0.4}_{-0.3}$ & $0.05^{+0.01}_{-0.02}$ & $0.075^{+0.005}_{-0.005}$ \\
& 400 & $14.2^{+18.6}_{-16.5}$ & $0.0011^{+0.0006}_{-0.0001}$ & $0.83^{+0.03}_{-0.49}$ & $2.4^{+1.6}_{-1.9}$ & $-4.8^{+1.0}_{-0.4}$ & $0.05^{+0.01}_{-0.03}$ & $0.075^{+0.007}_{-0.012}$ \\
& 
\\[-4.5mm]
\hline 
&
\\[-4.mm]
\multirow{6}{*}{B2} 
& 20 & $-1.8^{+3.6}_{-0.7}$ & $0.0012^{+0.0010}_{-0.0006}$ & $0.5^{+0.4}_{-0.5}$ & $1.0^{+1.7}_{-0.9}$ & $-3.5^{+0.7}_{-0.4}$ & $0.03^{+0.03}_{-0.03}$ & $0.075^{+0.009}_{-0.042}$ \\
& 40 & $-1.2^{+5.0}_{-1.2}$ & $0.0012^{+0.0011}_{-0.0005}$ & $0.6^{+0.3}_{-0.6}$ & $1.2^{+1.7}_{-1.0}$ & $-3.7^{+0.9}_{-0.5}$ & $0.04^{+0.03}_{-0.03}$ & $0.077^{+0.007}_{-0.041}$ \\
& 100 & $4^{+12}_{-6}$ & $0.0011^{+0.0008}_{-0.0004}$ & $0.82^{+0.09}_{-0.65}$ & $1.9^{+1.5}_{-1.5}$ & $-4.2^{+1.2}_{-0.4}$ & $0.04^{+0.02}_{-0.03}$ & $0.078^{+0.006}_{-0.023}$ \\
& 200 & $33.1^{+18.1}_{-23.5}$ & $11493^{+4}_{-4} 10^{-7}$ & $0.83^{+0.01}_{-0.14}$ & $3.2^{+1.1}_{-1.3}$ & $-4.8^{+0.4}_{-0.1}$ & $0.03^{+0.01}_{-0.02}$ & $0.076^{+0.006}_{-0.004}$ \\
& 300 & $42.6^{+18.2}_{-24.0}$ & $11492.2^{+2.3}_{-1.8} 10^{-7}$ & $0.834^{+0.007}_{-0.010}$ & $3.2^{+1.1}_{-1.0}$ & $-4.9^{+0.3}_{-0.2}$ & $0.03^{+0.01}_{-0.01}$ & $0.076^{+0.005}_{-0.004}$ \\
& 400 & $47.0^{+17.6}_{-24.6}$ & $11493.0^{+1.6}_{-2.1} 10^{-7}$ & $0.830^{+0.007}_{-0.008}$ & $3.0^{+1.4}_{-1.0}$ & $-4.9^{+0.3}_{-0.3}$ & $0.03^{+0.01}_{-0.02}$ & $0.075^{+0.006}_{-0.004}$ \\
&
\\[-4.5mm]
\hline 
&
\\[-4.mm]
\multirow{6}{*}{B3} 
& 20 & $7^{+12}_{-8}$ & $0.0011^{+0.0011}_{-0.0001}$ & $0.5^{+0.5}_{-0.4}$ & $2.6^{+1.5}_{-1.6}$ & $-3.7^{+0.7}_{-0.2}$ & $0.056^{+0.008}_{-0.030}$ & $0.079^{+0.005}_{-0.008}$ \\
& 40 & $11^{+15}_{-9}$ & $110.4^{+1.3}_{-1.4} 10^{-5}$ & $0.5^{+0.5}_{-0.4}$ & $3.1^{+1.8}_{-1.6}$ & $-4.0^{+0.6}_{-0.3}$ & $0.058^{+0.005}_{-0.012}$ & $0.075^{+0.007}_{-0.006}$ \\
& 100 & $13.9^{+12.8}_{-12.0}$ & $109.8^{+1.3}_{-1.0} 10^{-5}$ & $0.5^{+0.4}_{-0.5}$ & $2.7^{+2.0}_{-1.1}$ & $-4.0^{+0.4}_{-0.5}$ & $0.059^{+0.004}_{-0.006}$ & $0.072^{+0.008}_{-0.004}$ \\
& 200 & $10^{+10}_{-9}$ & $109.4^{+1.8}_{-1.2} 10^{-5}$ & $0.6^{+0.4}_{-0.5}$ & $2.3^{+1.8}_{-1.1}$ & $-4.1^{+0.5}_{-0.6}$ & $0.058^{+0.004}_{-0.007}$ & $0.071^{+0.008}_{-0.004}$ \\
& 300 & $12.6^{+11.2}_{-10.3}$ & $109.2^{+1.3}_{-1.1} 10^{-5}$ & $0.6^{+0.4}_{-0.5}$ & $2.2^{+1.6}_{-1.1}$ & $-4.2^{+0.5}_{-0.7}$ & $0.058^{+0.003}_{-0.006}$ & $0.070^{+0.007}_{-0.004}$ \\
& 400 & $7^{+12}_{-7}$ & $109^{+50}_{-2} 10^{-5}$ & $0.6^{+0.3}_{-0.5}$ & $2.0^{+2.1}_{-1.0}$ & $-4.3^{+0.6}_{-0.8}$ & $0.058^{+0.004}_{-0.011}$ & $0.070^{+0.007}_{-0.004}$ \\
&
\\[-4.5mm]
\hline 
&
\\[-4.mm]
\multirow{6}{*}{B4} 
& 20 & $13^{+11}_{-9}$ & $0.0011^{+0.0007}_{-0.0005}$ & $0.5^{+0.5}_{-0.4}$ & $1.4^{+1.1}_{-0.4}$ & $-3.0^{+0.3}_{-0.7}$ & $0.04^{+0.01}_{-0.01}$ & $0.0837^{+0.0007}_{-0.0197}$ \\
& 40 & $5^{+8}_{-5}$ & $0.0012^{+0.0009}_{-0.0005}$ & $0.5^{+0.5}_{-0.5}$ & $1.1^{+1.6}_{-0.5}$ & $-3.2^{+0.5}_{-0.9}$ & $0.03^{+0.02}_{-0.02}$ & $0.080^{+0.004}_{-0.027}$ \\
& 100 & $1^{+6}_{-2}$ & $0.0012^{+0.0011}_{-0.0005}$ & $0.4^{+0.6}_{-0.4}$ & $0.9^{+1.5}_{-0.6}$ & $-3.5^{+0.8}_{-1.0}$ & $0.03^{+0.03}_{-0.03}$ & $0.07^{+0.01}_{-0.03}$ \\
& 200 & $-1.3^{+4.1}_{-1.0}$ & $0.0013^{+0.0010}_{-0.0006}$ & $0.5^{+0.5}_{-0.4}$ & $0.7^{+1.6}_{-0.5}$ & $-3.7^{+0.9}_{-1.2}$ & $0.03^{+0.03}_{-0.02}$ & $0.07^{+0.02}_{-0.03}$ \\
& 300 & $-1.9^{+2.0}_{-0.7}$ & $0.0013^{+0.0009}_{-0.0007}$ & $0.5^{+0.5}_{-0.4}$ & $0.7^{+1.2}_{-0.5}$ & $-3.9^{+1.1}_{-1.1}$ & $0.03^{+0.03}_{-0.02}$ & $0.07^{+0.02}_{-0.03}$ \\
& 400 & $-2.0^{+1.2}_{-0.6}$ & $0.0014^{+0.0008}_{-0.0007}$ & $0.5^{+0.5}_{-0.4}$ & $0.6^{+1.2}_{-0.5}$ & $-4.1^{+1.2}_{-1.1}$ & $0.02^{+0.03}_{-0.02}$ & $0.06^{+0.02}_{-0.03}$ \\
\hline\hline

\end{tabular}}
\caption{Search results for the four benchmarks with the old likelihood. The first column denotes the median value and the symmetric 90\% credible interval of the log Bayes factor distribution. The remaining columns indicate the median values and the 90\% credible regions \cite{LIGOScientific:2013yzb} of the six search parameters from the overall posterior distributions.
}
\label{tab:BMresults_old}
\end{center}
\end{table}

\begin{figure}[!h]
	\centering
   \includegraphics[width=16cm]{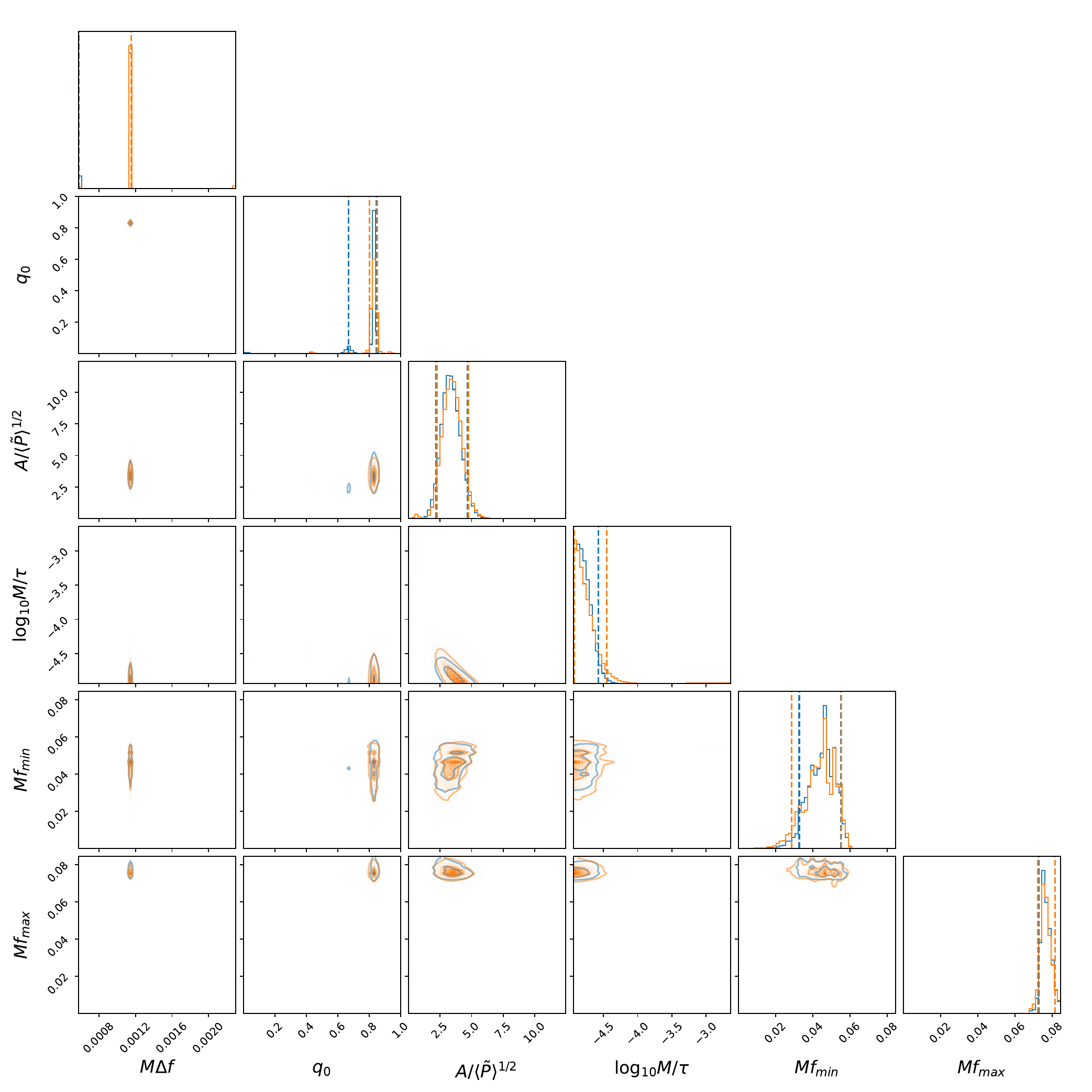} 
\caption{\label{fig:B1cornerAll} {
The corner plot for the overall posterior distributions of all six search parameters of the injected B1 model with $T\Delta f_{\rm max}=200$. The full prior ranges are shown. The contours in the off-diagonal panels denote the $1\sigma$ and $2\sigma$ ranges of the 2D posteriors. The dashed vertical lines in the diagonal panels denote the 68\% credible intervals.}
}
\end{figure}

To provide a comprehensive analysis, we also present additional results for the Bayesian search of the four benchmarks discussed in Sec.~\ref{sec:searchbenchmark}. The full search results for the new and old likelihoods are summarized in Tabs.~\ref{tab:BMresults_new} and \ref{tab:BMresults_old}, respectively. As a demonstration, Fig.~\ref{fig:B1cornerAll} shows the corner plot of all six search parameters for one case with a high detection probability. Both likelihoods allow us to accurately determine the spacing $\Delta f$, width $1/\tau$, relative shift $q_0$, and amplitude $A$ in general. The spacing, in particular, can be measured with very high precision because of the large number of QNMs captured. In contrast, the frequency band is less accurately determined due to its minor impact on the total SNR. 
Furthermore, we did not observe significant degeneracy among the six parameters, indicating their effectiveness in capturing the essential features of the dominant QNMs associated with echoes.
These general discussions are consistent with our previous results obtained using the old likelihood in ~\cite{Ren:2021xbe}.

\clearpage
\newpage

\bibliography{References_GWEphase}{}
\bibliographystyle{apsrev4-1}

\end{document}